\documentclass[sigconf]{acmart}
\usepackage{blindtext}
\usepackage{tikz}

\usepackage{soul}

\usepackage{dsfont}
\usepackage{enumitem}

\usepackage{times}
\renewcommand\footnotetextcopyrightpermission[1]{} 

\usepackage{booktabs} 
\usepackage{mdframed}
\usepackage{framed}

\usepackage{hyphenat}
\usepackage{fnpct}
\usepackage{algpseudocode}
\usepackage{amsmath,epsfig}
\usepackage[boxed,ruled,vlined,linesnumbered]{algorithm2e}
\usepackage{amsthm}
\usepackage{setspace}
\usepackage{enumitem}
\usepackage{multirow}
\usepackage{hhline}
\usepackage{caption}
\usepackage{subcaption}
\usepackage{graphicx}
\usepackage{wrapfig}
\usepackage{amsmath,epsfig}
\usepackage{amsthm}

\newtheoremstyle{mystyle}
  {}
  {}
  {\itshape}
  {}
  {\bfseries}
  {.}
  { }
  {}

\theoremstyle{mystyle}
\newtheorem{theorem}{Theorem}

\setcopyright{rightsretained}

\usepackage{balance}

\usepackage{csquotes}
\usepackage{enumitem}
\usepackage{tablefootnote}
\usepackage{fnpct}
\usepackage{bbm}
\usepackage{lipsum}
\usepackage{setspace}

\makeatletter
\def\thm@space@setup{\thm@preskip=0pt
\thm@postskip=0pt}
\makeatother

\usepackage{array}
\usepackage{makecell}

\usepackage{tikz}

\newcommand{\pie}[1]{%
  \begin{tikzpicture}
    \draw (0,0) circle (1ex);
    \fill (1ex,0) arc (0:#1:1ex) -- (0,0) -- cycle;
  \end{tikzpicture}%
}

\usepackage{tikz}
\usetikzlibrary{shapes}

\tikzset{
  filledCircle/.style={
    circle, draw, fill=black, minimum size=1em
  },
  emptyCircle/.style={
    circle, draw, fill=white, minimum size=1em
  },
  halfCircle/.style={
    path picture={
      \fill[black] (path picture bounding box.south west)
        rectangle (path picture bounding box.north);
    },
    circle, draw, fill=white, minimum size=1em
  }
}

\newcommand{\fcircle}{\tikz[baseline=-0.5ex]\node[filledCircle] {};}
\newcommand{\ecircle}{\tikz[baseline=-0.5ex]\node[emptyCircle] {};}
\newcommand{\hcircle}{\tikz[baseline=-0.5ex]\node[halfCircle] {};}

\DeclareRobustCommand{\fcircle}{\tikz[baseline=-0.5ex]\node[filledCircle] {};}
\DeclareRobustCommand{\hcircle}{\tikz[baseline=-0.5ex]\node[halfCircle] {};}
\DeclareRobustCommand{\ecircle}{\tikz[baseline=-0.5ex]\node[emptyCircle] {};}

\newcommand{\B}{\vspace*{-\smallskipamount}}
\newcommand{\BB}{\vspace*{-\medskipamount}}
\newcommand{\BBB}{\vspace*{-\bigskipamount}}

\usepackage{longtable}
\usepackage{mdframed}
\usepackage{colortbl}

\usepackage[hang,flushmargin]{footmisc}

\usepackage{natbib}
\newcommand*\wrapletters[1]{\wr@pletters#1\@nil}
\def\wr@pletters#1#2\@nil{#1\allowbreak\if&#2&\else\wr@pletters#2\@nil\fi}
\usepackage{tikz}
\usetikzlibrary{automata,matrix,shapes,arrows,positioning,chains,calc}
\usetikzlibrary{snakes}
\usetikzlibrary{arrows,scopes}
\usetikzlibrary{positioning,chains,fit,shapes,calc}
\usepackage{caption}
\usepackage{subcaption}
\usepackage{amsmath}
\usepackage{lineno}

\definecolor{lightyellow}{HTML}{FFFFFF}
\sethlcolor{lightyellow}

\definecolor{myblue}{HTML}
{BF40BF}

\mdfdefinestyle{MyFrame}{%
    linecolor=white,
    outerlinewidth=2pt,
    innertopmargin=4pt,
    innerbottommargin=4pt,
    innerrightmargin=4pt,
    innerleftmargin=4pt,
        leftmargin = 4pt,
        rightmargin = 4pt,
        backgroundcolor=yellow!15!white}

\begin{document}

\setlength{\belowdisplayskip}{0pt}
\setlength{\belowdisplayshortskip}{0pt}
\setlength{\abovedisplayskip}{0pt}
\setlength{\abovedisplayshortskip}{0pt}

\onecolumn
\fancyhead{}
 \title{
\textsc{Doc}$^\star$: Access Control for Information-Theoretically Secure Key-Document Stores
}

\author{Yin Li,$^1$ Sharad Mehrota$^2$, Shantanu Sharma$^3$, and Komal Kumari$^3$ \\ }
\affiliation{\institution{$^1$Dongguan University of Technology, China. $^2$UC, Irvine, USA. $^3$New Jersey Institute of Technology, USA. \\}} 

\setlength{\belowdisplayskip}{0pt}
\setlength{\belowdisplayshortskip}{0pt}
\setlength{\abovedisplayskip}{0pt}
\setlength{\abovedisplayshortskip}{0pt}

\begin{abstract}

This paper\footnote{An extended abstract of this version has been accepted in VLDB 2025.} presents a novel key-based access control technique for secure outsourcing key-value stores where values correspond to documents that are indexed and accessed using keys.   
The proposed approach adopts Shamir's secret-sharing that offers unconditional or information-theoretic security. It supports keyword-based document retrieval while preventing leakage of the data, access rights of users, or the size (\textit{i}.\textit{e}., volume of the output that satisfies a query). 
The proposed approach allows servers to detect (and abort)  malicious clients from gaining unauthorized access to data, and prevents malicious servers from altering data undetected
while ensuring efficient access -- it takes 231.5ms over 5,000 keywords across 500,000 files.

\end{abstract}

 \maketitle
\medskip
\section{Introduction}
\label{sec:Introduction}
While secure data outsourcing and query over encrypted data has been widely studied over the past two decades~\cite{DBLP:conf/sp/SongWP00,DBLP:conf/sigmod/HacigumusILM02}, the problem supporting access control over ciphertext has received little attention. 
This paper focuses on access control in the context of key-document (KD) stores --- a type of KV store wherein the value corresponds to a document. Such a document may contain additional keys that may also be used to index the document, \textit{e}.\textit{g}., a document (value) may consist of a doctor's note, and it may be indexed based on tags extracted from the text in the note. 
Such KD stores often store data in the form of an inverted index of doc-ids of documents associated with the key. As in regular KV stores, access to documents is based on keys.
Popular KV stores such as Redis~\cite{redis} allow storage of KD stores in addition to regular KV pairs. 

We consider the problem of outsourcing when the inverted index of doc-ids/file-ids based on keywords, 
 the set of documents/files, and the access control rights are outsourced in ciphertext to the cloud. A query for a keyword $k$ over the outsourced database would retrieve all documents containing $k$ for which the user has been granted access rights by the database owner (DBO).

In our model, neither the cloud nor the clients are trusted.  Clients access only data to which they have access rights. 
The technique ensures that the cloud does not learn cleartext data, the client queries, or which clients have access rights to which data.

\medskip
\medskip
\noindent
\textbf{Access Control in Key-Value Stores.} 
\label{subsec:Access Control in Key-Value Stores}
Access control in regular KV stores can be specified at either the \emph{\textbf{record-level}} or at the \emph{\textbf{key-level}}. At the record-level, a DBO specifies who can have what type of 
access to which record (\textit{i}.\textit{e}., a KV pair).  In contrast, at the key-level, 
if a client is allowed/denied access to a key $k$, then the client is allowed/denied access to all KV pairs for the key $k$. 
In key document (KD) stores, as in regular KV database, access control can be specified at the \emph{\textbf{key-level}} (a client is allowed/denied access to documents containing a given key) or at the \emph{\textbf{document-level}} when users are explicitly allowed/denied access to certain documents.

While record/document-level access control is more popular, systems such as Redis support key-level access control as well~\cite{redis_access}, since key-based access can often be much easier to specify, easier to implement, and scales well with a large number of documents and a large number of clients. Consider, for example, a DBO with hundreds of thousands of documents. It is often easy for DBO to define access policies based on a limited set of keywords, \textit{e}.\textit{g}., permit Alice to access all documents containing the keyword ``Urgent'' but not those containing ``Finance.''  Also, for the system perspective, it becomes easier to check user's access rights for a key compared to checking access-rights of many files during query execution. Specifying and managing policies at the document-level, especially, when there are many clients, becomes difficult.

Key-level access control, nonetheless, adds complexity since now if a client has access to a keyword $k_1$ and not $k_2$, then a document $d$ containing both $k_1$ and $k_2$ must not be returned, while documents containing $k_1$ but not $k_2$ should be returned.\footnote{\scriptsize {
\textsc{Doc}$^\star$ uses a policy model that permits only explicitly allowed actions, due to its stronger security guarantees as discussed in~\cite{DBLP:journals/pieee/SaltzerS75},  
ensuring that if DBO accidentally omits an access permission, the default is set to be denial, thereby preventing unauthorized access. Although such a denial might be inconvenient for the client, it does not compromise the system's security. }}

At first glance, one could convert a key-level access control specification to a document-level representation for which solutions have been explored in the literature~\cite{metal,Titanium}. However, such a conversion 
 is challenging when there are a large number of documents and clients, and, furthermore, the access control policies may dynamically change.  
 For instance, say a DBO wishes to change Alice's policy to allow temporary access to documents containing the keyword ``Finance'' in addition to those containing ``Urgent.'' If access control implementation was achieved through a document-level control, then DBO will need to determine a set of outsourced documents that contain the keyword (possibly tens of thousands), update the access control of these documents individually, and have to change the policies again when the temporary access is to be revoked. In contrast, with key-level access control, DBO would only need to update the policy for the relevant keywords, making the process significantly more efficient.

\begin{figure*}
\BBB\BBB\BBB
\begin{center}
\includegraphics[width=17cm,height=3.4cm]{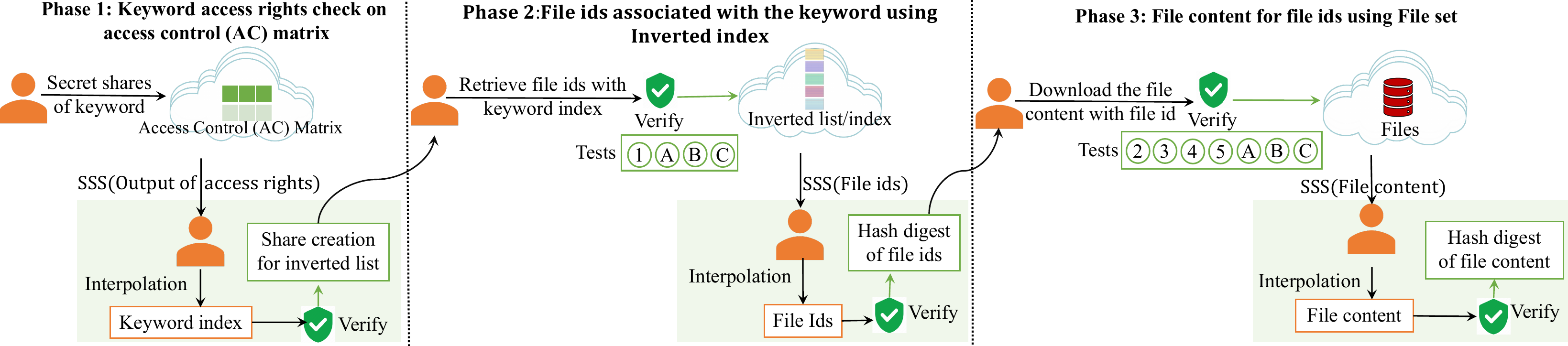}
\BBB
 \caption{\textbf{Execution of three phases of \textsc{Doc}$\boldsymbol{^\star}$.}}
 \label{figs:high-level_overview}
\end{center}
\BBB
\end{figure*}

\medskip
\medskip
\noindent
\noindent
\textbf{\textsc{Doc}$^\star$: A Key-based Access Control Mechanism.} 
This paper focuses on the key-level access control in the context of key-document (KD) stores.  Other types of access control, \textit{e}.\textit{g}., document-level access control in KD stores, or record-level access control in KV stores are relatively simpler and have also been 
studied in the previous literature~\cite{BigSecret,cong,metal} and the (simpler-version of the) solution we develop can also be applied to the key-level access control in KV stores. 

We develop an access control mechanism, entitled \textsc{Doc}$^\star$ ({where $\star$ refers to STorage with Access control and secuRity})
when secret-sharing is used to outsource data. Secret-sharing, unlike encryption mechanisms
which are computationally secure, is unconditionally or \emph{information-theoretically secure}, regardless of the adversary's computational capabilities. 
Several additional benefits of  secret-sharing (over encryption-based techniques) are well recognized in the literature: 
\emph{distributed trust} (moving trust from a single server to multiple servers) to avoid insider attacks~\cite{insiderattack1,insiderattack2}; 
\emph{computational efficiency} to perform addition and multiplication on ciphertext; 
avoiding the risk of a single point of failure~\cite{myspace}; 
avoiding the risk of data theft~\cite{datastolen1,datastolen2}; and 
tolerating malicious servers. 
Secret-sharing has gained popularity for data processing~\cite{DBLP:journals/isci/EmekciMAA14,janapdas,DBLP:journals/cj/ArcherBLKNPSW18,DBLP:conf/eurosys/VolgushevSGVLB19,smcql} and multi-party computation~\cite{DBLP:conf/stoc/CanettiFGN96,DBLP:conf/tcc/DamgardFKNT06,DBLP:conf/icisc/HamadaKICT12,DBLP:conf/nordsec/BogdanovLT14,DBLP:conf/fc/MakriRVW21} by both academia and industries. 
While we develop our access control mechanism using secret-sharing for outsourcing, homomorphic encryption~\cite{seal}, which also offers addition and multiplication over ciphertexts, can also be used. 

\begin{mdframed}[style=MyFrame,nobreak=true,align=center]
 \noindent
 {\textbf{{Code of the paper is  available in~\cite{fullversions2d}}}}. Appendices provide: 
(\textit{i})~dynamic operations--adding/deleting new files with new/existing keywords, 
(\textit{ii})~access grant/revocation, 
(\textit{iii})~methods for enhancing Phase~3, 
(\textit{iv})~client-side result verification methods, 
(\textit{v})~formal security proofs with proofs of theorems given in this paper, 
(\textit{vi})~degree reduction method, and 
(\textit{vii})~six additional experiments.
\end{mdframed}

\medskip
\section{Preliminary} 
\label{subsec:Preliminary}
This section overviews our model, secret-sharing, and the desired security requirements.

\setlength{\columnsep}{6pt}

\subsection{Model}
\label{subsec:model}
Our model 
consists of three entities: 

\medskip
\medskip
\noindent
(\textit{i}) {\textbf{Database Owner (DBO)}}: outsources ciphertext data along with access rights in ciphertext form to the cloud. DBO defines access rights for different users based on keywords and outsources them in the form of an \textbf{\emph{Access Control (AC) matrix}} of $\alpha\times \beta$, where $\alpha$ is the number of clients and $\beta$ is the number of keywords. A row of AC matrix represents a capability list for a client.
Additionally, DBO outsources an \emph{\textbf{inverted index/list}}  with $\beta$ rows, one corresponding to each keyword $k_i$ containing doc-ids/file-ids for documents/files containing $k_i$. AC matrix, inverted index/list, and documents/files are outsourced using secret-sharing to a (set of) servers.

\medskip
\medskip
\noindent
(\textit{ii}) \textbf{Servers}. A set of $c>1$ servers stores data and executes queries. Particularly, an $i^{\mathit{th}}$ server stores the $i^{\mathit{th}}$ shares of \emph{{AC matrix, inverted index, and documents}}. Upon receiving a query keyword $k$ from a client, the servers return (a subset of) documents in which $k$ appears that do not contain any keyword $k_{\!j}$ for which 
the client does not have access rights. A minority (see next subsection for the details on this) of the servers can also be malicious and may interfere with the execution of the protocol to prevent correct execution. Malicious servers may collude amongst themselves and/or with the clients. 

\medskip
\medskip
\noindent
(\textit{iii}) \textbf{Clients}: send queries to retrieve all documents/files associated with a specified keyword. Clients can be honest or malicious. An honest client follows the protocol correctly. A malicious client can try to download documents/files/data to which they may not have access. Malicious clients can collude with malicious servers.

\medskip
\subsection{Shamir's Secret-Sharing (SSS)}
\label{subsec:SSS}
We use 
SSS~\cite{DBLP:journals/cacm/Shamir79}. 
Let $S$ be a secret. Let $p$ be a prime number. Let $\mathbb{F}_p$ be a finite field of order $p$. 
SSS requires a secret owner to distribute $S$ into $c{>}{c^{\prime}}$ shares by randomly selecting a polynomial of degree $c^{\prime}$ with $c^{\prime}$ random coefficients, s.t. 
$f(x){=} S{+}a_1x{+}\cdots{+}a_{c^{\prime}}x^{c^{\prime}}$, where $f(x){\in} \mathbb{F}_p[x]$ and $a_i{\in} \mathbb{N}$ ($1{\leq} i{\leq} c^{\prime}$). 
Secret $S$ is distributed into $c{>}{c^{\prime}}$ shares, by computing $f(x)$ for $x{=}1,2,\ldots, c$. An $i^{\mathit{th}}$ share is placed at the $i^{\mathit{th}}$ server. $S$ can be reconstructed using Lagrange interpolation~\cite{corless2013graduate} over any $c^{\prime}{+}1$ shares. An adversary can reconstruct $S$, iff they collude with $c^{\prime}{+}1$ servers. Thus, by choosing $c^{\prime}$ as the degree of the polynomial,  the scheme remains secure even if $c^{\prime}$ servers can collude. 
Further, $3c^{\prime}{+}1$ shares are needed to tolerate $c^{\prime}$ malicious servers and ensure error detection and correction (as proved in~\cite{DBLP:conf/stoc/Ben-OrGW88}). 

SSS is \textit{additive homomorphic}~\cite{DBLP:journals/iacr/AsharovL11}: 
the sum of shares at servers produces the sum of cleartext value after interpolation. 
{\color{black} SSS is, also, \textit{multiplicative homomorphic}~\cite{DBLP:journals/iacr/AsharovL11}: servers can locally multiply shares, and the result can be constructed at the \emph{owner} if having enough shares, as each multiplication increases the polynomials' degree. The multiplication result reconstruction, on a different entity that does not possess the original secret shares, requires an additional degree reduction step at the server, which comes with additional cost.} We use terms `\textit{multiplicative secret-sharing/ multiplicative shares}' and `\textit{Shamir's secret-sharing/  secret-shares}' interchangeably.
\footnote{\scriptsize While SSS requires more than one server, 
this assumption is supported by various factors, including economic incentivization (as collusion goes against their financial interests), legal regulations, 
and jurisdictional boundaries. Such servers can be selected on different clouds, making the assumption more realistic. The proliferation of independent cloud vendors over the years has led organizations to adopt multi-cloud solutions to mitigate risks, \textit{e}.\textit{g}., vendor lock-in, and to enhance fault-tolerance~\cite{multicloud1,multicloud2,ibmmulticloud,multicloud4,multicloud5}. Leveraging multi-cloud environments enables organizations to outsource shares more effectively.}

\medskip
\subsection{Security Requirements }
\label{subsec:Trust Assumptions and Security Properties}
We list the security requirements 
for our problem setting below: 

\noindent
(\textit{i})~\emph{\textbf{Confidentiality:}} Servers (even if servers collude) must never
learn cleartext data outsourced by DBO, access control rights given by DBO to a client, or the query keyword the client wishes to retrieve.

\medskip
\noindent
(\textit{ii})~\emph{\textbf{Read obliviousness.}} Servers must not be able to distinguish between two or more queries based on which data is returned to clients, \textit{i}.\textit{e}., access-patterns/identity of the object is hidden from servers. 

\medskip
\noindent
(\textit{iii})~\emph{\textbf{Restricted access to the client.}} The client cannot fetch any documents that contain a keyword to which they do not have access. Note that such documents may contain other keywords for which a client has access. Nonetheless, the presence of a restrictive keyword should prevent client from gaining access to such a document.

\medskip
\section{High-level Overview of \textsc{Doc}$^\star$}
\label{subsec:doc_overview}
A client's query is processed in three phases. 
In the first phase, the servers, given a query for a keyword, check AC matrix and return to the client a vector of size $\beta$ with one number for each keyword. The returned vector contains a zero (in SSS form) if the client has access to the keyword, else a random value. In the second phase, the client can then use the index of the returned value zero to pose a query to retrieve all file-ids containing the keyword. Finally, based on the file-ids, the client fetches the files in the third phase. We illustrate \textsc{Doc}$^\star$  strategy using an example.

\medskip
\medskip
\noindent
\textbf{Example of \textsc{Doc}$^\star$.} Consider the following files/documents:

{\begin{center}
1: The King of Torts is a suspense novel written by John Grisham.

2: Stephen King is known as the King of Horror.

3: $\ldots$
\end{center}}

A DBO first constructs \emph{\textbf{access control (AC) matrix}} by creating columns with \emph{searchable keywords} -- the keywords using which a client can search for files (words such as  ``a,'' ``is,'' and ``hello'' are removed since clients do not search based on such words). 
AC matrix in \textsc{Doc}$^\star$ is a $\alpha\times \beta$ matrix, where $\alpha$ corresponds to the number of clients/{\color{black}groups in an organization} and $\beta$ is the number of keywords. A row of AC matrix represents a capability list for a client. 
Suppose, two (searchable) keywords are King and Horror. DBO creates capability lists for a client, Alice, see below, where $\mathbb{M}(\ast)$ denotes a secret-share. This shows Alice is allowed to search for the keyword King (indicated by $\mathbb{M}(0)$), not Horror (indicated by a random number). All values except the client name in AC matrix are secured using SSS, (the method is explained below~{\color{blue}\S\ref{sec:Data Outsourcing}}). 
Second, DBO outsources an \emph{\textbf{inverted list/index}} using SSS, see below, where the gray part is not outsourced to the cloud and is written to show the row corresponding to the keyword. Finally, DBO outsources all files using SSS. 


\begin{minipage}[h]{0.55\linewidth}
\centering
\scriptsize
 \begin{tabular}{|l|l|l|l|l|l|@{}p{0.29cm}|@{}p{0.29cm}|p{0.3cm}|p{0.5cm}|l|l|}    \hline
   Keywords $\rightarrow$ & $\mathbb{M}$(King) &  $\mathbb{M}$(Horror) & $\ldots$ \\ \hline\hline
   Alice & $\mathbb{M}$(0) & 
   $\mathbb{M}$(19) &$\ldots$ \\\hline
   
 $\vdots$ & $\vdots$ & $\vdots$ & $\vdots$\\\hline
  \end{tabular}
\captionof*{table}{Access control (AC) matrix.}
\BBB\BB
\end{minipage}
\begin{minipage}[h]{0.40\linewidth}
\centering
\scriptsize
 \begin{tabular}{|l||l|l|p{.7cm}|@{}p{2.4cm}|p{0.18cm}|p{0.18cm}|p{0.23cm}|p{0.5cm}|l|l|}    \hline
 {\cellcolor[HTML]{B8B8B8}{Positions/Row-id}}  & File-ids \\\hline\hline
  
{\cellcolor[HTML]{B8B8B8}{1 (King)}} & $\mathbb{M}$(1), $\mathbb{M}$(2) \\\hline
 
{\cellcolor[HTML]{B8B8B8}{ 2 (Horror)}} & $\mathbb{M}$(2) \\\hline

{\cellcolor[HTML]{B8B8B8}{ $\vdots$}} & $\vdots$ \\\hline
   
\end{tabular}
\captionof*{table}{Inverted index.}
\BB
\end{minipage}

Suppose, Alice searches for the keyword Horror. \textsc{Doc}$^\star$ will execute the following three phases/rounds (as depicted in {\color{blue}Figure~\ref{figs:high-level_overview}}). 

\medskip
\noindent
\textbf{\textit{Phase~1}:} Alice will send to servers, the keyword Horror using SSS, 
servers will perform a computation \emph{obliviously} (\textit{i}.\textit{e}., without knowing access-patterns -- memory address/identity of the object) over the capability row of Alice in AC matrix and return a vector of size equal to the number of keywords in AC matrix. 
Alice interpolates the values and obtains all random numbers, showing either she does not have access to the keyword Horror or the keyword Horror is not present, and then, she terminates the protocol. 
Now, suppose, Alice searches for the keyword King, then she will obtain a vector as $\langle$0, random number, $\ldots\rangle$, after interpolation, where the position of zero refers to the row/index of the inverted index, containing all file-ids associated with the keyword King. 

\medskip
\noindent
\textbf{\emph{Phase~2}:} Alice, to fetch the desired row of the inverted list, sends a vector $\langle1,0,\ldots,0\rangle$ of size $\beta$ (the number of rows in the inverted list) with all positions zeros and only the desired position with one using SSS. {\color{black} Servers verify the correctness of the vector (\textit{i}.\textit{e}., containing all zeros except a single one at the desired position), then perform a dot product between the vector} and the inverted index and return the result. 
Then, Alice learns file-ids~1 and~2, after interpolation.

\medskip
\noindent
\textbf{\emph{Phase~3}:} Alice tries to fetch files~1 and~2. The servers will obliviously return only file~1, not file~2, due to having the keyword Horror, which she is not allowed to search. 
At the end of Phase~3, Alice learns file~1 in cleartext after interpolation, not file~2, indicating that file~2 contains at least one keyword without search permission to her.~$\blacksquare$

\medskip
\noindent
\textbf{Information leakage.} 
In an ideal setting, a client should be able to retrieve a file containing the queried keyword only if that file does not include any keyword to which the client is denied search access. Also, the client should not learn any extra information, such as the presence or absence of keywords in AC matrix, the queries made by other clients, or the number of retrievable files relative to the total number of files that match the queried keyword. \textsc{Doc}$^\star$ prevents all such information leakage, with one exception during Phase~3. 

{\color{black}
In Phase~3, if a file is not retrieved, the client can infer that it contains query keyword (since its file-ids were retrieved in Phase~2) and includes at least one keyword to which client is denied access (since the file itself was not returned). Although the client cannot access file's content, they may attempt multiple queries across all keywords to infer the set of keywords associated with that file. If the client knows and has access to all the keywords, except one $k_{\!j}$, this will reveal that file contains $k_{\!j}$. In all other scenarios, client cannot determine which keywords with disallowed access appear in the file.

Such a multi-query leakage, when the client knows and has access to all keywords but one, can be prevented by randomizing the file-ids returned in Phase~2, for example, by adding to file-ids a random number derived from a query-specific seed or by padding with fake file-ids (which are excluded in Phase~3). These techniques can help prevent the client from inferring the presence of restricted keywords in the file.

}

\medskip
\section{A Baseline \& Challenges} 
\label{subsec:A Baseline Solution}
We establish a \textbf{baseline solution} for secure key-based access control over secret-shared data to identify the specific challenges that \textsc{Doc}$^\star$ needs to address and to compare performance against \textsc{Doc}$^\star$. The baseline offers weaker security, as it assumes the clients to be honest that \textsc{Doc}$^\star$ does not. As such, this baseline can be viewed as being unfair to \textsc{Doc}$^\star$ given \textsc{Doc}$^\star$ stronger security. Nonetheless, it suffices since, as the experiments will show,  \textsc{Doc}$^\star$ significantly outperforms the baseline. We choose such a baseline since it is easy to implement using state-of-the-art secret-sharing tools MP-SPDZ~\cite{mp-spdz}.

In the baseline solution, we store AC matrix using Shamir's secret-sharing (SSS) 
and execute state-of-the-art secret-sharing-based MP-SPDZ~\cite{mp-spdz} to check the client's access rights for searching a query keyword. On success, the servers return all the doc-ids associated with the query keyword to the client using the {{inverted index}}. In the baseline: 
(\textit{i})~{AC matrix} is created for a single client where columns correspond to one of 5,000 keywords and cell values correspond to the access rights of the client, and 
(\textit{ii})~inverted index contains 5,000 rows, one for every 5,000 keywords in the same order as they appear in AC matrix, and each row contains all doc-ids associated with the keyword. While \textsc{Doc}$^\star$ will fetch secret-shared files also, we leave this step aside from the baseline solution. This baseline solution to answer a client request took 24.23 seconds. 
The  reasons of this inefficiency 
will be discussed in Challenge~1 below. 

\bgroup
\def\arraystretch{1}
\begin{table}[!t]
\BB
  \centering
  \scriptsize
  \begin{tabular}{|l|l|}\hline
  \textbf{Notations} & \textbf{Meaning} \\\hline

    $\alpha$ & \# clients or the group allowed to perform search operation \\\hline
    $\beta$ & \# unique keywords across all files/documents \\\hline
    $\gamma$ & The maximum number of files associated with a keyword (known to every entity)\\\hline
    $\delta$ & \# files/documents ($\delta\geq \gamma$) \\\hline
    $p$ & A prime number used as modulo in secret-sharing\\\hline
    $\mathbb{M}(x)$ & Multiplicative shares or Shamir's secret-share (SSS) of $x$ \\\hline
$\mathit{sw}$ & A keyword at a server in AC matrix\\ \hline
$\mathit{uw}$ & A keyword at a client \\ \hline
$A \odot B$ & Dot product between $A$ and $B$, where $A\odot B=\sum_{0 < i< {n+1}} A[i]\times B[i]$ \\\hline 
$\mathbb{M}(\mathit{AC})[i]$ & The $i^{\mathit{th}}$ value of access control matrix, denoted by AC \\\hline
$ACT\_pos$ & The third row having hash digest of the positions in AC matrix
\\\hline
     \end{tabular}
  \caption{Frequently used notations in this paper.}
  \label{tab:notations}
 \BBB\BBB
\end{table}
\egroup

{\color{black}
To understand the summary of operations supported by \textsc{Doc}$^\star$, we classify them in terms of the challenges we addressed. Later sections will provide details of all these operations.}

\medskip\medskip
\noindent\textbf{Challenge~1: Efficient query processing.} The baseline solution took 24.23s, where the offline/preprocessing and online/computation phases of multi-party computation (MPC)-based processing took 19.77s and 4.46s. Let us focus only on the online phase {\color{black}of MPC}. For checking access rights, MP-SPDZ~\cite{mp-spdz} uses the equality test of~\cite{DBLP:conf/PKC/NishideO07}. The online phase took $2{\times}8$ rounds of communication among servers, where 8 rounds were used to check the keyword and another 8 rounds were used to check the access right. It incurred $\approx$50MB dataflow among servers, while the size of ACT matrix was only 157KB. The reason of the overwhelming dataflow and the number of rounds is as follows: In each online round, $81\ell$ bits flow among servers for the multiplication protocol involved in the equality check, where $\ell$ is the number of bits. Thus, to check access right over 5,000 keywords that are represented as 64 bits each, it requires $2{\times}8{\times}81{\times}64{\times}5000\approx$51.84MB. Finally, a dot product between the output of the access check and the inverted index is performed, resulting in all the doc-ids containing the keyword. 

\textbf{Our solution ({\color{blue}\S\ref{subsec:phase1}}).} \textsc{Doc}$^\star$ develops an 
\textbf{efficient} 
protocol for obliviously (\textit{i}.\textit{e}., without revealing the identity of the keyword in AC matrix) checking the access rights of the client and returning the doc-ids. 
To bring efficiency for the \emph{same operation}, our protocols \emph{do not need} communication among servers if clients are assumed to be trusted or use two rounds of communication, each transmitting a few integers, when clients are not assumed to be trusted, to obliviously verify clients' behavior at the server. This makes our protocol at least 45 times faster than MP-SPDZ --- while the online phase in the baseline took 4.46s, our new method took only 95ms (16ms for checking the access rights and 79ms for performing the dot product).

\medskip
\medskip
\noindent
\textbf{Challenge~2: Obliviously returning only files having no keywords for which access is denied.} On receiving the file-ids, the client retrieves the file content. But now, the client may behave maliciously and try to fetch any file. Further, the file may contain a keyword for which the client does not have search access, and such a file should not be returned by the server. Of course, MP-SPDZ can address this; however, it will incur significant computational overhead.

\textbf{Our solution ({\color{blue}\S\ref{subsec:phase3}}).} \textsc{Doc}$^\star$ develops a \textbf{verifiable, oblivious} file retrieval protocol. 
To fetch a file, the client creates a vector containing all zeros except for one at the desired position. {\color{black}We develop oblivious algorithms to verify the correctness of the vector at the server and oblivious algorithms to ensure the file has no keyword to which access is disallowed.} 
The entire process does not reveal to  \emph{{servers which file the client is fetching and whether the file contains a keyword to which the client does not have access rights or not}} by always returning a file (real/dummy).  

\medskip
\medskip
\noindent
\textbf{Challenge~3: Randomization of polynomials after multiplication --- producing irreducible polynomials.} 
When servers perform multiplication on the secret-shared data and the query, the resulting polynomial becomes reducible (\textit{i}.\textit{e}., not fully random, because such a polynomial can be factorized and potentially found by exhaustive search), which may, hence, reveal some information about the secret-shared data to the client. The famous BGW protocol~\cite{DBLP:conf/stoc/Ben-OrGW88} can make the polynomial irreducible, with a high cost among the servers due to the communications required by interpolation and resharing.


\noindent
\textbf{Our solution ({\color{blue}\S\ref{subsec:Random Number Generation and Verification}} and  {\color{blue}\S\ref{subsubsec:algo in phase1}}).} \textsc{Doc}$^\star$ develops a secure and computationally efficient method compared to BGW to make polynomials irreducible and fully randomized by eliminating the need for interpolation and resharing among servers for randomization. {\color{black} For this, distributed secret-shared random numbers are generated at the server before the query execution and used to randomize a polynomial.}

\medskip
\medskip
 \noindent
 \textbf{Challenge~4: Dealing with malicious clients colluding with a minority of the servers.} MP-SPDZ, when using SSS, does not deal with a case when malicious clients can collude with a minority of the (malicious) servers. This collusion can reveal additional information to the client, \textit{e}.\textit{g}., nullifying the impact of making the polynomial irreducible, as discussed above. 

 \textbf{Our solution.} \textsc{Doc}$^\star$ protocols deal with malicious clients colluding with a minority of the servers, making it impossible for the client to deduce any information about the secret. For example, the \emph{{presence of a minority of the servers colluding with malicious clients does not impact the process of randomization of polynomials}}. {\color{black}In \textsc{Doc}$^\star$, the client sends a one-hot bit vector in shared form to the servers. However, a malicious client may create incorrect vectors by either placing one at the wrong places or having non-binary values. To detect such malicious behavior of clients by the server, \textsc{Doc}$^\star$  
 develops three Tests:~A,~B, and~C ({\color{blue}\S\ref{Sec:Verification of the Client's Vector}}).}

\begin{figure*}
\BBB\BBB\BBB
\hspace{-2cm}
\begin{minipage}[t]{0.17\linewidth} 
\centering
\scriptsize
 \begin{tabular}{|p{0.65cm}|p{1.05cm}|p{0.7cm}|p{0.70cm}|p{0.5cm}|p{0.5cm}|}   \hline
    File-ids & File content \\ \hline\hline
    1 & How are you\\ \hline
    2 & Are you Ana\\ \hline
    3 & Fig is a fruit \\ \hline
  \end{tabular}
\captionof{table}{Cleartext files.}
\label{tab:cleartext_files}
\end{minipage}
\begin{minipage}[t]{0.287\linewidth}
\centering
\scriptsize
 \begin{tabular}{|p{2.77cm}|@{}p{0.43cm}|@{}p{0.43cm}|@{}p{0.29cm}|p{0.3cm}|p{0.5cm}|l|l|}    \hline
   Keywords $\rightarrow$ & Are &  Ana & Fig & Fake\\ \hline\hline

  {\cellcolor[HTML]{B8B8B8}{Starting address in inverted index }}  &  {\cellcolor[HTML]{B8B8B8}{1}} &  {\cellcolor[HTML]{B8B8B8}{2}} &  {\cellcolor[HTML]{B8B8B8}{3}} &  {\cellcolor[HTML]{B8B8B8}{0}}\\ \hline
 
   Hash digest $\rightarrow$  & {\cellcolor[HTML]{BAC8FE}{$\mathsf{H}(1)$}} & {\cellcolor[HTML]{BAC8FE}{$\mathsf{H}(2)$}} & {\cellcolor[HTML]{BAC8FE}{$\mathsf{H}(3)$}} & {\cellcolor[HTML]{BAC8FE}{$\mathsf{H}(0)$}}\\ \hline
   \multicolumn{5}{|l|}{Clients' information $\downarrow$   }
      \\ \hline
   Lisa & {\cellcolor[HTML]{FFFF33}{\textbf{0}}} & 
   {\cellcolor[HTML]{FFFF33}{\textbf{1}}} & {\cellcolor[HTML]{FFFF33}{\textbf{2}}} 
   & {\cellcolor[HTML]{FFFF33}{\textbf{0}}}\\\hline
   
   Ava  & {\cellcolor[HTML]{FFFF33}{\textbf{3}}} & 
   {\cellcolor[HTML]{FFFF33}{\textbf{0}}} & {\cellcolor[HTML]{FFFF33}{\textbf{0}}} & {\cellcolor[HTML]{FFFF33}{\textbf{0}}}\\\hline
  \end{tabular}
\captionof{table}{Access control (AC) matrix.}
\label{tab:ACT_table}
\BBB\BBB
\end{minipage}
\begin{minipage}[t]{0.25\linewidth}
\centering
\scriptsize
 \begin{tabular}{|p{.7cm}|@{}p{2.4cm}|p{0.18cm}|p{0.18cm}|p{0.23cm}|p{0.5cm}|l|l|}    \hline
   {\cellcolor[HTML]{B8B8B8}{Positions}}  & File-ids \\\hline\hline
  
 {\cellcolor[HTML]{B8B8B8}{1 (are)}} & 1, 2, 0,
 $\mathsf{H}(2,\mathsf{H}(1,\mathsf{H}(\textnormal{are})))$ \\\hline
 
  
   {\cellcolor[HTML]{B8B8B8}{2 (Ana)}} & 2, 0, 0, $\mathsf{H}(0,\mathsf{H}(2,\mathsf{H}(\textnormal{ana})))$ \\\hline
   
   {\cellcolor[HTML]{B8B8B8}{3 (Fig)}} & 3, 0, 0, $\mathsf{H}(3,\mathsf{H}(\textnormal{Fig}))$ \\\hline

   {\cellcolor[HTML]{B8B8B8}{0 (Fake)}} & 0, 0, 0, $\mathsf{H}(0,\mathsf{H}(\textnormal{Fake}))$ \\\hline
   
  \end{tabular}
\captionof{table}{Inverted index.}
\label{tab:inverted_list}
\BBB\BBB
\end{minipage}
\begin{minipage}[t]{0.22\linewidth}
\centering
\scriptsize
 \begin{tabular}{|p{0.64cm}|@{}p{1.30cm}|@{}p{3.15cm}|p{0.65cm}|p{0.70cm}|p{0.5cm}|p{0.5cm}|}   \hline
    {\cellcolor[HTML]{ffffff}{file\_ids}} & AP list & File content  with digest\\ \hline\hline
 {\cellcolor[HTML]{ffffff}{1}}&1,0,$\mathsf{H}(1)$ & How are you, $\mathsf{H}(\textnormal{how are you},\mathsf{H}(1))$\\ \hline

    {\cellcolor[HTML]{ffffff}{2}} & 1,2,$\mathsf{H}(1){+}\mathsf{H}(2)$ &Are you Ana, $\mathsf{H}(\textnormal{are you Ana},\mathsf{H}(2))$\\ \hline
    
 {\cellcolor[HTML]{ffffff}{3}} & 3,0,$\mathsf{H}(3)$ & Fig is a fruit, $\mathsf{H}(\textnormal{Fig is a fruit},\mathsf{H}(3))$ \\ \hline  
 
    {\cellcolor[HTML]{ffffff}{0}} & 0,0,$\mathsf{H}(0)$ & Dummy, $\mathsf{H}(\textnormal{Dummy},\mathsf{H}(0))$ \\\hline 

  \end{tabular}
\captionof{table}{Files.}
\label{tab:ss_file}
\BBB\BBB
\end{minipage}

\end{figure*}

\begin{figure*}
\BBB
\hspace{-2cm}
\begin{minipage}[t]{0.22\linewidth}
\centering
\scriptsize
\begin{tabular}{|@{}p{0.65cm}|p{0.7cm}|p{0.7cm}|p{0.7cm}|p{0.7cm}|p{0.5cm}|l|l|}    \hline




    Keywords  & 112816 &  112412 & 161918 \\ \hline\hline

   Hash  & {\cellcolor[HTML]{BAC8FE}{$\mathbb{M}(\mathsf{H}(1))$}} & {\cellcolor[HTML]{BAC8FE}{$\mathbb{M}(\mathsf{H}(2))$}} & {\cellcolor[HTML]{BAC8FE}{$\mathbb{M}(\mathsf{H}(3))$}}  \\ \hline
  
   Lisa & {\cellcolor[HTML]{FFFF33}{\textbf{1}}} & 
   {\cellcolor[HTML]{FFFF33}{\textbf{2}}} & {\cellcolor[HTML]{FFFF33}{\textbf{3}}} \\\hline
   
   Ava  & {\cellcolor[HTML]{FFFF33}{\textbf{4}}} & 
   {\cellcolor[HTML]{FFFF33}{\textbf{1}}} & {\cellcolor[HTML]{FFFF33}{\textbf{1}}} \\\hline
  \end{tabular}
\captionof{table}{Share1 of AC matrix.}
\label{tab:ACT_table_share1}
\BBB\BBB\BBB\BBB
\end{minipage}
\begin{minipage}[t]{0.24\linewidth}
\centering
\scriptsize
\begin{tabular}{|@{}p{0.65cm}|p{0.7cm}|p{0.7cm}|p{0.7cm}|p{0.7cm}|p{0.5cm}|l|l|}    \hline

    Keywords  & 112817 &  112413 & 161919 \\ \hline\hline 
     Hash  & {\cellcolor[HTML]{BAC8FE}{$\mathbb{M}(\mathsf{H}(1))$}} & {\cellcolor[HTML]{BAC8FE}{$\mathbb{M}(\mathsf{H}(2))$}} & {\cellcolor[HTML]{BAC8FE}{$\mathbb{M}(\mathsf{H}(3))$}}  \\ \hline
   Lisa & {\cellcolor[HTML]{FFFF33}{\textbf{2}}} & 
   {\cellcolor[HTML]{FFFF33}{\textbf{3}}} & {\cellcolor[HTML]{FFFF33}{\textbf{4}}} \\\hline
   
   Ava  & {\cellcolor[HTML]{FFFF33}{\textbf{5}}} &
   {\cellcolor[HTML]{FFFF33}{\textbf{2}}} & {\cellcolor[HTML]{FFFF33}{\textbf{2}}} \\\hline
  \end{tabular}
\captionof{table}{Share2 of AC matrix.}
\label{tab:ACT_table_share2}
\end{minipage}
\begin{minipage}[t]{0.22\linewidth}
\centering
\scriptsize
\begin{tabular}{|@{}p{0.65cm}|p{0.7cm}|p{0.7cm}|p{0.7cm}|p{0.7cm}|p{0.5cm}|l|l|}    \hline

    Keywords  & 112818 &  112414 & 161920 \\ \hline\hline 
   Hash  & {\cellcolor[HTML]{BAC8FE}{$\mathbb{M}(\mathsf{H}(1))$}} & {\cellcolor[HTML]{BAC8FE}{$\mathbb{M}(\mathsf{H}(2))$}} & {\cellcolor[HTML]{BAC8FE}{$\mathbb{M}(\mathsf{H}(3))$}}  \\ \hline
   Lisa & {\cellcolor[HTML]{FFFF33}{\textbf{3}}} & 
   {\cellcolor[HTML]{FFFF33}{\textbf{4}}} & {\cellcolor[HTML]{FFFF33}{\textbf{5}}} \\\hline
   
   Ava  & {\cellcolor[HTML]{FFFF33}{\textbf{6}}} &
   {\cellcolor[HTML]{FFFF33}{\textbf{3}}} & {\cellcolor[HTML]{FFFF33}{\textbf{3}}} \\\hline
  \end{tabular}
\captionof{table}{Share3 of AC matrix.}
\label{tab:ACT_table_share3}
\end{minipage}
\begin{minipage}[t]{0.1\linewidth}
\centering
\scriptsize
 \begin{tabular}{|p{1.05cm}|}    \hline
   File-ids \\\hline\hline
  
2, 3, $\mathbb{M}(x_1)$ \\\hline
 

3, 1, $\mathbb{M}(x_2)$ \\\hline

4, 1 $\mathbb{M}(x_3)$ \\\hline
  \end{tabular}
  \BB
  \captionsetup{width=3.4\columnwidth}
 \captionof{table}{Three shares of inverted index.}
 
\label{tab:share1 of IL}
\BBB\BBB\BBB\BBB
\end{minipage}
\begin{minipage}[t]{0.08\linewidth}
\centering
\scriptsize
 \begin{tabular}{|p{1.05cm}|p{0.18cm}|p{0.18cm}|p{0.23cm}|p{0.5cm}|l|l|}    \hline
   File-ids \\\hline\hline
  
3, 4, $\mathbb{M}(y_1)$ \\\hline
 

4, 2, $\mathbb{M}(y_2)$ \\\hline

5, 2 $\mathbb{M}(y_3)$ \\\hline
  \end{tabular}
 
\label{tab:share2 of IL}
\BBB\BBB\BBB\BBB
\end{minipage}
\begin{minipage}[t]{0.01\linewidth}
\centering
\scriptsize
 \begin{tabular}{|p{1.05cm}|p{0.18cm}|p{0.18cm}|p{0.23cm}|p{0.5cm}|l|l|}    \hline
   File-ids \\\hline\hline
  
4, 5, $\mathbb{M}(z_1)$ \\\hline
 

5, 3, $\mathbb{M}(z_2)$ \\\hline

6, 3, $\mathbb{M}(z_3)$ \\\hline
  \end{tabular}

\label{tab:share3 of IL}
\BBB\BBB\BBB
\end{minipage}
\BBB
\end{figure*}

\medskip
\section{Data Outsourcing in \textsc{Doc}$\boldsymbol{^\star}$}
\label{sec:Data Outsourcing}
This section develops a method for DBO to outsource a set of files using SSS to servers, by first, creating three data structures: an access control matrix, an inverted index/list, and a file set, and then, creating shares. 
Below, we explain the three data structures: 
\begin{enumerate}[nolistsep,noitemsep,leftmargin=0.01in]

\medskip\medskip
\item \emph{\textbf{Access-control (AC) matrix}:}  contains access control information for each keyword and each client/group of clients (\textit{e}.\textit{g}., groups can be CS, EE, and ME departments). Let $\alpha$ be the number of clients/groups. Let $\beta$ be the total number of (searchable) keywords in all the documents. AC matrix contains $\alpha{+}2$ rows and $\beta{+}2$ columns (see {\color{blue}Table~\ref{tab:ACT_table}}). 
  Each row corresponds to a client/group, and each column corresponds to a keyword. %
  Each row contains zero or a random number in each column, where a zero in the $(i,j)^{\mathit{th}}$ cell indicates that the client $i$ is allowed to search for the $j^{\mathit{th}}$ keyword, while a random number indicates otherwise. Random numbers are selected carefully, see the end of this subsection. 
  
One of the additional rows contains keywords, and another row contains the hash of the row-id corresponding to the keyword in the inverted index.\footnote{\scriptsize The hash digest will be used in Phase~3 to ensure that the client does not retrieve a file if it contain a keyword to which access is denied.} One of the additional columns contains 
clients/groups name, (or the client's provable identity that is sent with a query) in cleartext, and another column contains a fake keyword with allowed access to conceal volume (\textit{i}.\textit{e}., the count of the files) during Phase~3. 

\medskip\medskip  \item \emph{\textbf{Inverted list/index:}} contains, for each keyword, doc-ids/file-ids in which the keyword appears. 
For enabling \emph{verification by clients}, each row of the 
index includes a hash digest over the doc-ids associated with the keyword.\footnote{\scriptsize Hash digest can be computed by chaining the digests of the doc-ids and the keyword; \textit{e}.\textit{g}., the hash digest for a keyword appearing in files $f_1$ and $f_2$ would be: $\mathsf{H}(f_2,(\mathsf{H}(f_1,\mathsf{H}(\textnormal{keyword}))))$, where $\mathsf{H}$ is a secure hash function~\cite{dworkin2015sha}. This chaining process allows the addition/deletion of the doc-ids to/from the inverted index, without recomputing the hash digest for all file-ids associated with a keyword (explained in detail in~{\color{black}Appendix~\ref{app_subsubsec:Case 1: Adding Files having Existing keywords}} in~\cite{fullversions2d}).} To make each entry of the inverted index of an identical length, fake file-ids (say zero/random numbers greater than real doc-ids) are added  
(to reduce space and computation overheads by adding fake doc-ids,~{\color{blue}\S\ref{sec:Inverted Index Optimization}} develops a method). One row is added with a fake keyword and fake doc-ids to hide allowed/ denied access from servers; see the last paragraph on this issue in~{\color{blue}\S\ref{subsec:discussion_outsource}}.

\medskip
\medskip
\item \emph{\textbf{File (set)/Documents.}} Each file contains a file-id, the file content, a hash digest over the file-id and the content (enabling verification of the file content by the client), and an AP (absence/presence of keywords) list. The AP list prevents servers from sending a file containing at least one keyword to which the client has no access. Two possibilities for storing AP lists are: 
(\textit{i})~AP list is of size either $\beta$, each value with 0 or 1, showing the absence or presence of each of the $\beta$ keywords in the file, where $0$ refers absence; otherwise, 1.  
(\textit{ii})~AP list contains the column number of AC matrix corresponding to the keywords that appear in the file, along with the \emph{sum of the hash digest} of the positions (see {\color{blue}Table~\ref{tab:ss_file})}. 
The first alternative of AP list offers full security, while the second reveals the number of keywords appearing in a file to the client -- we will discuss this in~{\color{blue}\S\ref{subsubsec:Ensuring Correctness of the Vector}}. Finally, DBO adds a dummy/fake file that is used to hide the volume from servers during Phase~3 of file retrieval. 


\end{enumerate}

\medskip
\medskip
\noindent\textbf{Creating and outsourcing shares.} 
DBO generates \emph{\textbf{four multiplicative shares}} of the AC matrix, inverted list, and files using random \emph{\textbf{polynomials of degree one}}.\footnote{\scriptsize {\color{black}We selected polynomials of degree one under the assumption that no cloud server will collude, hence two shares are enough. However, we perform one multiplication at the server, so three shares are needed. The fourth share is used for client-side result verification purposes.}} DBO \emph{does not create shares of the client's name in the AC matrix}. Shares of number are created straightforwardly using SSS. To create shares of English keywords, they are converted to numbers (using letter positions or ASCII codes), equalized in length by adding a random number, and then multiplicative shares are generated. Each $i^{\mathit{th}}$ share is outsourced to the $\mathcal{S}_i$ server.

\medskip
\medskip
\noindent
{\textbf{Selecting random numbers for AC matrix.}} Suppose two keywords in AC matrix are `AB' and `ABC.' Their numerical representations could be `0102' and `010203,' respectively. Suppose only 26 numbers are needed. We can add any number greater than 26 and smaller than 100 to AB (\textit{e}.\textit{g}., 010299), making its length identical to ABC's length. 
Random numbers in the cells of AC matrix for denied access are always greater than the numerical representation of any string; \textit{e}.\textit{g}., if the largest string in numerical form is 2727 (\textit{i}.\textit{e}., ZZ), random numbers in any cell must be greater than 2727 (the reason will be clear in \textsc{Step} 2 of Phase 1 in~{\color{blue}\S\ref{subsec:phase1}}).

\medskip
\subsection{\textbf{Example of Data Outsourcing}}
\label{subsec:example_outsource}
\noindent
\textbf{Cleartext files and AC matrix.} 
{\color{blue}Table~\ref{tab:cleartext_files}} shows three files in cleartext. 
Suppose, the three files have three keywords: \texttt{are}, \texttt{ana}, \texttt{fig}. Based on these keywords, an AC matrix (see {\color{blue}Table~\ref{tab:ACT_table}}) is created for two clients, \texttt{Lisa} and \texttt{Ava}. The first row of AC matrix keeps the keywords. The second row (gray-colored) keeps the row-id of the inverted list in which the keywords appear. The third row (blue-colored) is for hash digest for those row-ids. 
Yellow-colored part shows the capability list of clients. $\langle 0,1,2,0\rangle$ 
indicates that \texttt{Lisa} has access to those files containing \texttt{are} and \texttt{fake} keywords, while \texttt{Ava} can access files containing \texttt{ana}, \texttt{fig}, and \texttt{fake} keywords.\footnote{\scriptsize {\color{black} \textsc{Doc}$^\star$ can handle multiple keywords connected by conjunctions and disjunctions by considering them a single composite keyword.}} {\textbf{\emph{{
For the purpose of simplicity, we put 1,2,3, as random numbers, in AC matrix}}}} and do not show shares of the fake keyword in {\color{blue}Tables~\ref{tab:ACT_table_share1},\ref{tab:ACT_table_share2},\ref{tab:ACT_table_share3}}. 
The gray part in the AC matrix indicates a connection with the inverted list and is not outsourced.

\medskip\medskip
\noindent
\textbf{Inverted index/list.}
{\color{blue}Table~\ref{tab:inverted_list}} shows the inverted index for the three keywords with the hash digest over the file-ids. A fake file-id 0 has been added to the keywords \texttt{Ana} and \texttt{Fig}, thereby all keywords have an identical number of file-ids. Another zero is added to all the keywords, and this zero shows an empty space for handling the insertion of the new files. 
DBO can add such zeros multiple times, depending on the insertion workload. A row with fake file-ids is also added. In the inverted list, the gray-colored part, which is written for the purpose of explanation, is \emph{\textbf{not}} outsourced. 

\medskip \medskip%
\noindent
\textbf{File data structure.}
{\color{blue}Table~\ref{tab:ss_file}} shows the file-ids, the content of files, the hash digest over the file content and its id, and AP list. The last row shows a dummy file-id with its dummy content. The fake keyword, fake file-id, and fake file do not need to be at the end of the data structures. They can be placed at any place. 

\medskip
\medskip%
\noindent\textbf{Share creation.} DBO creates SSS of AC matrix, inverted list, and the files. DBO represents keywords as: \texttt{are} as \texttt{11,28,15}, 
\texttt{ana} as \texttt{11,24,11}, and \texttt{fig} as \texttt{16,19,17}, and creates shares of such numbers. {\textbf{\textit{{For the purpose of simplicity, we select a single polynomial ($\boldsymbol{f(x)=(x{+}s){\bmod} p}$, where $\boldsymbol{p=500009}$, $\boldsymbol{s}$ is a secret, and $\boldsymbol{x{\in}\{1,2,3\}}$) to show the shares of AC matrix and inverted list.}}}} 
However, in real deployments and in our experiments, \textsc{Doc}$^\star$ selects different random polynomials for every secret.
{\color{blue}{Tables~\ref{tab:ACT_table_share1}}}, {\color{blue}{\ref{tab:ACT_table_share2}}}, and {\color{blue}{\ref{tab:ACT_table_share3}}} show three shares of AC matrix of {\color{blue}{Table~\ref{tab:ACT_table}}}. Three shares of the inverted list of {\color{blue}{Table~\ref{tab:inverted_list}}} are shown in {\color{blue}{Table~\ref{tab:share1 of IL}}}, where 
$\mathbb{M}(\ast)$, where $\ast{\in}\{x_i,y_i,z_i\}$ ($i$=1,2,3) indicates multiplicative shares/SSS of the hash digest. We do not show shares of fake items, and the remaining shares of AC matrix, inverted list, and files ({\color{blue}Table~\ref{tab:ss_file}}) can be created similarly, but are not shown here due to space limitations. 

\medskip
\subsection{\textbf{Discussion}}
\label{subsec:discussion_outsource}

\noindent
\textbf{Information leakages from the secret-shared data.} 
Since DBO selects random polynomials to create shares, a keyword/file-id appearing at multiple places (either inverted list, AC matrix, or files) will look different in ciphertext; 
preventing an adversary from learning information by looking at AC matrix, inverted list, and files. 
~
Leakage that occurs from AC matrix is the number of keywords and the length of the keyword. 
The inverted list may reveal the maximum number of files with a keyword. 
The files and AP list may reveal their size. These leakages can also be prevented by padding, \textit{e}.\textit{g}., padding keywords to the same length, padding dummy keywords to AC matrix, padding dummy file-ids to the inverted list, or adding dummy content to the files. Padding strategy increases the data size, and so, the processing time. {\color{black} After interpolation, the client discards dummies.
A common way is to use a predefined dummy value, \textit{e}.\textit{g}., {-99}, known to all participating entities. As we are working on English letters, a dummy value of -99 or 27 will work; for a larger character set like Unicode, we need to select a larger dummy, \textit{e}.\textit{g}., -9999999999. }

\medskip\medskip%
\noindent
\textbf{Why padding is secure?} 
Padding makes two keywords ``Jo'' and ``John'' appear as ``12,15,dummy,dummy'' and ``12,15,08,14'' in cleartext, and DBO will create shares of such numbers. Since shares are generated with random polynomials, repeated appearances of the same number appear different, preventing adversaries from distinguishing between real and fake keywords or deducing their lengths.

\medskip\medskip
\noindent
{{\textbf{Why keywords are not encoded in AC matrix?}} 
We represent keywords in letter positions/ASCII codes and pad each to the maximum length---increasing the size of AC matrix. Using encoding, \textit{e}.\textit{g}., a hash map, to allocate numbers to keywords, we could reduce keyword size but introduce issues: how to avoid false positives, how clients know the position in the map, how to handle collisions, and how to insert new keywords. Thus, we did not use encoding methods.

\medskip
\medskip
\noindent
{{\textbf{Different access rights for keywords.}}
Servers may learn the query, if they can distinguish keywords. Suppose, there are only two keywords, $k{\neq} k^{\prime}$, a client has access to search only $k$, and \emph{{all such are known to servers}}. 
Now, if servers return files after checking access rights, they will learn that the query is for $k$. \textsc{Doc}$^\star$ can also avoid this by returning fake files for $k^{\prime}$ (\textit{i}.\textit{e}., executing Phase~2 and Phase~3, even if access is not allowed).
To handle this, DBO inserts two fake entries with allowed/disallowed access rights in all data structures, (we show one of them with allowed access in~{\color{blue}Tables~\ref{tab:ACT_table},\ref{tab:inverted_list},\ref{tab:ss_file}}). 

\medskip
\section{Query Processing in \textsc{Doc}$\boldsymbol{^\star}$}
\label{sec:query processing}

{\color{black}
\textsc{Doc}$^\star$ has three interrelated phases to fetch the file containing the desired keyword. 
In the following, we develop protocols where the clients do not verify the results obtained from the server, while the server verifies the client queries in Phase~2 and Phase~3. Below, for 
simplicity, we use three servers and assume one of them is malicious.} 
{\color{blue}Table~\ref{tab:notations}} shows frequently used notations.

{\color{black}
\medskip
\subsection{\textbf{A Building Block: Distributed Secret-Shared Random Number Generation}}
\label{subsec:Random Number Generation and Verification} 
The primary purpose of using random numbers in \textsc{Doc}$\boldsymbol{^\star}$ query processing is to preserve data confidentiality throughout various computational steps. For instance, in Phase~1 (Step~2), the servers obliviously evaluate whether a client has access to a queried keyword; if access is denied, servers return a random number. 
Another example is Phase~3 (Step~8), where servers obliviously add random numbers to the file content if it contains a keyword to which the client has no search access, while the client tries to retrieve the file. 
}

{\color{black}
These random numbers also serve a second critical function: constructing the constant term \emph{\textbf{$\boldsymbol{\mathbb{M}(0)}$ of degree two}} that is used to randomize the shared polynomial after multiplication. 
Particularly, servers 
perform a single multiplication between a query $q$ and one of their data structures $\mathbb{M}(a)$, say $\mathbb{M}(c)\leftarrow \mathbb{M}(a){\times}\mathbb{M}(q)$.
Here, we observe that a single multiplication can obscure the true value $\mathbb{M}(a)$. However, the polynomial corresponding to $\mathbb{M}(c)$ 
is reducible, which may potentially lead to information leakage. To mitigate this, we propose the addition of a quadratic term $\mathbb{M}(0)$ to $\mathbb{M}(c)$, which introduces uncertainty in the reducibility of the resulting polynomial while preserving the polynomial's constants (the secrets).}\footnote{\scriptsize 
{A well-known method for polynomial randomization is BGW~\cite{DBLP:conf/stoc/Ben-OrGW88}. However, BGW method has certain limitations that preclude its direct application in our scenario. Firstly, it necessitates that all servers select random polynomials with zero constants and aggregate these polynomials to the original polynomial without any form of verification, thereby failing to prevent malicious behavior by servers. 
In contrast, our method allows the servers to verify $\mathbb{M}(0)$; see details below on verification. 
Secondly, BGW method can only randomize the polynomial during the query processing, which introduces additional communication and computational overhead. In contrast, our method generates  $\mathbb{M}(0)$ before query execution and uses it during query processing. }}

\medskip
\medskip
{\color{black}\noindent\textbf{Design objectives.}
Generating these random numbers is independent of a query and should be done before query execution for efficiency. The na\"{i}ve way of generating random numbers using a common seed at all the servers is not viable, as colluding servers could reveal the seed, defeating randomness's purpose. Instead, random numbers are generated in a distributed manner, so servers will have only their share of the random number at the end of the computation, but do not know the actual random number.

Distributed random number generation introduces another challenge:  a malicious server can potentially compromise the correctness of the output, \textit{i}.\textit{e}., the shares of the random numbers at each server. To address this, non-malicious servers, first, verify the generated random numbers before using them in further computations.

Reusing the same random number across multiple queries can leak information to the client. Thus, \emph{for each query}, new {\hl{$\mathbb{M}(\mathsf{RN})[]$}} and $\mathbb{M}(0)$ are generated in advance.}

}

\medskip
\medskip
\noindent
\textbf{Method.} 
We assume that each server $\mathcal{S}_{z\in\{1,2,3\}}$ generates, say $q$, non-zero random numbers, creates shares of those using \emph{polynomial of \textbf{degree one}}, and distributes appropriate shares to the other servers.  On receiving shares from the other servers, $\mathcal{S}_{z}$ aggregates them. {\hl{$\mathbb{M}(\mathsf{RN})[]$}} denotes such random numbers in share form at each server. The value of $q$ could be the number of keywords in AC matrix if the random numbers are used in Phase~1 or the size of a file if they are used in Phase~3.

Meanwhile, $\mathcal{S}_z$ generates $\mathbb{M}(0)$ \emph{\textbf{of degree two}} by multiplying $\mathbb{M}(\mathsf{RN})$ with $c_z$ locally (where $c_z$ is the input for SSS, \textit{i}.\textit{e}., $f(x{=}c_z)=(ac_z{+}b)c_z$ and  $f(x)=ax{+}b$ corresponds to the polynomial of $\mathbb{M}(\mathsf{RN})$).  

{\color{black}
\medskip\medskip
\noindent 
\textbf{Verification of $\boldsymbol{\mathbb{M}(\mathsf{RN})[]}$.}
Before using {\hl{$\mathbb{M}(\mathsf{RN})[]$}} 
in query execution, non-malicious servers verify them, since malicious servers can distribute any number as share, contaminating {\hl{$\mathbb{M}(\mathsf{RN})[]$}} and $\mathbb{M}(0)$. To verify, $\mathcal{S}_z$ does: 
$a_z{\leftarrow} \sum_{1\leq i \leq q}(PRG(seed)\times\mathbb{M}(\mathsf{RN})[i])\bmod{p}.$

In other words, each server first generates $q$ new different random numbers using a common seed and then performs a dot product between the new random number and $\mathbb{M}(\mathsf{RN})$. Then, each two servers interpolate $\langle a_z,a_{z+1}\rangle$, $\langle a_{z+1},a_{z+2}\rangle$, and $\langle a_{z+2},a_{z}\rangle$. 
This approach leverages the observation that legal random number shares correspond to a polynomial of degree one, whereas illegal ones correspond to a polynomial of any other degree. Thus, a polynomial of degree one leads to consistent interpolation results between any two servers, a property that does not hold for a polynomial of any other degree.

\begin{theorem}
\textbf{If a server creates shares of random numbers incorrectly, $LI(a_{z},a_{z+1})$,
$LI(a_{z+1},a_{z+2})$, and $LI(a_{z+2},a_{z})$ at $\mathcal{S}_z$ will produce different results, where $LI(*)$ means Lagrange interpolation.} 
\end{theorem}


}

\medskip
\subsection{Phase~1: Access Control Check over AC Matrix}
\label{subsec:phase1}
\medskip\noindent
\textbf{High-level idea.} 
{\color{black}
Phase~1 works over AC matrix ({\color{blue}Table~\ref{tab:ACT_table}}) and enables the client to determine their search access rights for queried keywords. If the client has access to a queried keyword, they learn the row-id of the inverted list, which contain all file-ids associated with the keyword. If access is denied, the client learns nothing --- specifically, they cannot distinguish whether the keyword is not present in AC matrix or whether access is denied.} E.g.,  Phase~1 allows the client Lisa to learn whether she is allowed to search for a keyword or not, but she gains no further information. Even in the presence of collusion between Lisa and a minority of servers, she remains unable to infer whether keywords such as Ana or Fig are present in the AC matrix or if access to them is denied; see {\color{blue}Table~\ref{tab:ACT_table}}.

A client sends keywords in SSS form. Servers operate obliviously over AC matrix to find the keyword and the access right, and then, return a vector of SSS form to the client. After interpolation at the client, the vector contains \emph{{a zero, if the client is allowed to search the keyword}}; otherwise, all random numbers.

\medskip
\medskip
\noindent
{\color{black}\textbf{Algorithm design objectives.} We need an algorithm in Phase~1 to:

\noindent
(\textit{i})~\emph{Check access rights over ciphertext.} Since both access rights and the client's query are in ciphertext, the server needs to find the access rights without knowing the query keyword in cleartext and the access decision in cleartext. 

\noindent
 (\textit{ii})~\emph{Prevent information leakages.} The server must not learn additional information, such as which keyword the client is searching for, the client's access right, and the content of the access control matrix. 
 The client must not learn additional information,  such as access rights of other clients and queries executed by other clients. }

\medskip
\subsubsection{\textbf{Algorithms in Phase~1:}} work as follows:
\label{subsubsec:algo in phase1}

\medskip
\medskip%
\noindent
\textbf{\textsc{Step} 1: \textit{Client:}} sends their (provable) identity (\textit{e}.\textit{g}., name) and a keyword ($\mathit{uw}$) in SSS form to  three servers $\mathcal{S}_{z\in\{1,2,3\}}$. Secret-shares of the keyword are created using the same strategy as used by DBO to create shares of keywords (mentioned in~{\color{blue}\S\ref{sec:Data Outsourcing}}). Note that the client can select any polynomial to create shares.

\medskip
\medskip
\noindent
\textbf{\textsc{Step} 2: \textit{Servers:}} performs the following:

\medskip%
\noindent
\textbf{\emph{1. Obliviously checking access rights and polynomial randomization.}} Servers find a desired row corresponding to the client in AC matrix based on the client's identity and perform the following:
\noindent
{\color{myblue}
\begin{equation*}
\begin{split}
\mathbb{M}(\mathit{ans}\mathcal{S}_{z})[i] \leftarrow [(&(\mathbb{M}(\mathit{sw})[i]{-}\mathbb{M}(\mathit{uw}) + \mathbb{M}(AC)[i]) \times {\textnormal{\hl{$\mathbb{M}(\mathsf{RN})[i])$}}} \\[-0.7ex] 
&  \quad\quad\quad\quad + \mathbb{M}(0)] 
{\bmod} p, \forall i \in \{1,\ldots,\beta\}
\end{split}
\end{equation*}}

\noindent
$\mathcal{S}_{z\in\{1,2,3\}}$ subtracts each keyword ($\mathbb{M}(\mathit{sw})[i]$ --- the first row of {\color{blue}Table~\ref{tab:ACT_table}}) of AC matrix with the keyword $\mathbb{M}(\mathit{uw})$, received from the client, and adds the corresponding access control value $\mathbb{M}(\mathit{AC}[i])$. 

Note that $\mathbb{M}(\mathit{sw})[i]{-}\mathbb{M}(\mathit{uw})+\mathbb{M}(AC)[i]$ may provide additional information  (\textit{e}.\textit{g}., random numbers associated with denied access or which keywords are present) to the client, $\mathcal{S}_{z}$ multiplies \hl{$\mathbb{M}(\mathsf{RN})[]$} to each output 
(to make the client unable to distinguish keywords).
 Further, 
the servers need to \emph{\textbf{randomize}} the whole result by adding $\mathbb{M}(0)$ (as produced in~{\color{blue}\S\ref{subsec:Random Number Generation and Verification}}) with different secret polynomials of degree two for each query to avoid factorization of the polynomial at clients ($\mathbb{M}(0)$ with polynomial degree one can be recovered by a single server). We discuss the security analysis later.

\medskip%
\noindent
\textbf{\emph{2. Sending results.}}
$\mathcal{S}_{z}$ sends a vector, having $\beta$ numbers in SSS form (where AC matrix contains $\beta$ keywords) to the client.
\emph{{If the client is allowed to search for the queried keyword, the above equation will produce zero in SSS form; otherwise, a random number}}.

\medskip
\medskip
\noindent
\textbf{\textsc{Step 3a}: \textit{Client:}} performs Lagrange interpolation (denoted as $\mathit{LI}$) over the vectors received from servers, \textit{i}.\textit{e}.,  $${\mathit{vecR1}[i]\leftarrow \mathit{LI}(\mathbb{M}(\mathit{ans}\mathcal{S}_{1})[i],\mathbb{M}(\mathit{ans}\mathcal{S}_{2})[i],\mathbb{M}(\mathit{ans}\mathcal{S}_{3})[i])}.$$ If $\mathit{vecR1}[]$ contains only random numbers, it means that the client is \emph{not} allowed to search for the keyword. 
Otherwise, $\mathit{vecR1}[]$ contains \emph{only one zero at the position corresponding to the keyword's position in AC matrix or inverted list}.


\medskip
\subsubsection{\textbf{Example of Phase~1:}} is given in 
{\color{blue}Table~\ref{tab:Example of Phase1}}. \emph{For simplicity, we do not add $\mathbb{M}(0)$ in the example.}

\medskip
\subsubsection{\textbf{Correctness.}}
SSS is somewhat homomorphic and does not affect the final result of the expression. 
According to the formulation of $\mathit{ans}\mathcal{S}_{z}$ presented in \textsc{Step 2}, only if the query keyword matches the $i^{\mathit{th}}$ keyword in AC matrix and the corresponding related access control number in the capability list of the client $AC[i]$ is zero, $\mathit{ans}\mathcal{S}_{z}$ will be zero. Otherwise, the client always obtains nonzero numbers, indicating that the keyword is not in AC matrix or the client has no right to search it. Further, the non-zero random number of $\mathbb{M}(AC)$ (allocated using the method of~{\color{blue}\S\ref{sec:Data Outsourcing}}) prevent false positives,  \textit{i}.\textit{e}., $\mathbb{M}(\mathit{sw})[i]{-}\mathbb{M}(\mathit{uw}) \neq \mathbb{M}(AC)[i]$.


\medskip
\subsubsection{\textbf{Security Discussion.}}

\noindent\emph{\textbf{Servers:}} (\textit{i}) receive from a client a keyword $\mathit{uw}$ of SSS form; thus, servers cannot learn the keyword in cleartext; (\textit{ii})  perform identical operations on each keyword of AC matrix, making it impossible for them to learn anything from the operations they perform; thereby access-patterns are hidden from servers; (\textit{iii}) always return a vector of length $\beta$; thus, volume of the answer is also hidden from servers.

\noindent
\textbf{\emph{Clients:}}~cannot learn information about the random numbers used to indicate denied access for a keyword, as servers multiply random numbers \hl{$\mathbb{M}(\mathsf{RN})[]$} to obfuscate the subtraction results.
Even if a minority of the servers collude with a malicious client, they cannot learn $\mathbb{M}(AC)[i]$ of a keyword with denied access. Also, a malicious client colluding with malicious servers cannot learn the access right information of other clients or queries by them in cleartext due to not having enough shares. 
~
Furthermore, a client that does not knowing a keyword $\mathbb{M}(\mathit{sw})[i]$ in AC matrix cannot distinguish between the absence or presence of the keyword or disallowed access to keywords, due to multiplying  \hl{$\mathbb{M}(\mathsf{RN})[]$}; see the following theorems.\footnote{\scriptsize{This is important when revealing the presence/absence of highly sensitive keywords (\textit{e}.\textit{g}., nuclear).}} 

\begin{table}[!t]
\BBB
    \centering
    \begin{tabular}{|p{8.4cm}|}\hline
\texttt{Lisa} wants to fetch files containing a keyword \texttt{are}, represented as {112815}. Note that \texttt{Lisa} is allowed to search for \texttt{are}; see~{\color{blue}{Table~\ref{tab:ACT_table}}}. 
 Suppose \hl{$\mathbb{M}(\mathsf{RN}_{1})=[4,5,6]$},
 \hl{$\mathbb{M}(\mathsf{RN}_{2})=[5,6,7]$},
 \hl{$\mathbb{M}(\mathsf{RN}_{3})=[6,7,8]$}. \emph{For simplicity, we do not add $\mathbb{M}(0)$ here.} \\

\textbf{\textsc{Step} 1: \textit{Client.}}
\texttt{Lisa} creates SSS of \texttt{are}, using a polynomial $f(x)=(5x{+}s){\bmod} p$, where $p=500,009$: for $x{=}1: 112820$, which is sent $\mathcal{S}_1$, for $x{=}2: 112825$, which is sent $\mathcal{S}_2$, and for $x{=}3: 112830$, which is sent $\mathcal{S}_3$.
\\

\textbf{\textsc{Step} 2: \textit{Servers:}} 
Based on Table \ref{tab:ACT_table_share1},
$\mathcal{S}_1$ performs the following: 

{\scriptsize
{\centerline{$(112816\!-\!112820 + 1) \times 4 \bmod p 
= 499997$}}
{\centerline{$(112412\!-\!112820 + 2) \times 5\bmod p 
=497979$}}
{\centerline{$(161918\!-\!112820 + 3) \times 6\bmod p
=294606$}}
}
$\mathcal{S}_1$ sends 499997, 499997, 294606 to \texttt{Lisa}. Based on {\color{blue}Table~\ref{tab:ACT_table_share2}},  $\mathcal{S}_2$ performs the following: \\ 

{\scriptsize
\centerline{$(112817{-}112825 + 2) \times 5\bmod p 
= 499979$}
{\centerline{$(112413\!-\!112825 + 3)\times 6\bmod p
=497555$}}
{\centerline{$(161919\!-\!112825+4)\times 7\bmod p
=343686$}}}
$\mathcal{S}_2$ sends 499979, 497555, 343686 to \texttt{Lisa}. Based on {\color{blue}Table~\ref{tab:ACT_table_share3}}, $\mathcal{S}_3$ performs the following:

{\scriptsize
{\centerline{$(112818\!-\!112830 + 3)\times 6 \bmod p = 499955$}}
{\centerline{$(112414\!-\!112830 + 4)\times 7\bmod p =497125$}}
{\centerline{$(161920\!-\!112830 + 5)\times 8\bmod p =392760$}}}

$\mathcal{S}_3$ sends 499955, 497125, 392760 to \texttt{Lisa}.\\

\textbf{\textsc{Step 3a}: \textit{Client:}} \texttt{Lisa} performs interpolation and obtains the results: $[0,498397,245520]$ that indicates that \texttt{Lisa} is allowed to search the keyword \texttt{are}, which appears at the first position.         \\\hline
    \end{tabular}
    \caption{Example of Phase~1.}
    \label{tab:Example of Phase1}
\BBB\BBB\B
\end{table}

\begin{theorem}
\label{th:single query}
\textbf{If a malicious client colludes with a minority of malicious servers, such malicious entities cannot deduce {\hl{$\mathsf{RN}$}} to obtain additional information, except for $\mathit{ans}\mathcal{S}_{z}$, in Phase~1.}
\end{theorem}

Multiplication between two shares may lead to related secret polynomial reducible. Also, the client may create 
$\mathbb{M}(\mathit{uw})$ twice or more using the same polynomial, then deduce $\mathit{sw}[i]{+}AC[i]$ by finding the common divisor between two secret polynomials related to $(\mathbb{M}(\mathit{sw})[i]{-}\mathbb{M}(\mathit{uw}){+}\mathbb{M}(AC)[i]){\times} \mathbb{M}(\mathsf{RN})[i]$ for two queries. Randomization of $\mathbb{M}(\mathit{uw})$, \textit{i}.\textit{e}., adding $\mathbb{M}(0)$, avoids this. 

\begin{theorem}
\label{th:multi-query}
\textbf{Randomization of $\mathbb{M}(\mathit{uw})$ prevents malicious clients from deducing $\mathit{sw}[i]+AC[i]$ over multiple queries, even colludes with a minority of malicious servers.}
\end{theorem}



\medskip
\subsubsection{\textbf{Cost Analysis.}}
Computation cost at a server and the client is $\mathcal{O}(\beta)$.  Communication cost between a server and a client is $\mathcal{O}(\beta)$.

\medskip
\subsection{Phase 2: Finding File-Ids from Inverted List}
\label{subsec:phase2}

\medskip
{\color{black}\noindent\textbf{High-level idea.} 
Phase~2 works on the inverted list to allow clients to retrieve the file-ids associated with the keyword they queried in Phase~1, provided that the client has search access to the keyword. E.g., if Lisa has searched for the keyword `Are' in Phase~1, then after Phase~2, she will learn that files~1 and~2 contain the keyword `Are.'

A client sends a vector, containing all zeros except for a single one, in SSS form to retrieve file-ids associated with the keyword they searched in Phase~1. The one is placed in the vector according to the row-id, revealed to the client in Phase~1.  
Before query execution, servers obliviously verify the \textbf{correctness of the vector} (\textit{i}.\textit{e}., containing all zeros except a single one at the desired position)}, ensuring that clients do not fetch file-ids associated with a keyword that is not allowed to be searched, \textbf{even if they skip Phase~1}.
If the vector is \emph{correct}, servers perform a dot product between the vector and the inverted index and send the output of the dot product (\textit{i}.\textit{e}., file-ids) to the client, who learns the file-ids after interpolation.

\medskip
\medskip%
{\color{black}
\noindent
\textbf{Algorithm design objectives.} An algorithm in Phase~2 needs to: 

\noindent(\textit{i})~\emph{Handle a malicious client.} Although the output of Phase~1 informs the client whether they are authorized to proceed to Phase~2 and, if so, reveals the specific row index in the inverted list, a malicious client may attempt to fetch an arbitrary row of the inverted list. Thus, the servers must be able to obliviously verify that the client's request corresponds to a row of the inverted list that is associated with a keyword the client is authorized to search. This requirement presents a significant challenge, as the query keyword, the Phase~1 result, and Phase~2's query vector are all in the ciphertext at the server.

\noindent(\textit{ii})~\emph{Prevent information leakages.} The servers must not learn any additional information, such as the keyword of Phase~1 and the requested row-id of the inverted list. }

\begin{table}[!t]
\BBB
    \centering
    \begin{tabular}{|p{8.4cm}|}\hline
We continue with the example of Phase~1, given in {\color{blue}Table~\ref{tab:Example of Phase1}}.

\noindent
\textbf{\textsc{Step 3b}: \textit{Client:}} \texttt{Lisa} after \textsc{Step 3a} creates a vector $\langle 1,0,0\rangle$, since the keyword \texttt{are} appears at the first index in the inverted list 
(see~{\color{blue}Table~\ref{tab:inverted_list}}). Finally, \texttt{Lisa} creates three multiplicative shares of the vector using $f(x){=}(10x{+}s){\bmod}p$, where $p{=}500009$: 
$\langle 11,10,10 \rangle$ sent to $\mathcal{S}_1$,
$\langle 21,20,20\rangle$ sent to $\mathcal{S}_2$,
$\langle 31,30,30\rangle$ sent to $\mathcal{S}_3$.

\noindent
\textbf{\textsc{Step 4}: \textit{Servers:}}, first performs the three tests of~{\color{blue}\S\ref{Sec:Verification of the Client's Vector}} to ensure the vector has only one and all zeros, and then the following Test~1 to ensure \texttt{Lisa}'s access to the keyword:


{\scriptsize
{\centerline{$\mathcal{S}_1$: 
$[(1,2,3)\odot (11,10,10) {\bmod} p] =61$}}
{\centerline{$\mathcal{S}_2$: 
$[(2,3,4)\odot (21,20,20) {\bmod} p] =182$}}
{\centerline{$\mathcal{S}_3$: 
$[(3,4,5)\odot (31,30,30) {\bmod} p]=363$}}}

Finally, each server sends the output of the test to other servers and interpolates them. The final answer is $\langle0\rangle$, showing \texttt{Lisa} has created the vector correctly (\textit{i}.\textit{e}., 
she has access to search the keyword). Afterward, servers perform a dot product between the vector and  three share tables given in {\color{blue}Table~\ref{tab:share1 of IL}}:

{\scriptsize
{\centerline{$\mathcal{S}_1: \langle(2,3),(3,1),(4,1)\rangle \odot (11,10,10){\bmod} p=92, 53$}}
{\centerline{$\mathcal{S}_2: \langle(3,4),(4,2),(5,2)\rangle \odot (21,20,20){\bmod} p=243, 164$}}
{\centerline{$\mathcal{S}_3: \langle(4,5),(5,3),(6,3)\rangle \odot (31,30,30){\bmod} p=454,335$}}
}

\textbf{\textsc{Step 5a}: \textit{Client:}} \texttt{Lisa} interpolates: $\{(1, 92), (2, 243), (3, 454)\}$, $\{(1, 53), (2, 164), (3, 335)\}$ and obtains the secret 1 and 2, which correspond to the file-ids in row one of the inverted list.
\\\hline
    \end{tabular}
    \caption{Example of Phase~2.}
    \label{tab:Example of Phase2}
\BBB\BBB\BB
\end{table}

\medskip
\subsubsection{\textbf{Algorithms in Phase~2:}} work as follows:

\label{subsubsec:Algorithms in Phase2}


\medskip
\smallskip
\noindent
\textbf{\textsc{Step 3b}: \textit{Client:}} creates a vector $v$ of length $\beta$ containing zeros except for a single one at the $i^{\mathit{th}}$ position that corresponds to the position of the single zero in $\mathit{vecR1}$ of \textsc{Step 3a}. The client generates SSS of this vector $v$, denoted by $\mathbb{M}(v)$, and sends them to 
$\mathcal{S}_{z\in\{1,2,3\}}$. 


\medskip
\medskip
\noindent
\textbf{\textsc{Step 4}: \textit{Server:}} has three objectives: 
(\textit{i})~obliviously verifying the  \emph{\ul{correctness}} of the client's vector, 
(\textit{ii})~obliviously checking the client's access rights for the keyword, 
and
(\textit{iii})~obliviously returning the file-ids associated with the keyword, if the vector is correct.

\medskip%
\noindent
\emph{\textbf{1. Correctness of the client’s vector.}}  
To achieve the first objective, Tests~A and~B of {\color{blue}\S\ref{Sec:Verification of the Client's Vector}} are executed and  \emph{prevents malicious clients from generating a wrong type of vector} $v$, \textit{e}.\textit{g}., $v{=}\{0,0,\ldots,0,0\}$, $v{=}\{1,0,\ldots,1,0\}$, or $v{=}\{0,0,\ldots,10,-9\}$. 

\medskip%
\noindent
\emph{\textbf{2. Ensuring allowed access to the client for the keyword.}}
After that, for the second objective, servers obliviously check the client's access to the keyword via the following Test~1: 

 {\color{myblue}
{\centerline{\noindent
 Test~1: $\mathbb{M}(\mathit{test}_1) \leftarrow \mathbb{M}(AC)\odot \mathbb{M}(v)
   $
}}}

Test~1 ensures the client has access rights for searching the keyword. $\mathcal{S}_z$ performs a dot product between the received vector $\mathbb{M}(v)$ and the capability list of the client. 
If the \emph{{vector $\mathbb{M}(v)$ is correct and the client has search access to the keyword, then $\mathbb{M}(\mathit{test}_1){=}\mathbb{M}(0)$}}; otherwise, a random number, which corresponds to the no access right value. 
To obtain the values of $\mathbb{M}(\mathit{test}_1)$ in cleartext, $\mathcal{S}_{z}$ sends the output of the test, which is in share form, to other servers. Then, 
each server interpolates the values. Note that at this step, the malicious client/server will learn the value corresponding to the no access right. However, \emph{this does not enable the malicious client to learn $\mathit{sw}$, due to Theorem~\ref{th:multi-query}}. 
Of course, we can also hide this value too, if it is required using the method given in Appendix~\ref{app_sec:Hiding Test1 Output for Incorrect Vector} in~\cite{fullversions2d}.

\medskip%
\noindent
\textbf{\emph{3. Returning file-ids.}}  
If the interpolated value is $\langle 0\rangle$\footnote{\scriptsize If one of the three servers does not perform computation correctly, then Test~1 and the subsequent Tests of Phase~3 will fail. This malicious behavior can be detected by non-malicious servers by involving the fourth server and  
executing the computation (\textit{i}.\textit{e}., the tests) at all four servers. Suppose $\mathcal{S}_1$ is non-malicious and $\mathcal{S}_3$ is malicious. To learn the output of the tests, $\mathcal{S}_1$ performs interpolation over values of $\langle\mathcal{S}_1,\mathcal{S}_2,\mathcal{S}_3\rangle$, $\langle\mathcal{S}_1,\mathcal{S}_2,\mathcal{S}_4\rangle$, and $\langle\mathcal{S}_1,\mathcal{S}_3,\mathcal{S}_4\rangle$. Now, all these interpolated values will not produce identical results, showing malicious behavior by one of the servers, (and non-malicious servers may terminate the protocol).} for Test~1, then, $\mathcal{S}_{z}$ performs a dot product between the vector $\mathbb{M}(v)$ and 
the inverted list, and \emph{{this results in all file-ids, denoted by $\mathbb{M}(\mathit{fid\mathcal{S}}_{z})[]$, that are associated with the keyword $\mathit{uw}$}}.\footnote{\scriptsize{This keyword could be $\mathit{uw}$, used in Step~1 or a keyword to which the client has search access. We can also make sure that the file-ids are only those associated with the keyword $\mathit{uw}$ used in Phase~1, by adding the following test: $\mathbb{M}(\mathit{test})\leftarrow 
(\mathbb{M}(sw){\odot} \mathbb{M}(v)){-} \mathbb{M}(uw)  
 $, and $\mathbb{M}(\mathit{test}){=}0$ ensures this.}} 
 $\mathbb{M}(\mathit{fid\mathcal{S}}_{z})[]$ is sent to the client.

\medskip
\medskip
\noindent
\textbf{\textsc{Step 5a}: \textit{Client:}} interpolates the file-ids received from servers and learns a set of file-ids, say $\mathit{fid}[]$, associated with the query keyword.


\medskip
\subsubsection{\textbf{Example of Phase~2:}} is given in
{\color{blue}Table~\ref{tab:Example of Phase2}}
.

\medskip
\subsubsection{\textbf{Correctness.}} \label{subsubsec:discussion_phase2}
The approach works at servers since an $i^\mathit{th}$ position of the vector $v$ having one indicates that $i^\mathit{th}$ keyword of AC matrix has the search permission for the client. 
The formulation of Test~1 ensures the client has the search right to the $i^\mathit{th}$ keyword; otherwise, $\mathbb{M}(\mathit{test}_1){\neq}\mathbb{M}(0)$.
Thus, \textsc{Step~4} produces only file-ids associated with the $i^{\mathit{th}}$ keyword to which search permission is allowed.


\medskip
\subsubsection{\textbf{Security Discussion.}}

\noindent\emph{\textbf{Servers:}} 
(\textit{i}) Servers receive a vector of SSS form. Also, the output of Test~1 contains 0 if the vector is correct, regardless of the keyword or its position in AC matrix (\textit{e}.\textit{g}., an $i^{\mathit{th}}$ keyword in which the client is interested). Thus, servers cannot learn the keyword, its position, and/or the query keyword in \textsc{Step~4}. 
(\textit{ii})~Since servers communicate with others to exchange the output of the test, even in the presence of a minority of malicious servers colluding with the client, the test cannot fail. This prevents the client from fetching file-ids that are not associated with the keyword to which the client has no access. The reason is that malicious entities cannot generate the vector $v$ and shares of the test's output; thereby, interpolated values result in zero at non-malicious servers unless malicious entities know the keywords in AC matrix and their random numbers for non-access. 
(\textit{iii})~Servers perform identical operations on the inverted list, hiding access-patterns from servers. Servers always return the maximum number of file-ids in which a keyword can appear regardless of the query keyword; thus, volume is also hidden from servers.


\noindent
\textbf{\emph{Clients:}}
(\textit{i})~Firstly, Tests A and B verify that client generated a legal vector (consisting of all zeros except a single one). But clients may generate a wrong vector to know file-ids, which are associated with a keyword to which search access is disallowed. In this case, 
Test~1 will fail only if the client possesses knowledge of the index of the keyword in AC matrix and their random number for non-access. However,  a client cannot have such information. 
(\textit{ii})~Clients learn the maximum number of file-ids in which a keyword can appear. 



\medskip
\subsubsection{\textbf{Cost Analysis.}}

Communication cost from a client to a server is $\mathcal{O}(\beta)$, and from a server to the client is $\mathcal{O}(\gamma)$, where $\beta$ is the number of searchable keywords and $\gamma$ is the maximum number of files associated with a keyword. Communication cost among servers involves only a few numbers. Computation cost at a server is $\mathcal{O}(\beta\gamma)$ and at a client is $\mathcal{O}(\gamma)$.

\medskip
\subsection{Phase 3: Retrieving Documents/Files}
\label{subsec:phase3}

{\color{black}
\medskip\noindent
\textbf{High-level idea.} 
Phase~3 allows a client to retrieve the files corresponding to the keyword queried in Phase~1, based on the file-ids they learned in Phase~2. If a file contains even a single keyword for which the client lacks search access, that file is \emph{not} returned to the client, and in this case, servers will obviously return a fake file. 

For example, Lisa, who searched for the keyword `Are' during Phase~1, learned that files~1 and~2 both match the keyword `Are' after Phase~2. In Phase~3, the server returns to her the actual file~1, as well as a fake file in place of file~2. Although both files contain the keyword `Are,' file~2 also contains the keyword `Ana,' to which Lisa does not have search access (see~{\color{blue}Tables~\ref{tab:ACT_table},\ref{tab:inverted_list},\ref{tab:ss_file}}). This reveals to Lisa that the keyword `Are' appears in two files, but she is only authorized to access one of them. Importantly, she does not learn which specific unauthorized keyword prevents access to file~2, thereby preserving the confidentiality of other keywords.


A client creates a new vector of length $\delta$ (where $\delta$ is the number of files) containing all zeros and only one at the position corresponding to one of the file-ids obtained in Phase~2 and sends this vector in share form to the servers. Servers operate over the file data structure ({\color{blue}Table~\ref{tab:ss_file}}) and, first, obliviously verify the \textbf{\emph{vector's correctness}}, like Phase~2. If the vector is correct, servers obliviously ensure that the file does not have a keyword to which the client does not have search access and then obliviously send the desired file to the client.

\medskip\medskip
\noindent
\textbf{Algorithm design objective.} 
An algorithm in Phase~3 needs to:

\noindent(\textit{i})~\emph{Prevent sending an unauthorized file to a client.} A  client must not obtain a file $F$ that contains a keyword $k$ that the client has searched in Phase~1 for which the client has allowed search access, and the file $F$ also contains a keyword $k^{\prime}$ that the client is disallowed to search.


\noindent(\textit{ii})~\emph{Prevent information leakages.} The protocol must guarantee that neither the client nor the server learns additional information during query execution. Particularly, the server must not learn the queried keyword from Phase~1, the requested row-id of the inverted list, and whether any specific file is not returned in Phase~3 due to access restrictions. Similarly, the client must not infer the existence of keywords for which it lacks access, nor determine which specific keyword within a file has led to access denial.  }

\medskip
\subsubsection{\textbf{Algorithms in Phase Three:}} work as follows:
\label{subsubsec:algo_in_phase3}

\medskip
\medskip%
\noindent
\textbf{\textsc{Step 5b}: \textit{Client:}} creates $x{\leq} \gamma$ (a keyword appears in at most $\gamma$ files) vectors $v_x$, each of size $\delta$, to fetch $x$ files containing the keyword $\mathit{uw}$. Recall that the client learns $x$ file-ids, $\mathit{fid}[]$, in \textsc{Step~5a}. 
Each such vector contains zeros, except one at the $i^{\mathit{th}}$ file-id position, based on $\mathit{fid}[]$. 
Client creates SSS of such vectors ($\mathbb{M}(v_x)$) and sends them to $\mathcal{S}_{z\in\{1,2,3\}}$. Note that if a keyword appears in $x{<} \gamma$ files, then the $\gamma{-}x$ vectors will fetch a fake file with zeros. Below, \emph{for simplicity, we consider a case, when the client fetches only a single file by sending a vector $\mathbb{M}(v)$ of length $\delta$ to $\mathcal{S}_{z\in\{1,2,3\}}$.}


\subsubsection*{\textbf{Ensuring Correctness of the Vector.}}
\label{subsubsec:Ensuring Correctness of the Vector}
 \noindent

\medskip
\medskip%
\noindent
\textbf{\textsc{Step 6}: \textit{Server:}} receive $\mathbb{M}(v)$ vector from the client and has three objectives to ensure:
(\textit{i})~$\mathbb{M}(v)$ contains all zero and except a single one; 
(\textit{ii})~$\mathbb{M}(v)$ contains one at the position of the file-id that was sent in Phase~2 by servers; and 
(\textit{iii})~the requested file does not contain a keyword to which the client has no access. 

\medskip%
\noindent
\emph{\textbf{1. Ensuring $\mathbb{M}(v)$ having all zeros and a single one.}} Test~A and Test~B of~{\color{blue}\S\ref{Sec:Verification of the Client's Vector}} achieve the first objective. 

\medskip%
\noindent
\emph{\textbf{2. Ensuring the correct position of one in $\mathbb{M}(v)$.}} The following Test~2 achieves the second objective. 

    {\centerline{\color{myblue}
    Test 2: $\mathbb{M}(\mathit{test}_2)\leftarrow 
    (\mathbb{M}(\mathit{file\_id}){\odot} \mathbb{M}(v))-\mathbb{M}(\mathit{fid\mathcal{S}_z})[i] 
     $}}



 On the success of Tests~A and~B, $\mathcal{S}_z$ performs Test~2 that executes a dot product between the received vector $\mathbb{M}(v)$ and file-ids $\mathbb{M}(\mathit{file\_id})$ (see the first column of {\color{blue}Table~\ref{tab:ss_file}}). This produces the $i^{\mathit{th}}$ file-id that is subtracted from one of the file-ids $\mathbb{M}(\mathit{fid\mathcal{S}_z})[i]$, sent by servers in \textsc{Step 4} of Phase~2. The client informs the file-ids used for subtraction. \emph{If the  $\mathbb{M}(v)$ is correct, $\mathbb{M}(\mathit{test}_2)=\mathbb{M}(0)$}. $\mathcal{S}_{z}$ communicates with other servers to receive their shares of this computation and performs interpolation to know the value of $\mathit{test}_2$ in cleartext. 

\medskip%
\noindent
\textbf{\emph{3. Ensuring the file excludes keywords restricted to the client.}} 
Test~3 and subsequent \textsc{Steps~7-9} achieve the third objective.

{\centerline{\color{myblue}
Test 3: $\mathbb{M}(\mathit{test}_3)[]\leftarrow 
\mathbb{M}(v)\odot \mathbb{M}(AP) 
 $}}

Test~3, on the success of Test~2, performs a dot product between $\mathbb{M}(v)$ and the AP list (denoted by $\mathbb{M}(AP)$; see the second column of {\color{blue}Table~\ref{tab:ss_file}}). 
This results in the position of the keywords appearing in the file and their hash digest --- all in SSS form of degree two. $\mathcal{S}_{z}$ sends all the positions to the client, 
while keeping the hash digest (denoted by $\mathbb{M}(\mathsf{H}(AP))$, \textit{i}.\textit{e}., the last value of AP list) at their end.

\subsubsection*{\textbf{Return the File.}}
 \noindent

\medskip\medskip
\noindent
\textbf{\textsc{Step 7}: \textit{Client:}} interpolates the received values to learn the position of keywords in the file---\emph{the client does not learn the keyword.} An honest client wishes to enable the server to know the access right for all these keywords by creating a keyword position vector, say $\mathit{kpv}$, filled with zeros except for the position of the keywords returned in \textsc{Step 6} to be one. The client creates shares of $\mathit{kpv}$, denoted by $\mathbb{M}(\mathit{kpv})$, and sends them to servers. 
We will argue how servers will detect malicious behavior of the client in creating $\mathbb{M}(\mathit{kpv})$. 

\medskip
\medskip
\noindent
\textbf{\textsc{Step 8}: \textit{Server:}}  
(\textit{i})~ensures $\mathbb{M}(\mathit{kpv})$ contains only zeros and ones using Test~C of {\color{blue}\S\ref{Sec:Verification of the Client's Vector}}; 
(\textit{ii})~ensures $\mathbb{M}(\mathit{kpv})$ is created for positions of keywords sent in \textsc{Step 6}; 
(\textit{iii})~returns the file. 

{\centerline{\color{myblue} 
Test 4: $\mathbb{M}(\mathit{test}_4)\leftarrow 
(\mathbb{M}(\mathit{kpv}){\odot} \mathbb{M}(\mathsf{H}(\mathit{ACT\_pos})))-\mathbb{M}(\mathsf{H}(\mathit{AP})) 
 $}}
Test~4 verifies the second condition. 
Here, $\mathbb{M}(\mathsf{H}(\mathit{ACT\_pos}))$ denotes 
the hash digest for the positions in AC matrix (see the blue-colored third row of {\color{blue}Table~\ref{tab:ACT_table}}) and $\mathbb{M}(\mathsf{H}(\mathit{AP}))$ is the hash digest, computed in \textsc{Step~6}. 
\emph{If $\mathbb{M}(kpv)$ is correct, Test~4 produces zero.}

Now, servers will send the file. Note that the \emph{objective is to ensure that the \textbf{file return operation is oblivious}}, \textit{i}.\textit{e}., the server cannot determine whether the file that the client is accessing contains a keyword to which access is allowed or not. The server always returns a file---either a real or a garbage file---without distinguishing between the two. Further, if a file contains even a single keyword for which the client does not have access rights, the client must not be able to learn the file's content, even if the file is retrieved. To do so, $\mathcal{S}_{z}$ performs the following: 

{\centerline{\color{myblue} Test 5: 
$\mathbb{M}(\mathit{test}_5)\leftarrow 
\mathbb{M}(\mathit{kpv})\odot \mathbb{M}(\mathit{AC}) 
 $}}


  $\mathcal{S}_{z}$ performs a dot product between $\mathbb{M}(kpv)$ and the access capability list  $\mathbb{M}(AC)$ of the client (see yellow-colored rows of {\color{blue}Table~\ref{tab:ACT_table}}). 
  It is important to note that \emph{zero in SSS form, being the result of Test~5, indicates that the client has access to all keywords, appearing in the file.} Otherwise, Test~5 will produce a secret-shared random number equal to the sum of the non-access right value, as in Test~1.

Finally,  $\mathcal{S}_z$ selects {\hl{$\mathbb{M}(\mathsf{RN}_z)$}} equals to the size of a file, multiplies {\hl{$\mathbb{M}(\mathsf{RN}_z)$}} with the output of Test~5, and adds this to the file before sending it to the client. Before multiplication, servers reduce the degree of the polynomial of the output of Test~5 from two to one.\footnote{\scriptsize Servers can reduce the degree using an existing method~\cite{DBLP:conf/stoc/Ben-OrGW88} or our method of {\color{blue}Appendix~\ref{app_sec:Hiding Test1 Output for Incorrect Vector}} in~\cite{fullversions2d}. }

\medskip
\medskip
\noindent
\textbf{\textsc{Step} 9: \textit{Client:}} interpolates the file content,  obtained from servers.

\medskip
\subsubsection{\textbf{Correctness.}} Obviously, \textsc{Step 6} verifies if the client generates the correct vectors to fetch the $i^{\mathit{th}}$ file, and its correctness is similar to what we analyzed in~{\color{blue}\S\ref{subsubsec:discussion_phase2}}. An important aspect to discuss is how the algorithm can detect malicious behavior of the client in {\textsc{Step 8a}}. The reason is: a dot product between a wrong vector, created by the client in {\textsc{Step}~7}, with the respective vectors of hash digest ($\mathbb{M}(\mathsf{H}(\mathit{ACT\_pos})))$), and then the given subtraction will not result in a value of zero of Test~4 and Test~5.

\medskip
\subsubsection{\textbf{Security Discussion.}}

\emph{\textbf{Servers:}}
(\textit{i})~Servers receive vectors in SSS 
\noindent form; thus cannot learn the file-id and the number of real files requested by the client.
(\textit{ii})~Servers execute an identical operation to check the vector's correctness and to send files to the client. Thus, servers cannot learn anything based on access-patterns.
(\textit{iii})~The algorithm is designed to hide the volume. If $x{<}\gamma$, then $\gamma{-}x$ files containing $\mathbb{M}(0)$ will be sent to the client; thus, hiding the volume from servers. While this comes with additional computational cost at servers and communication cost, this does not increase the space overhead, since DBO adds a single dummy file; see~{\color{blue}Table~\ref{tab:ss_file}}. Servers cannot learn the number of times they send the dummy file, since access-patterns are hidden from servers. Also, this does not reveal to the client any other file, which does not contain the queried keyword.

\noindent
\textbf{\emph{Client:}} only receives the desired file and cannot fetch files, 
containing at least a keyword to which access is disallowed. \emph{\textsc{Step}~7 reveals the number of keywords and their positions in AC matrix.\footnote{\scriptsize {\color{blue}Appendix~\ref{app_subsec:phase3 Fully Secure Operation}} provides a way to hide from a client number of keywords, their positions in AC matrix.}} 

\medskip
\subsubsection{\textbf{Cost Analysis.}} 

Communication cost from a client to a server is $\mathcal{O}(\gamma\delta)$, and from a server to the client is $\mathcal{O}(\gamma\eta)$, where $\eta$ is the size of a single file. Communication cost among servers is $\mathcal{O}(\delta)$. Computation cost at a server is $\mathcal{O}(\gamma\eta\delta)$ and at the client is $\mathcal{O}(\gamma\eta)$.\footnote{\scriptsize
{\color{blue}Appendix~\ref{app_subsec:phase3 Fetching Multiple Files}} extends the above method to fetch multiple files with less communication cost.}


\medskip
\subsection{\textbf{The Tests A, B, and C}}
\label{Sec:Verification of the Client's Vector}
{\color{black}During different phases of the query processing, the client sends a bit vector in shared form to the servers. However, a malicious client may create incorrect vectors by either placing one at the wrong places or having non-binary values. }
The servers' objective is to ensure that the vector $v$ is valid, \textit{i}.\textit{e}., the vector contains
(\textit{i})~all zeroes and a single one, or 
(\textit{ii})~many zeros and many ones only. 
To do so, $\mathcal{S}_{z}$ performs the following tests, which we have used in Phases~1-3: 

{\centerline{\color{myblue}
Test A: $\mathbb{M}(\mathit{test}_A) \leftarrow \textstyle \sum \mathbb{M}(v)
$}}
{\centerline{\color{myblue}Test B: $\mathbb{M}(\mathit{test}_B) \leftarrow \textstyle \sum \mathbb{M}(v^2)
$}}
{\centerline{\color{myblue}Test C: $\mathbb{M}(\mathit{test}_C) \leftarrow  \textstyle \mathbb{M}(v_i^2) - \mathbb{M}(v_i); \: \forall i\in\{1,|v|\}$  
}}

Test~A and Test~B achieve the first objective.  
These tests were executed in \textsc{Step}~4 of {\color{blue}\S\ref{subsec:phase2}} and \textsc{Step}~6 of {\color{blue}\S\ref{subsec:phase3}}. In Test~A, $\mathcal{S}_z$ adds the values of the vector. In Test~B, $\mathcal{S}_z$ multiplies the value itself, and then adds all the values. 
If the client has created a correct vector, {Test~A and Test~B produce one}.
Test~C achieves the second objective and 
is executed in \textsc{Step}~8A of {\color{blue}\S\ref{subsec:phase3}} to prevent a malicious client from accessing files with keywords to which the client has no access. In Test~C, $\mathcal{S}_z$ computes values-wise subtraction between the square of the value and the value itself. If the client has created a correct vector, {Test~C produces a vector with all zeroes}. $\mathcal{S}_z$ learns the results of these tests in cleartext by exchanging shares with other servers.

\medskip
\section{Optimization of the Inverted Index}
\label{sec:Inverted Index Optimization}
\medskip
\noindent
\textbf{Objectives and high-level idea.} 
The method of~\color{blue}\S\ref{subsec:phase2}} has the space and computational overheads, due to the size of the inverted list, in which each entry is padded to the same maximum size to make them identical in size (see~{\color{blue}\S\ref{sec:Data Outsourcing}}). 
For example, if a keyword $k_1$ appears in 1000 files, while all the remaining keywords appear only in a few files, then fake file-ids are added to all the remaining keywords, thereby each row of the inverted list has 1000 file-ids,  and this huge size inverted list results in the overhead. 
To overcome the overhead, below, we develop an algorithm based on two new data structures and explain the high-level idea of the algorithm.

\medskip
\noindent\textbf{New data structures.} We create two arrays: 

\begin{enumerate}[noitemsep,nolistsep,leftmargin=0.01in]
    \item
\medskip%
 \textbf{\textsf{{AddrList}}} ({\color{blue}Table~\ref{table:cleartext_addr}}). Each entry in \textsf{AddrList} contains the starting index position (SiP) of the file-ids associated with the keyword in the second array, the count (CuT) for the files-ids having the keyword, the hash digest ({\color{purple}HD}) of the row, and the hash digest ({\color{orange}HdV}) that is the sum of the hash digest of $p$ positions of the file-ids associated with the keyword (where SiP${<}p{<}$SiP ${+}$ CuT) and is used by servers for client's query verification. We use two colors for differentiating {\color{purple}HD} and {\color{orange}HdV}. \textsf{AddrList} stores keywords according to the positions in AC matrix; see the second row of 
{\color{blue}Table~\ref{tab:ACT_table}}.

\medskip
\item \textbf{\textsf{{OptInv}}} ({\color{blue}Table~\ref{table:cleartext_OptInv}}). \textsf{OptInv} is a single-dimensional array and stores the file-ids associated with the keyword and the hash digest. We allocate all file-ids associated with a keyword in adjacent slots and hash digests over the file-ids in a slot after the last file-ids. To handle new documents, \textsf{{OptInv}} has some empty slots, filled with zeros or random numbers) for each keyword (see ``empty' slots in~{\color{blue}Table~\ref{table:cleartext_OptInv}}). The decision on the number of empty/fake slots depends on DBO or the frequency of the keywords. Of course, these empty/fake slots will be occupied as new files are inserted. {\color{blue}Appendix~\ref{dynamic_operation:Add Operation}}  develops a method to extend the \textsf{OptInv} when it has no free slots.

\end{enumerate}

\medskip%
\noindent\textbf{Outsourcing.} DBO outsources \textsf{AddrList} and \textsf{OptInv} using SSS. 

\medskip%
\noindent\textbf{Example.} \textsf{AddrList} and \textsf{OptInv} are created for cleartext files of {\color{blue}{Table~\ref{tab:cleartext_files}}} and AC matrix of {\color{blue}Table~\ref{tab:ACT_table}}. 
Gray parts of {\color{blue}Tables~\ref{table:cleartext_addr}} and{~\color{blue}\ref{table:cleartext_OptInv}} is written for the purpose of explanation and is not outsourced.

\begin{table}[h!]
\centering
\begin{tabular}{|l|l|l|l|l|l|}  \hline
{\cellcolor[HTML]{B8B8B8}{Keywords}} & \textbf{SiP} & \textbf{CuT} & \textbf{\color{purple}HD} & \textbf{\color{orange}HdV} \\\hline
 {\cellcolor[HTML]{B8B8B8}{are}} & 1 & 3 & ${\color{purple}H_1}=\mathsf{H}(1,3,\textnormal{are})$ & ${\color{orange}h_1}=\mathsf{H}(1) + \mathsf{H}(2) + \mathsf{H}(3)$ \\\hline 
 {\cellcolor[HTML]{B8B8B8}{ana}} & 4 & 3 & ${\color{purple}H_2}=\mathsf{H}(4,3,\textnormal{ana})$ & ${\color{orange}h_2}=\mathsf{H}(4) + \mathsf{H}(5) + \mathsf{H}(6)$  \\\hline 
 {\cellcolor[HTML]{B8B8B8}{how}} & 9 & 2 & ${\color{purple}H_3}=\mathsf{H}(9,2,\textnormal{how})$ & ${\color{orange}h_3} = \mathsf{H}(9) + \mathsf{H}(10)$\\\hline 
\end{tabular}
\captionof{table}{Cleartext \textsf{AddrList} array.}
\label{table:cleartext_addr}
\BBB
\end{table}

\begin{table}[!h]
\centering
\begin{tabular}{|l|l|l|l|l|l|l|l|l|l|}
  \hline

  {\cellcolor[HTML]{B8B8B8}{1}} & {\cellcolor[HTML]{B8B8B8}{2}} & {\cellcolor[HTML]{B8B8B8}{3}} & {\cellcolor[HTML]{B8B8B8}{4}} & {\cellcolor[HTML]{B8B8B8}{5}} & {\cellcolor[HTML]{B8B8B8}{6}} & {\cellcolor[HTML]{B8B8B8}{7}} & {\cellcolor[HTML]{B8B8B8}{8}} & {\cellcolor[HTML]{B8B8B8}{9}} & {\cellcolor[HTML]{B8B8B8}{10}} \\\hline
  
  $f1$ & $f2$ & $hd_1$ & 
  $f2$ & $f3$ & $hd_2$ & empty & empty & $f1$ & $hd_3$\\\hline
\end{tabular}
\captionof{table}{Cleartext \textsf{OptInv} array. {\scriptsize \textbf{Notations.} ${hd_1=\mathsf{H}(f2,(\mathsf{H}(f1,\mathsf{H}(\textnormal{are})))}$, $hd_2=\mathsf{H}(f3,\mathsf{H}(f2,\mathsf{H}(\textnormal{ana})))$, $hd_3=\mathsf{H}(f1,\mathsf{H}(\textnormal{how}))$}}
\label{table:cleartext_OptInv}
\BBB\BB
\end{table}

\medskip
\subsection{Details of  Query Execution}
\label{subsec:Inverted Index Optimization query exection}

%
{\textbf{Objective.}} 
This section explains query execution on \textsf{AddrList} ({\color{blue}Table~\ref{table:cleartext_addr}}) and \textsf{OptInv} ({\color{blue}Table~\ref{table:cleartext_OptInv}}). 
Let $\gamma$ be the maximum number of file-ids associated with a keyword, and let $n$ be the size of \textsf{OptInv}. $\gamma$ and $n$ are known to each entity in \textsc{Doc}$^\star$. 
The client's objective is to fetch the desired file-ids from \textsf{OptInv} for the keyword they have searched in Phase~1, without revealing access-patterns and volume 
to servers. 
While returning $\gamma$ file-ids hides volume from servers, the goal of \textsc{Doc}$^\star$ is to allow the client to learn only the file-ids associated with the keyword; nothing else. 
Also, \textsc{Doc}$^\star$ must allow the server to, first, verify that the client is only fetching the file-ids associated with the keyword to which they have search permission. 

\medskip
\medskip
\noindent
\textbf{High-level idea.} \noindent
\textbf{Query execution.} 
Phase~1 (\textsc{Step~1}-\textsc{Step~3a}) is executed on AC matrix without any modification. Then, Phase 2 is executed with the new steps (developed below) on \textsf{AddrList} and \textsf{OptInv}. 
After executing Phase~1, the client learns which row of \textsf{AddrList} needs to be fetched. Then, \textsc{Step 3b} - \textsc{Step 5a} of~{\color{blue}\S\ref{subsubsec:Algorithms in Phase2}} are executed, resulting in SiP, CuT, and {\color{purple}HD} of the desired row of \textsf{AddrList}. Note that the servers keep {\color{orange}HdV} value of the returned row, for verification at a later stage. 
Next, the client executes the following \textsc{Step~A} - \textsc{Step~C}, to learn the file-ids from \textsf{OptInv}. 

\medskip
\medskip
\noindent
\textbf{\textsc{Step~A}: \emph{Client --- transmitting two vectors}.} Clients want to fetch~$\gamma$ file-ids, regardless of the number of files associated with a keyword. To do so, the client creates two vectors: $\mathit{row\_vec}$ and $\mathit{pos\_vec}$. 

\noindent
$\boldsymbol{\mathit{row\_vec}}$: The client interprets \textsf{OptInv} as a  $x{\times} y{=}n$ matrix 
 and computes the row number of the matrix that will contain file-ids, starting from SiP to SiP+CuT for the keyword. Based on the row number, the client creates a \emph{row vector}, denoted by $\mathit{row\_vec}$. We assume $y \geq \gamma$; thus, up to $\gamma$ file-ids associated with a keyword may span at most two rows in the matrix. 
 For simplicity, we assume all $\gamma$ file-ids to be fetched appear in a single, $i^{\mathit{th}}$, row of the matrix, and thus, $\mathit{row\_vec}$ contains $x$ elements,  with $x-1$ zeros and one at the $i^{\mathit{th}}$ position.

\smallskip
\noindent
$\boldsymbol{\mathit{pos\_vec}}$: The client constructs a position vector (denoted $\mathit{pos\_vec}$) of length $y$, where all elements are ones except for $\Delta {\leq} \gamma$ zeros at the specified positions from SiP to SiP+CuT (where a keyword may be associated with $\Delta {\leq} \gamma$ file-ids). Shares of $\mathit{row\_vec}$ and $\mathit{pos\_vec}$ are then created by the client and sent to servers.

\medskip
\medskip%
\noindent
\textbf{\textsc{Step~B}: \emph{Servers --- returning $y$ file-ids}.} The objective of servers is to send only the desired $\Delta$ file-ids to the client. To do so, servers: 
(\emph{i})~obliviously selects the desired row of the matrix --- by organizing \textsf{OptInv} in the form of $x\times y$ matrix, multiply all values of the $i^{\mathit{th}}$ row of the matrix by $\mathit{row\_vec}[i]$, and finally, adds all the values of each column, resulting in 
$y\geq \gamma$ file-ids, \textit{i}.\textit{e}., 

{\color{myblue}
$\mathbb{M}(\mathit{fids})[i]\leftarrow \sum_{1\leq j \leq y}(\mathbb{M}(\textnormal{\textsf{OptInv}})[i][j] \times \mathbb{M}(\mathit{row\_vec})[i])$}

\noindent
(\emph{ii})~To prevent sending extra file-ids, 
obliviously remove the file-ids not associated with the keyword 
by position-wise multiplying \textnormal{\hl{$\mathbb{M}(\mathsf{RN})[i]$}} with  $\mathit{pos\_vec}$, and then, position-wise adding the output to  $\mathbb{M}(\mathit{fids})[]$ file-ids, $i\in \{1,y\}$, \textit{i}.\textit{e}., 

\noindent
{\color{myblue}
$\mathbb{M}(\mathit{final\_fids})[i] \leftarrow \mathbb{M}(\mathit{fids})[i] + (\textnormal{\hl{$\mathbb{M}(\mathsf{RN})[i]$}}[i] {\times} \mathbb{M}(\mathit{pos\_vec})[i])$}


\medskip
\medskip
\noindent
\textbf{\textsc{Step~C}: \emph{Client --- interpolation for knowing file-ids}.} The client interpolates $\mathbb{M}(\mathit{final\_fids})$. Since servers have added random numbers to non-desired file-ids (\textit{i}.\textit{e}., $\textnormal{\hl{$\mathbb{M}(\mathsf{RN})[i]$}}\times \mathbb{M}(\mathit{pos\_vec})[i]$, as $\mathit{pos\_vec}[i]=0$ for the desired position; otherwise one), the client will not learn any additional non-desired file-ids.


\medskip
\subsubsection{\textbf{Example of Query Example}}
\label{subsec:example_opt_inv}
We continue from the example of~{\color{blue}Table~\ref{tab:Example of Phase2}}. In \textsc{Step 3b}, the client \texttt{Lisa} creates a vector $\langle 1,0,0\rangle$ and sends it to three servers in share form. 
\emph{{For the purpose of 
understanding, below, we do not show operations over shares}}.

\medskip
\medskip
\noindent
\textbf{\textsc{Step 4}: \textit{Servers:}} perform a dot product of the received vector $\langle \mathbb{M}(1),\mathbb{M}(0),\mathbb{M}(0)\rangle$ and \textsf{AddrList} 
({\color{blue}Table~\ref{table:cleartext_addr}}) and returns $\langle\mathbb{M}(1), \mathbb{M}(3), {\color{purple}\mathbb{M}(H_1)}\rangle$ to the client. Servers \emph{\textbf{{ keep ${\color{orange}\mathbb{M}(h_1)}$ at their ends}}}. All such shares are of degree two

\medskip
\medskip
\noindent
\textbf{\textsc{Step~A}: \textit{Client:}} interpolates the values and learns SiP to be 1 and CuT to be 3. Also, ${\color{purple}H_1}$ verifies the correctness of SiP and CuT. {Client} learns based on SiP and CuT that they need to fetch index 1, 2, and 3 from \textsf{OptInv}. 
By interpreting \textsf{OptInv} as a $3{\times}4$ matrix, 
indexes 1-3 will appear in the first row.  Thus, the client creates $\mathit{row\_vector}{=}[1,0,0]$,  $\mathit{pos\_vec}{=}[0,0,0,1]$, and outsources them to three servers in the share form.

\medskip
\medskip
\noindent
\textbf{\textsc{Step~B:} \textit{Servers:}} 
organize \textsf{OptInv} in a $3{\times 4}$ matrix (see {\color{blue}Table~\ref{table:computation stepA}}) and compute the following: 

\begin{figure}[h]
\begin{minipage}[t]{0.4\linewidth}
 \begin{tabular}{l|l|l|l|l|p{0.95cm}|p{0.95cm}|p{0.95cm}|p{0.95cm}|}  \hline
 $\mathbb{M}(f_1)$ & $\mathbb{M}(f_2)$ & $\mathbb{M}(\mathit{hd}_1)$ & $\mathbb{M}(f_2)$ \\\hline 

  $\mathbb{M}(f_3)$ &  $\mathbb{M}(\mathit{hd}_2)$ & $\mathbb{M}(\mathit{empty})$ & $\mathbb{M}(\mathit{empty})$ \\\hline

  $\mathbb{M}(f_1)$ &  $\mathbb{M}(\mathit{hd}_3))$ & $\mathbb{M}(\mathit{empty})$ & $\mathbb{M}(\mathit{empty})$ \\\hline

\end{tabular}
\label{table:optinv_in_matrix_form}
\end{minipage}
\hspace{2.35cm}
\begin{minipage}[t]{0.045\linewidth}
 \begin{tabular}{l}    \\
    $\odot$ \\ 
        \end{tabular}
\end{minipage}
\begin{minipage}[t]{0.25\linewidth}
 \begin{tabular}{|l|}    \hline
   
    {\cellcolor[HTML]{33abd3}$\mathbb{M}(1)$} \\ \hline
    {\cellcolor[HTML]{33abd3}$\mathbb{M}(0)$} \\ \hline
    {\cellcolor[HTML]{33abd3}$\mathbb{M}(0)$} \\ \hline
     \end{tabular}
\label{tab:correct_row_vec}
\end{minipage}

\begin{minipage}[h]{\linewidth}
 \begin{tabular}{|l|p{1.12cm}|p{1.35cm}|p{1.35cm}|p{0.95cm}|p{0.95cm}|p{0.95cm}|p{0.95cm}|}  \hline
 {\cellcolor[HTML]{fff333}$\mathbb{M}(f_1)$} & 
 {\cellcolor[HTML]{fff333}$\mathbb{M}(f_2)$} & 
 {\cellcolor[HTML]{fff333}$\mathbb{M}(\mathit{hd}_1)$} & 
 {\cellcolor[HTML]{fff333}$\mathbb{M}(f_2)$} \\\hline 
\end{tabular}
\end{minipage}
\captionof{table}{Left most: \textsf{OptInv}. Right most: a row vector in cyan. Bottom part is the output {\scriptsize{($\sum_{j=0}^{j=y}(\mathbb{M}(\textnormal{\textsf{OptInv}})[i,j] {\times} \mathbb{M}(\mathit{row\_vec})[i])$) in yellow.}}}
\label{table:computation stepA}
\BBB
\end{figure}

Assume \textnormal{\hl{$\mathbb{M}(\mathsf{RN})[i]$}} ${=}\langle$\textnormal{\hl{$\mathbb{M}(r_1^z)$}},
\textnormal{\hl{$\mathbb{M}(r_2^z)$}},
\textnormal{\hl{$\mathbb{M}(r_3^z)$}},
\textnormal{\hl{$\mathbb{M}(r_4^z)$}}$\rangle$.
Servers do the following computation and send their output to the client:


{\centerline{
$\textnormal{\hl{$\mathbb{M}(r_1^z)$}}{\times}\mathbb{M}(0)+\mathbb{M}(f_1) = \mathbb{M}(f_1), \quad \textnormal{\hl{$\mathbb{M}(r_2^z)$}}{\times} \mathbb{M}(0)+\mathbb{M}(f_2)= \mathbb{M}(f_2)$}}

{\centerline{
$\textnormal{\hl{$\mathbb{M}(r_3^z)$}}{\times} \mathbb{M}(0) + \mathbb{M}(\mathit{hd}_1) {=} \mathbb{M}(\mathit{hd}_1)$,\footnote{\scriptsize{We multiply very large random numbers also to hide the location of hash digests, if the returning row contains several hash digests corresponding to other keywords/file-ids.}} \quad
$\textnormal{\hl{$\mathbb{M}(r_4^z)$}}{\times} \mathbb{M}(1){+}\mathbb{M}(f_2){=}\mathbb{M}(\mathit{Random})$}}

\medskip
\medskip
\noindent
\textbf{\textsc{Step~C:} \textit{Client:}} interpolates and learns the desired file-ids: $f_1$, $f_2$, and their hash digests, which is used to verify the correctness of the file-ids. The random number will not provide any information about the last file-id, (though it is the second desired file-id $f_2$).

\medskip
\subsection{Server Size Verification}
We have shown how the client can fetch $\gamma$ file-ids, without incurring space overhead in maintaining the optimized inverted list in~{\color{blue}\S\ref{sec:Inverted Index Optimization}}. Below, we develop a method for servers to verify $\mathit{row\_vec}$ and $\mathit{pos\_vec}$ used to fetch data from \textsf{OptInv}. 

\medskip
\medskip%
\noindent
\textbf{Objectives and high-level idea.} The client may try to retrieve random file-ids and/or more than the desired file ids by sending \emph{incorrect $\mathit{row\_vec}$ and $\mathit{pos\_vec}$ (\emph{both in SSS form}}). For instance, in the example of~{\color{blue}\S\ref{subsec:example_opt_inv}}, {\color{blue}Table~\ref{tab:Example of Phase1}}, Lisa may create $\mathit{row\_vec}=[0,1,0]$ and $\mathit{pos\_vec}=[0,0,0,0]$ to learn the entire second row of the matrix; see {\color{blue}Table~\ref{table:computation stepA}}. Thus, the servers need to verify such vectors against the SiP and CuT values, which were sent in \textsc{Step 4} (see {\color{blue}\S\ref{sec:Inverted Index Optimization}}) before sending file-ids to the client.
To do so, the servers compute hash digest on some numbers and compare the hash digest value of the {\color{orange}HdV} column of \textsf{AddrList}, particularly, the {\color{orange}HdV} value, which was obtained in \textsc{Step 4} (see~{\color{blue}\S\ref{subsec:Inverted Index Optimization query exection}}) and has not been sent to the client. The following steps will be executed before \textsc{Step~B} of~{\color{blue}\S\ref{subsec:Inverted Index Optimization query exection}}, \textit{i}.\textit{e}., before sending the file-ids to the clients.

\medskip
\noindent
\subsubsection{\textbf{Details of the Method}}

Servers perform the following:  
\begin{enumerate}[noitemsep,nolistsep,leftmargin=0.01in]
  \item 
  Subtract the $\mathbb{\mathit{pos\_vec}}$ from a vector containing $y$ one in cleartext, \textit{i}.\textit{e}., $\langle 1_1,\ldots, 1_y\rangle$. Note that the output will be in SSS form of degree one.

  \item 
  Compute hash digest over numbers 1 to $n$ (here, the hash digest will be in cleartext) and then, 
  organize such hash digests of 1 to $n$ numbers into a $x\times y$ matrix.

  \item 
  Multiply the $i^{\mathit{th}}$ value of $\mathbb{M}(\mathit{row\_vec})$ to all the values of the $i^{\mathit{th}}$ row of the matrix of hash digests and compute the sum over each of the $y$ columns. This results in a single row of the matrix in SSS form of degree one, since $\mathit{row\_vec}$ is in SSS form of degree one.

  \item 
  Perform a dot product of the output of the step ({1}) and the output of step ({3}), and this results in a value, say $\mathbb{M}(\mathit{val})$, in SSS form of polynomial degree two.

  \item 
  Subtract $\mathbb{M}(\mathit{val})$ with 
  {\color{orange}HdV} value computed in \textsc{Step 4} (see~{\color{blue}\S\ref{subsec:example_opt_inv}}), and this results in a value, say $\mathbb{M}(\mathit{op})$, which is 
  sent to all other servers. Note that \emph{if the client has created the correct $\mathit{row\_vec}$ and $\mathit{pos\_vec}$, then $\mathbb{M}(\mathit{op})=\mathbb{M}(0)$}. 

  \item Interpolate the values, after receiving shares from the other two servers. If the output value is zero, then servers send the output of \textsc{Step~B} of~{\color{blue}\S\ref{subsec:Inverted Index Optimization query exection}} to the client.
 
\end{enumerate}

\medskip
\subsubsection{\textbf{Example of Server-side Verification}}
\label{subsec:Example of server-side verification_opt_inv}

Suppose, the client learns SiP to be 1 and CuT to be 3 (see \textsc{Step~4} of~{\color{blue}\S\ref{subsec:example_opt_inv}}), but creates wrong $\mathit{row\_vec} = [0, 1, 0]$ and correct $\mathit{pos\_vec} = [0, 0, 0, 1]$ in SSS form. 
Below, we show with the help of an example, how servers can learn about the wrong  $\mathbb{M}(\mathit{row\_vec})$, without knowing it in cleartext. For the purpose of simplicity and understanding, we explain operations over cleartext.


Servers subtract $\mathbb{M}(\mathit{pos\_vec}){=}\mathbb{M}(0,0,0,1)$ from $y$ ones, as: $1111-\mathbb{M}(0)\mathbb{M}(0)\mathbb{M}(0)\mathbb{M}(1)=\mathbb{M}(1)\mathbb{M}(1)\mathbb{M}(1)\mathbb{M}(0)$. Next, servers compute hash digests on 1 to 10 numbers, organize them in a matrix form, and multiply $\mathbb{M}(\mathit{row\_vec})$; and this results in $\langle \mathbb{M}(\mathsf{H}(5)), \mathbb{M}(\mathsf{H}(6)),\mathbb{M}(\mathsf{H}(7)), \mathbb{M}(\mathsf{H}(8))\rangle$ (see {\color{blue}Table~\ref{table:computation step7_wrong}}).
Then, servers perform a dot product between $\langle \mathbb{M}(\mathsf{H}(5)),
\mathbb{M}(\mathsf{H}(6)),
\mathbb{M}(\mathsf{H}(7)),
\mathbb{M}(\mathsf{H}(8))\rangle$ and 
$\langle\mathbb{M}(1),
\mathbb{M}(1),
\mathbb{M}(1),
\mathbb{M}(0)\rangle$, and this produces  $\mathbb{M}(\mathit{val}){=}\mathbb{M}(\mathsf{H}(5)){+}\mathbb{M}(\mathsf{H}(6)){+}\mathbb{M}(\mathsf{H}(7))$.

\begin{figure}[h]
\begin{minipage}[t]{0.4\linewidth}
 \begin{tabular}{|p{0.95cm}|p{0.95cm}|p{0.95cm}|p{0.95cm}|}  \hline
 $\mathsf{H}(1)$ & $\mathsf{H}(2)$ & $\mathsf{H}(3)$ & $\mathsf{H}(4)$ \\\hline 

  $\mathsf{H}(5)$ & $\mathsf{H}(6)$ & $\mathsf{H}(7)$ & $\mathsf{H}(8)$ \\\hline 

   $\mathsf{H}(9)$ & $\mathsf{H}(10)$ & & \\\hline 
\end{tabular}
\label{table:hash_digest_of numbers}
\end{minipage}
\hspace{2cm}
\begin{minipage}[t]{0.08\linewidth}
 \begin{tabular}{l}    \\
    $\odot$ \\ 
        \end{tabular}
\end{minipage}
\begin{minipage}[t]{0.25\linewidth}
 \begin{tabular}{|l|}    \hline
   
    {\cellcolor[HTML]{33abd3}$\mathbb{M}(0)$} \\ \hline
    {\cellcolor[HTML]{33abd3}$\mathbb{M}(1)$} \\ \hline
    {\cellcolor[HTML]{33abd3}$\mathbb{M}(0)$} \\ \hline
     \end{tabular}
\label{tab:wrong_row_vec}
\end{minipage}

\begin{minipage}[h]{\linewidth}
\scriptsize
 \begin{tabular}{|p{0.95cm}|p{0.95cm}|p{0.95cm}|p{0.95cm}|}  \hline

 {\cellcolor[HTML]{fff333}$\mathbb{M}(\mathsf{H}(5))$} & 
 {\cellcolor[HTML]{fff333}$\mathbb{M}(\mathsf{H}(6))$} & 
 {\cellcolor[HTML]{fff333}$\mathbb{M}(\mathsf{H}(7))$} & 
 {\cellcolor[HTML]{fff333}$\mathbb{M}(\mathsf{H}(8))$} \\\hline 
\end{tabular}
\end{minipage}
\captionof{table}{Left most: Hash digest of 10 numbers. Right most: a wrong row vector in cyan. Bottom part is the output in yellow.}
\label{table:computation step7_wrong}
\BBB
\end{figure}

Then, servers subtract $\mathbb{M}(\mathit{val})$ from  the output of \textsc{Step 4} that is {\color{orange}$\mathbb{M}(h_1)$}, which equals to $\mathbb{M}(\mathsf{H}(1)) + \mathbb{M}(\mathsf{H}(2)) + \mathbb{M}(\mathsf{H}(3))$ (see~{\color{blue}Table~\ref{table:cleartext_addr}}). The output of the subtraction operation is sent to all servers, and after interpolation, the final output is not zero. This shows that 
the client has not created the vectors correctly.\footnote{\scriptsize
Recall that all these arithmetic operations are performed in $\mathbb{Z}_p$, the equation $h^{\prime}=h$ is in fact $h^{\prime}\bmod{p}=h\bmod{p}$. If $p$ is chosen large enough, the error probability can be ignored.} 

\medskip
\subsubsection{\textbf{Discussion on Security}} 

The basic idea of above this verification method is comparing the corresponding hash digest ${\color{orange}h}$ in {\color{blue}Table~\ref{table:cleartext_addr}} with the calculated value $h^{\prime}$ from {\color{blue}Table~\ref{table:computation step7_wrong}}. If the client creates wrong vectors, the server will obtain the wrong hash addition, which is not equal to the fixed one in {\color{blue}Table~\ref{table:cleartext_addr}}. The whole procedure is executed in SSS form. 
That is to say, servers cannot learn any extra information except the verification result (\textit{i}.\textit{e}., 0 or a random number). Meanwhile, the client’s wrong action will also be detected.



\medskip
\section{Experimental Evaluation}
\label{sec:experiments}
In this section, we evaluate the following: 

\begin{enumerate}[nolistsep,noitemsep,leftmargin=0.01in]
    
\item The size of secret-shared data produced by \textsc{Doc}$^\star$ --- Exp 1.

\item The end-to-end processing time for a query in \textsc{Doc}$^\star$ --- Exp 2.

\item Time on different sizes of data to show scalability of \textsc{Doc}$^\star$  --- Exp~3.


\item Time taken by different implementations of inverted list --- Exp 4.






\item Time taken by verification algorithms at the server and the client--- Exp~5.

\item  Performance of \textsc{Doc}$^\star$ in a wide-area network --- Exp~6.

\item Time taken by \textsc{Doc}$^\star$ and other systems --- Exp~7.
\end{enumerate}

\noindent\textbf{Machines.} 
We selected four \texttt{c7a.32xlarge} AWS machines, each with 128 cores and 256 GB RAM, playing the role of three servers. The client machine was with 32 cores and 64 GB RAM. These machines were located in different zones (which are connected over wide-area networks), of AWS Virginia region. 

\noindent\textbf{Data.} 
We use Enron dataset~\cite{enron}, a commonly used real data to evaluate techniques for document stores. This \emph{dataset was also used in other recent document store work~\cite{DBLP:conf/osdi/DautermanFLPS20,Titanium,metal}}. For all experiments, except  Exp~3 for scalability, we created a dataset, containing 5K searchable keywords in 500K files. 
The code is written in Java and contains more than 5000 lines. 
Time is calculated by taking the average of 20 program runs and shown in milliseconds (ms).

{\color{black}
\noindent
\textbf{Client-side storage.}
The maximum amount of data that a client stores is the number of file-ids to be retrieved from servers during Phase~3. In the worst case, if a keyword appears in all files, the space overhead equals the size of all file-ids. Since we selected 500K files, storing 500K file-ids requires a client to have 2MB of space.
}

\medskip
\subsection{Evaluation of \textsc{Doc}$\boldsymbol{^\star}$}
\label{subsec:DEvaluation}

\medskip
\noindent
\textbf{Exp 1: Share generation.}
We create shares of AC matrix, inverted lists, and files in the non-optimized version ({\color{blue}\S\ref{subsec:phase2}}) 
and of AC matrix, \textsf{AddrList}, \textsf{OptInv}, and files in the optimized version ({\color{blue}\S\ref{sec:Inverted Index Optimization}}).

\noindent
{\color{blue}Table~\ref{tab:Exp 1: Size of the shares and generation time}} shows the size of these data structures in SSS form at a single server for the dataset, containing 5K keywords in 500K files. Keywords appear on average in 38 files, and the median was 23. 
Skewness (\textit{i}.\textit{e}., the difference between the maximum and minimum number of files associated with a keyword) results in a large size of the inverted list ($\approx$1.6GB), due to padding each entry of the inverted list with fake values to have them 110K elements, as a keyword appears in at most 110K files. AC matrix contains access rights for 5K keywords. We created 4,096 clients, like existing work~\cite{metal,Titanium}.

\begin{table}[!h]
\BB
\begin{tabular}{|p{4cm}|l|l|l|l|}\hline
\centering
~    & AC matrix & Phase~2 data & Files \\\hline
Non-optimized, \textit{i}.\textit{e}., \textbf{inverted list-based \textsc{Doc}$^\star$}  & 618.7 MB  & 1.6 GB  & 3.5 GB \\\hline
Optimized, \textit{i}.\textit{e}., \textbf{\textsf{OptInv} \& \textsf{AddrList}-based \textsc{Doc}$^\star$}  & 618.7 MB  & 139.6 MB  & 3.5 GB   \\\hline
\end{tabular}
\caption{Exp 1: Size of the shares at a single server.}
\label{tab:Exp 1: Size of the shares and generation time}
\BBB
\end{table}

\begin{figure}[!t]
\BBB
\begin{center}
			\begin{minipage}{.99\linewidth}
				\centering
			
   \includegraphics[scale=0.6]{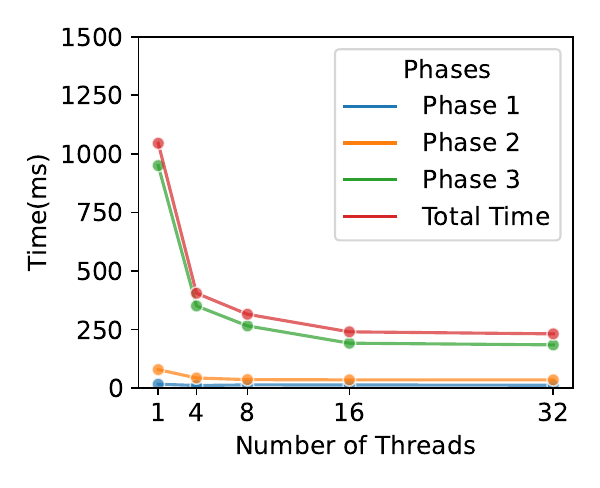}
   \BB\BBB
   \subcaption{Optimized implementation using \textsf{OptInv} and \textsf{AddrList}.}
				\label{fig:optimized parallelism}
			\end{minipage}
            
			\begin{minipage}{.99\linewidth}
				\centering
				\includegraphics[scale=0.6]{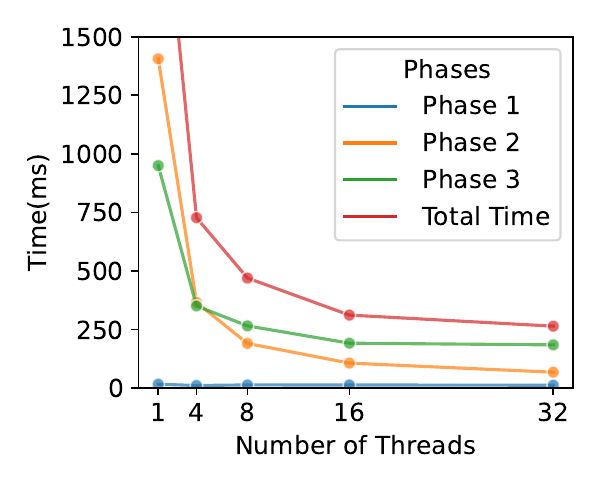}\BBB\B
				\subcaption{Non-optimized implementation using an inverted list.}
				\label{fig:non-optimized  parallelism}
			\end{minipage}			
			  \end{center}
		\BBB
		\caption{Exp 2: \textsc{Doc}$\boldsymbol{^\star}$ end-to-end processing time.}
		\label{fig:impact of thread parallelism}
	\end{figure}

\bgroup
\def\arraystretch{0.9}
\begin{table}[!h]
\begin{tabular}
{|l||l|l|l|l|}\hline
Entity     & Phase 1 & Phase 2  & Phase 3 & Total   \\\hline\hline

Client & 3 & 8.7 & 32.9 & 44.6 \\\hline

Server & 7.3 & 9.9 & 67.9 & 85.1  \\\hline

Network & 1.8 & 16.3 & 83.7 & 101.8 \\\hline
{\cellcolor[HTML]{C1FD8E}{\textbf{Total (milliseconds)}}} & {\cellcolor[HTML]{C1FD8E}{\textbf{12.1}}} & 
{\cellcolor[HTML]{C1FD8E}{\textbf{34.9}}} & 
{\cellcolor[HTML]{C1FD8E}{\textbf{184.5}}} & 
{\cellcolor[HTML]{00CC99}{\textbf{231.5}}}  \\\hline

Variation  & $\pm$2.18 & $\pm$2.5 & $\pm$3.42 & $\pm$8.1 \\\hline


\end{tabular}
\caption{Exp 2: Time on 5K keywords, 500K files, 32 threads.}
\label{tab:Exp 2:breakdown}
\BB
\end{table}
\egroup

\medskip
\medskip%
\noindent
\textbf{Exp 2: \textsc{Doc}$\boldsymbol{^\star}$ end-to-end processing time.} We implemented multi-threaded programs for all three phases of \textsc{Doc}$^\star$. These programs partition the data into multiple blocks of equal size, and each block is processed by a separate thread. 
{\color{blue}Figure~\ref{fig:impact of thread parallelism}} shows that as increasing the number of threads from 1 to 32, the processing time decreases for each phase. We implemented both versions of Phase~2, \textit{i}.\textit{e}., \textsf{OptInv} ({\color{blue}Figure~\ref{fig:optimized parallelism}}) and inverted list  ({\color{blue}Figure~\ref{fig:non-optimized  parallelism}}).

\textsf{OptInv}-based \textsc{Doc}$\boldsymbol{^\star}$, for all three phases, took $\approx$1 second using one thread, whereas took $\approx$231.5ms using 32 threads. \textsf{OptInv} performs better compared to the inverted list for 5K keywords in 500K files, since the size of \textsf{OptInv} is much smaller than the size of the inverted list (due to the absence of padding; as shown in {\color{blue}Table~\ref{tab:Exp 1: Size of the shares and generation time}}). {\color{blue}Table~\ref{tab:Exp 2:breakdown}} shows a breakdown of the time taken by the client, servers, and network for each phase of \textsc{Doc}$^\star$ with its optimized implementation that uses \textsf{AddrList} and \textsf{OptInv}. Important to note that \textbf{\emph{our access control method is also {highly efficient}, taking at most 12.1ms}} (see the first column). Further, the client's processing time is less than the server's in all three phases. 
Exp~4 will show a situation when the inverted list will work better compared to \textsf{OptInv}. Last row of {\color{blue}Table~\ref{tab:Exp 2:breakdown}} shows the variation of time in different phases. 

\begin{figure}[!h]
\B
\begin{center}
			\begin{minipage}{.99\linewidth}
				\centering
			\includegraphics[scale=0.6]{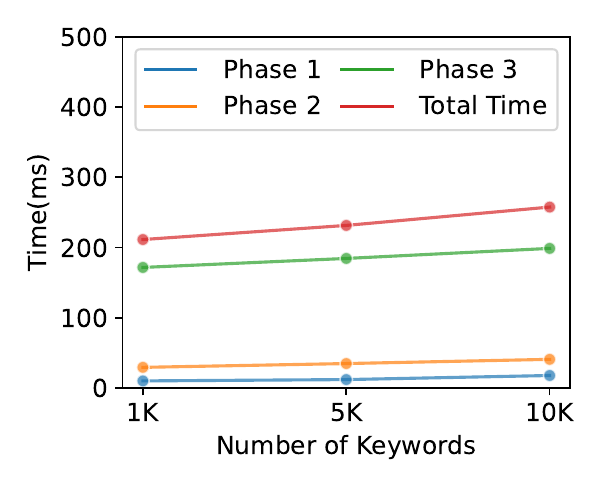}\BBB
				\subcaption{Varying number of keywords.}
				\label{fig:Varying number of keywords}
			\end{minipage}
            
			\begin{minipage}{.99\linewidth}
				\centering
				\includegraphics[scale=0.6]{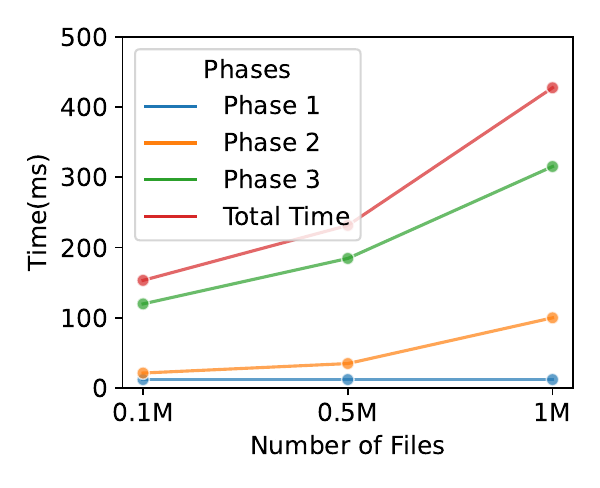}\BBB
				\subcaption{Varying number of files.}
				\label{fig:Varying number of file}
			\end{minipage}			
			  \end{center}
		\BBB
		\caption{Exp 3: Scalability test using 32 threads.}
		\label{fig:Sclability test using 32 threads}
		\BB
	\end{figure}

\medskip
\medskip%
\noindent
\textbf{Exp 3: Scalability of \textsc{Doc}$\boldsymbol{^\star}$.} To evaluate the scalability of \textsc{Doc}$^\star$, we created two types of datasets: 
(\textit{i})~varying the number of keywords from 1K, 5K, 10K in 500K files, and
(\textit{ii})~varying the number of files from 100K, 500K, and 1M with a fixed number of keywords to 5K. We run this and all the following experiments using 32 threads, since Exp 2 justifies the best performance with 32 threads. 
{\color{blue}Figure~\ref{fig:Sclability test using 32 threads}} shows that as the data increases, the entire computation time increases. Particularly, \textsc{Doc}$^\star$ took 257.6ms for 10K keywords in 500K files ({\color{blue}Figure~\ref{fig:Varying number of keywords}}), and 427.4ms for 5K keywords in 1M files ({\color{blue}Figure~\ref{fig:Varying number of file}}).

\begin{wrapfigure}{r}{0.30\textwidth}
\B
\begin{center}
\includegraphics[scale=0.5]{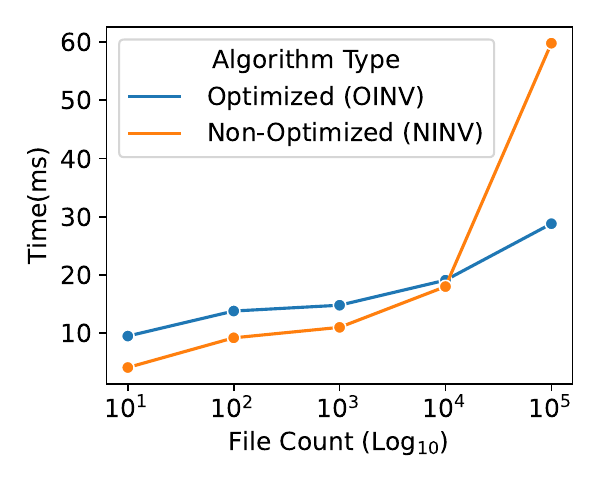}
\end{center}
\BBB
\caption{Exp 4: Different implementations of Phase~2.}
\label{fig:impact of file counts}
\BB
\end{wrapfigure}

\medskip
\medskip
\noindent
\textbf{Exp 4: Different implementation of the inverted list.} Phase~2 can be implemented using the inverted list, denoted by \textsc{NInv} below, ({\color{blue}Table~\ref{tab:inverted_list}}, {\color{blue}\S\ref{subsec:phase2}}) or an optimized approach using \textsf{AddrList} and \textsf{OptInv}, denoted by \textsc{OInv} below, ({\color{blue}Table~\ref{table:cleartext_addr}}, {\color{blue}Table~\ref{table:cleartext_OptInv}}, {\color{blue}\S\ref{sec:Inverted Index Optimization}}). 
\textsc{NInv} and \textsc{OInv} differ in the following two aspects: (\textit{i})~\textsc{NInv} takes one round of communication between a server and a client, while \textsc{OInv} takes two rounds, and (\textit{ii})~\textsc{NInv} adds fake file-ids to each keyword to make them identical in length in inverted list, while \textsc{OInv} does not add fake file-ids. This experiment investigates when we can use one of the methods for Phase~2. We implemented \textsc{NInv} and \textsc{OInv} on varying numbers of maximum numbers of files associated with a keyword. We considered four cases: a keyword can appear in at most 10, 100, 1,000, 10,000, and 100,000 files --- in other words, an entry of the inverted list will contain these many files, and we selected 5,000 keywords.   {\color{blue}Figure~\ref{fig:impact of file counts}} shows the results of this experiment, where $x$-\emph{axis is on a log scale and refers to these file numbers}. 


For the case of at most 1,000 files containing the same keyword (where on average a keyword appears in 528 files), \textsc{NInv} works best, due to less total processing time compared to the time of \textsc{OInv}. Particularly, in \textsc{NInv}, the processing time (9.5ms) at both servers and the client and transmission time (1.3ms), while \textsc{OInv} took 12.1ms processing time and 2.4ms for transmission. 
Here, the size of the inverted list (25MB) was larger than the size of \textsf{AddrList} and \textsf{OptInv} (22MB). 
\textsc{NInv} works similar to \textsc{OInv} in the case of 10K files. As we increase the number of keywords to 100K, \textsc{NInv} does not work best, due to significantly increasing the size of the inverted list by padding fake file-ids (1.52GB in \textsc{NInv} vs 257MB in \textsc{OInv}). The total time was 59.8ms for \textsc{NInv} compared to 28.8ms for \textsc{OInv}.

These observations highlight that \emph{\textsc{NInv} (inverted list-based) method is effective when the difference between the maximum and minimum number of files associated with a keyword is relatively small. In contrast, \textsc{OInv} optimized implementation performs well in scenarios, having a significant difference between the maximum and minimum number of files associated with a keyword}.

\medskip
\medskip
\noindent
\textbf{Exp 5: Impact of verification.} \textsc{Doc}$^\star$ provides verification options for both servers and clients. 
{\color{blue}Table~\ref{tab:Exp 5: verification_overhead}} compares the processing time of \textsc{Doc}$^\star$ with verification and without verification for 5K keywords in 500K files using 32 threads. 
Only Phase~2 takes more time in 
server-side verification for $\mathit{row\_vec}$ over \textsf{OptInv}, 
 which involves multiplication over hash digests of 160 bits for each value of \textsf{OptInv}.

\bgroup
\def\arraystretch{0.9}
\begin{table}[!h]
\B
\begin{tabular}
{|l||l|l|l|l|}\hline
     & Phase 1 & Phase 2  & Phase 3 & Total   \\\hline\hline
\multicolumn{5}{|c|}{Small data}\\\hline
{\cellcolor[HTML]{FFE4C4}{\textbf{Without verification}}} & {\cellcolor[HTML]{FFE4C4}{\textbf{12.1}}} & 
{\cellcolor[HTML]{FFE4C4}{\textbf{34.9}}} & 
{\cellcolor[HTML]{FFE4C4}{\textbf{184.5}}} & 
{\cellcolor[HTML]{FFE4C4}{\textbf{231.5}}}  \\\hline

{\cellcolor[HTML]{C1FD8E}{\textbf{With verification}}} & {\cellcolor[HTML]{C1FD8E}{\textbf{13.1}}} & 
{\cellcolor[HTML]{C1FD8E}{\textbf{90.0}}} & 
{\cellcolor[HTML]{C1FD8E}{\textbf{210.1}}} & 
{\cellcolor[HTML]{00CC99}{\textbf{313.20}}}  \\\hline



\end{tabular}
\caption{Exp 5: Verification time (ms) using 32 threads.}
\BBB
\label{tab:Exp 5: verification_overhead}
\BBB\BB
\end{table}
\egroup

\definecolor{kellygreen}{rgb}{0.3, 0.73, 0.09}
\bgroup
\def\arraystretch{1.0}
\begin{table*}[!t]
\BBB\BBB
\hspace{-2cm}

\begin{tabular}{|p{4.05cm}|l|l|l|p{2.0cm}|p{2.05cm}|p{3.05cm}|
p{0.95cm}|l|
p{1.0cm}
|p{1.32cm}|p{3cm}|l|l|}\hline
Systems  & 
\textbf{Sieve~\cite{Sieve16}} & \textbf{Ghoster
~\cite{Ghostor20}} &
\textbf{Dory~\cite{DBLP:conf/osdi/DautermanFLPS20}} & \textbf{Metal~\cite{metal}} & \textbf{Titanium~\cite{Titanium}} & \textbf{\textsc{Doc}}$\boldsymbol{^\star}$ \\\hline\hline


Offered security & 
\multicolumn{5}{c|}{Computational security} & {Unconditional security} \\\hline

Cryptographic techniques & 
Encryption & Encryption &
Encryption  & Encryption \& binary shares & Additive shares \& trusted proxy for access control (\S7 of~\cite{Titanium}) &  Shamir's shares \\\hline

Number of servers & 
1 
&1 & 2 & 2 &2 & 4 \\\hline


Access control & 
Attribute & 
File-ID & N/A & File-ID & File-ID & Keyword, Attribute, File-ID \\\hline

Granularity of access & 
{\cellcolor[HTML]{C1FD8E}{\textbf{Fine-grain}}} &  
{{\cellcolor[HTML]{FF6961}{\textbf{Coarse-grain}}}} & N/A
&
\multicolumn{2}{c|}{{\cellcolor[HTML]{FF6961}{\textbf{Coarse-grain}}}} & {\cellcolor[HTML]{C1FD8E}{\textbf{Fine-grain \& coarse-grain}}} \\\hline

Trusted proxy for access control & 

\multicolumn{2}{c|}{
{\cellcolor[HTML]{C1FD8E}{\textbf{No}}}} &
N/A &
{\cellcolor[HTML]{C1FD8E}{\textbf{No}}} 
& {\cellcolor[HTML]{FF6961}{\textbf{Yes}}} & {\cellcolor[HTML]{C1FD8E}{\textbf{No}}} \\\hline


Complexity of granting access & 
$\pie{0}$ & $\pie{0}$ out-of-band  & N/A & $\pie{360}$ out-of-band & $\pie{0}$ & $\pie{0}$\\\hline

Complexity of revoking access & 
$\pie{360}$ & $\pie{360}$ & N/A& \hcircle{} & $\pie{0}$ & $\pie{0}$\\\hline



False positives in returning answers & 
{\cellcolor[HTML]{C1FD8E}{\textbf{No}}} & {\cellcolor[HTML]{C1FD8E}{\textbf{No}}} & 
{\cellcolor[HTML]{FF6961}{Yes}} & {\cellcolor[HTML]{C1FD8E}{\textbf{No}}} & {\cellcolor[HTML]{C1FD8E}{\textbf{No}}} & {\cellcolor[HTML]{C1FD8E}{\textbf{No}}} \\\hline

Malicious servers' handling & 
{\cellcolor[HTML]{C1FD8E}{\textbf{Yes}}} & {\cellcolor[HTML]{FF6961}{{No}}} & 
{\cellcolor[HTML]{C1FD8E}{\textbf{Yes}}} & {\cellcolor[HTML]{FF6961}{\textbf{No}}} & {\cellcolor[HTML]{C1FD8E}{\textbf{Yes}}} & {\cellcolor[HTML]{C1FD8E}{\textbf{Yes}}} \\\hline

Malicious clients' handling & 
 {\cellcolor[HTML]{FF6961}{No}} & 
{\cellcolor[HTML]{FF6961}{No}} & 
N/A & {\cellcolor[HTML]{C1FD8E}{\textbf{Yes}}} & {\cellcolor[HTML]{C1FD8E}{\textbf{Yes via trusted proxy}}} & {\cellcolor[HTML]{C1FD8E}{\textbf{Yes}}} \\\hline

Information leakage from ciphertext & 
 {\cellcolor[HTML]{FF6961}{Yes$\sharp$}} & 
 {\cellcolor[HTML]{FF6961}{Yes$^\P$}} 
 &
 {\cellcolor[HTML]{C1FD8E}{\textbf{No}}} & {\cellcolor[HTML]{C1FD8E}{\textbf{No}}} &  {\cellcolor[HTML]{C1FD8E}{\textbf{No}}} &{\cellcolor[HTML]{C1FD8E}{\textbf{No}}} \\\hline 
 
Preventing access pattern and/or volume leakage & 
None  &  \multicolumn{4}{c|}{Only AP (\& volume leakage does not matter as fetching files based on ids)}   & AP, V \\\hline


\end{tabular}

\caption{Comparing \textsc{Doc}$\boldsymbol{^\star}$ against other secure document systems. {\color{kellygreen} Green color indicates good aspects of the systems.} {\color{red} Red color indicates negative aspects of the systems.}  Malicious servers can change data or collude with malicious clients, who want to access any file. Processing time for Dory is $>$2sec$\dag$), which is not included, since Dory does not offer access control and only supports finding file-ids.  Time for \textsc{Doc}$^\star$, Metal, and Titanium includes access control checking and file retrieval time. The dataset includes 5K searchable keywords in 500K files. $\ddag$: Sieve time for a single object of size 375KB was 0.44sec. $^\P$: reveals which object can be accessed by how many clients. $\sharp$: reveals attributes and which clients can access which attributes. Complexity of operations: \fcircle{}: hard, \hcircle{}: medium, \ecircle{}: easy.}

\BBB
\label{tab:d+comparing}
\end{table*}
\egroup

\bgroup
\def\arraystretch{1.0}
\begin{table*}[!t]
\begin{tabular}{|l|l|l|l|l|l|l|l|l|l|l|p{1.6cm}|p{0.55cm}|p{0.59cm}|p{1.35cm}|p{1.4cm}|p{0.9cm}|p{.65cm}|l|}\hline

\textbf{Systems}  & \textbf{MySQL} & \textbf{Baseline} & \textbf{\textsc{Doc}$\boldsymbol{^\star}$-$\boldsymbol{\mathsf{Leaky}}$} & 
\textbf{\textsc{Doc}$\boldsymbol{^\star}$-$\boldsymbol{\mathsf{Secure}}$} &
\textbf{Secrecy~\cite{DBLP:conf/nsdi/LiagourisKFV23}} &
\textbf{Dory~\cite{DBLP:conf/osdi/DautermanFLPS20}} \\\hline

Phase 1 time &0.8ms & \multirow{2}{1.24cm}{24.23sec \\ \S\ref{subsec:A Baseline Solution}} & 16.6ms & 16.6ms & 7ms & 
NA \\ \cline{1-2} \cline{4-7}

Phase 2 time & 2.2ms & & 4.2ms &  78.7ms &  $>$4sec & 
$>$2sec\\\hline

Avg. false positives & 0 & 0& 0 &0 &0&765\\\hline


\end{tabular}
\caption{Exp 7: Comparing \textsc{Doc}$\boldsymbol{^\star}$ \& other systems on 1 thread.}
\BBB\BBB
\label{tab:exp_8d+comparing}
\end{table*}
\egroup

{\color{black}
\medskip\medskip
\noindent
\textbf{Exp 6: Performance in a wide-area network.} We evaluate \textsc{Doc}$\boldsymbol{^\star}$ in a WAN setting, with servers located in three different AWS regions:  
Ireland, 
London, and 
Paris. The client was located in 
Frankfurt. 
Before data transmission, we compressed data at the servers and the client in all three phases. 
Over 5K keywords and 500K files, query execution took 1123.7ms, of which computation at the client and the server took 129.7ms, compression/decompression took 381ms, and data transmission took 613ms. This network time can
be reduced by using multiple sockets that enable full utilization of
the bandwidth between two different data centers.}

\begin{mdframed}[style=MyFrame,nobreak=true,align=center]
 \noindent
{\textbf{{Additional experiments are given in~\cite{fullversions2d}}}}. They investigate the impact of: 
(\textit{i})~different implementations of AP list in Phase~3, 
(\textit{ii})~granting/ revoking access rights, 
(\textit{iii})~adding new file-id(s) to an existing keyword, 
(\textit{iv})~deleting an existing file, 
(\textit{v})~deleting an existing keyword from AC matrix, and 
(\textit{vi})~increasing the number of servers.
\end{mdframed}

\medskip
\subsection{Comparing \textsc{Doc}$\boldsymbol{^\star}$ against Other Systems}
\label{subsec:comparingd_with_other}

\noindent
 \textbf{Exp~7: Time taken by \textsc{Doc}$\boldsymbol{^\star}$ and other systems.} We compare \textsc{Doc}$^\star$ against different systems on 5K keywords in 500K files:

\begin{enumerate}[noitemsep,nolistsep,leftmargin=0.01in]
   \medskip
    \item 
\emph{\textbf{A cleartext database system, MySQL}}: is used to store AC matrix, \textsf{AddrList}, and \textsf{OptInv} in a tabular format, with the data stored in cleartext. We execute SQL queries using the index supported by MySQL to know the desired file-ids associated with a keyword. 

 \medskip\item
 \emph{\textbf{The baseline method}}: is given~{\color{blue}\S\ref{subsec:A Baseline Solution}}. 
 
\medskip\item
\emph{\textbf{Secrecy~\cite{DBLP:conf/nsdi/LiagourisKFV23}}}: is a secret-sharing (SS)-based relational-data processing system (not a secure document storage system), and hence cannot be used to retrieve files). 
 We store AC matrix, \textsf{AddrList}, \textsf{OptInv} in Secrecy and execute SQL queries 
 to support Phase~1 and Phase~2 of \textsc{Doc}$^\star$. While we recognize that Secrecy offers complex operations, {\color{black}such systems are slower to support Phase~1 and Phase~2, due to their multiple rounds of communication among servers storing the shares to execute a search query --- particularly, for searching keywords over a set of $\beta$ keywords require $\mathcal{O}(\ell)$ communication rounds where $\ell$ is the maximum length of a word, and the total amount of information flow among servers (and between a server and a querier) can be $\mathcal{O}(\beta)$.}

\medskip\item
\emph{\textbf{\textsc{Doc}$^\star$-$\mathsf{Leaky}$}}: keeps AC matrix, \textsf{AddrList}, \textsf{OptInv}, and files in SS, and leaks access-patterns/volume during query execution.
(\textit{v})~\emph{\textbf{\textsc{Doc}$^\star$-$\mathsf{Secure}$}}: refers to the secure algorithms of all three phases, developed in this paper.

\medskip\item
\emph{\textbf{Dory~\cite{DBLP:conf/osdi/DautermanFLPS20}}}: is an encryption-based secure document store, which allows knowing only file-ids associated with a keyword and hides access-patterns and volume. 
While Dory does not provide the same security guarantees as \textsc{Doc}$^\star$, we select Dory to show the impact of computationally secure techniques. Dory does not support access control and returns file-ids not associated with a query keyword, \textit{i}.\textit{e}., \emph{false positive file-ids}. On the dataset of 5K keywords in 500 files, Dory returns 765 false positives on average, while the maximum, minimum, and median false positives were 27467, 0, 403, respectively.  {\color{black}Dory takes more time due to their technique, namely distributed point function~\cite{DBLP:conf/eurocrypt/GilboaI14} for hiding access-patterns, which requires executing a pseudo-random generator (PRG) for each row of the database. Executing PRG is more time-consuming compared to simple addition and multiplication operations as in \textsc{Doc}$^\star$.}

\medskip\item
\emph{\textbf{Metal~\cite{metal,metal_code} and Titanium~\cite{Titanium}}}: are recent conditionally secure document storage systems. The code of these systems is not available. 

\end{enumerate}

{\color{blue} Table~\ref{tab:d+comparing}} 
compares these systems on different parameters. {\color{blue} Table~\ref{tab:exp_8d+comparing}} shows the time for different systems. \emph{For this experiment, we do not include the time to fetch a file, since MySQL and Dory do not support such an operation.} 

\medskip
\section{Existing Work \& Their Limitations}
\label{subsec:related work}

\noindent\textbf{Access control for KV Stores.} We classify works on access control for KV and KD stores  into two classes: 
(\textit{i}) \emph{\textbf{Trusted proxy-based.}} BigSecret~\cite{BigSecret} and Titanium~\cite{Titanium} support key-level access over KV and document-level access KD stores, respectively, by using a trusted proxy to check  access rights.  
(\textit{ii}) \emph{\textbf{Cryptographic access control.}} Such works are either 
encryption-based~\cite{cong,Ghostor20}  that 
allows key-level access over KV and document-level access over KD stores, respectively, or 
secret-sharing with encryption-based (Metal~\cite{metal}) that offers document-level access over KD stores using secret-sharing for implementing access control and encryption for storing files. 

\noindent
\emph{{Limitations.}} These work deals with either a simple version of key-level access over KV store or document-level access over KD stores, inheriting all limitations of scalability and implementations, as mentioned in~{\color{blue}\S\ref{sec:Introduction}}. 
Other limitations are shown in red color in~{\color{blue}Table~\ref{tab:d+comparing}}. 

\medskip\medskip
\noindent
\textbf{Access control techniques.} 
Access control methods, \textit{e}.\textit{g}.,  query keyword-, attribute-, and role-based, have been developed~\cite{sandureview}. 
Cryptographic access control methods, \textit{e}.\textit{g}., 
attribute-based encryption (ABE)~\cite{DBLP:conf/eurocrypt/SahaiW05,DBLP:conf/sp/BethencourtSW07,DBLP:conf/crypto/GargGHSW13},  quorum secret-sharing-based access  (QSSAC)~\cite{DBLP:conf/ccs/NaorW96} and~\cite{fss-basedaccesscontrol} function secret-sharing (FSS)~\cite{DBLP:conf/eurocrypt/BoyleGI15},  have been proposed. 

\noindent
\emph{{Limitations.}} 
ABE 
is highly inefficient and harder for revocation~\cite{DBLP:journals/tdp/SarfrazNCB16}.  
{\color{black}
While related, traditional ABE approaches over encrypted data are not directly applicable to our setting, due to leaking access-patterns and not handling scenarios where access must be denied based on the presence of specific keywords a client does not have permission for. Extending ABAC-based techniques to mitigate such issues and apply them in our setting is an interesting direction for future work.} QSSAC~\cite{DBLP:conf/ccs/NaorW96} and~\cite{li2012efficient,DBLP:conf/edge/JoshiMJF17,DBLP:conf/nsdi/BurkhalterHVSR20,DBLP:conf/uss/ShafaghBRH20} have limitations: work only over encrypted databases, need access control and database servers as two different entities, keep access rights in cleartext, and/or reveal queries to access control servers. 
\cite{fss-basedaccesscontrol} provides function secret-sharing (FSS)~\cite{DBLP:conf/eurocrypt/BoyleGI15}-based access control techniques,  restricting clients from executing any functions. 
~\cite{fss-basedaccesscontrol} works over cleartext only. 
Titanium~\cite{Titanium} uses a trusted proxy to implement access control.
Furthermore, existing SS-based database systems (\textit{e}.\textit{g}.,~\cite{Sharemind,DBLP:conf/nsdi/LiagourisKFV23,DBLP:journals/cj/ArcherBLKNPSW18}) do not support integrated access control.


\medskip\medskip\noindent
\textbf{Secure document storage.}
MongoDB~\cite{mongodbqe,mongoqe}, GarbleCloud~\cite{GarbleCloud}, Virtru~\cite{virtru}, Enveil~\cite{Enveil}, and Proton~\cite{proton}  
are industrial document stores based on variants of searchable encryption~\cite{DBLP:conf/sp/SongWP00}.
None of them supports {key-based access control}.

\medskip\noindent
\textbf{{Secret-sharing (SS)-based systems.}} Several database systems (\textit{e}.\textit{g}.,  Sharemind~\cite{Sharemind}, Jana~\cite{DBLP:journals/cj/ArcherBLKNPSW18}, Conclave~\cite{DBLP:conf/eurosys/VolgushevSGVLB19}, Secrecy~\cite{DBLP:conf/nsdi/LiagourisKFV23}) uses secret-sharing (SS). None of these systems deal with access control and malicious clients. 
Such systems are also not efficient due to multiple communication rounds among servers during query processing; \textit{e}.\textit{g}., Jana~\cite{DBLP:journals/cj/ArcherBLKNPSW18} takes $\approx$475s for a selection query on 1M rows. We have also discussed Secrecy~\cite{DBLP:conf/nsdi/LiagourisKFV23} in~{\color{blue}\S\ref{subsec:comparingd_with_other}}.

\medskip\medskip
\noindent
 \textbf{{Access-pattern hiding techniques.}} 
 Techniques, \textit{e}.\textit{g}., Path-ORAM~\cite{DBLP:conf/ccs/StefanovDSFRYD13}, Private Information Retrieval (PIR)~\cite{DBLP:journals/jacm/ChorKGS98},  or Distributed Point Function (DPF)~\cite{DBLP:conf/eurocrypt/GilboaI14}, hide access-patterns. These techniques differ in the way they access/read the data and the amount of data they return to a querier. For instance, to fetch a single item/object, Path-ORAM accesses and returns the polylogarithmic size of data, while DPF accesses the entire data but returns only the desired answer. DPF is used in Dory~\cite{DBLP:conf/osdi/DautermanFLPS20}, and variants of ORAM are used in Metal~\cite{metal} and Titanium~\cite{Titanium}. However, ORAM techniques have several problems: 
(\emph{i})~The disclosure of additional data beyond the desired answer to a query by Path-ORAM poses a challenge when the client is not allowed to learn those extra answers. 
(\emph{ii})~ORAMs restrict the system throughput, due to their structures, see~\cite{DBLP:conf/sosp/DautermanFDCP21}. 
(\emph{iii})~When used to prevent volume leakage by sending the maximum number of files, say $\ell$, associated with a keyword, ORAM sends $\ell$ times polylogarithmic size of data. 
The use of such techniques comes with query inefficiency. For example, Dory~\cite{DBLP:conf/osdi/DautermanFLPS20} takes less than 0.1 seconds for \emph{{searching file-ids (\ul{not retrieving the file})}} associated with the desired keyword over $2^{20}$ files when \emph{{revealing access-patterns}} (Figure~9 of~\cite{DBLP:conf/osdi/DautermanFLPS20}); while 
Dory takes $\approx$10s for the same operation  
when hiding access-patterns 
and volume on   
 $2^{20}$ files.  
Titanium~\cite{Titanium} takes $\approx$7.2s \emph{to fetch a file} with access-patterns hidden on the same data. 
To fetch items based on an index-id of an object, PIR~\cite{DBLP:journals/jacm/ChorKGS98} or its several variants, \textit{e}.\textit{g}., PIR-by keyword~\cite{DBLP:journals/iacr/ChorGN98} or information-theoretically secure PIR~\cite{DBLP:conf/sp/Goldberg07}, can be used. However, 
such techniques do not deal with malicious client and server, and also reveals additional data to the client, \textit{e}.\textit{g}., the presence/absence of other keywords in using 
PIR-by-keywords. 

\medskip\medskip\noindent
\textbf{{Volume hiding techniques.}} 
~\cite{DBLP:conf/eurocrypt/KamaraM19,DBLP:conf/ccs/PatelPYY19,DBLP:conf/icdcs/00010LJ0ZWC020} provides volume-hiding techniques. These techniques degrade the system's performance in terms of space requirements due to storing a map and a stash at the data owner, making it harder to deal with new insertions, and reveal extra information to the clients than the desired answers. 

\medskip
\section{Conclusion} 
This paper develops an unconditionally-secure document/file storage system, \textsc{Doc}$^\star$, with query keyword-based access control. 
Operations at servers hide access-patterns and volume, and do not reveal any information to servers about the data. 
\textsc{Doc}$^\star$ exhibits 
efficient performance and 
takes 231.5ms over~5K keywords in 500K files. 

\section*{Acknowledgment}
We are thankful to Murat Kantarcioglu for his feedback on the baseline method of {\color{blue}\S\ref{subsec:A Baseline Solution}}. Y. Li was funded by National Natural Science Foundation of China Grant
62372107. S. Mehrotra was funded by NSF Grants  1952247, 2133391, 2032525, and 2008993. S. Sharma was funded by NSF Grant 2245374.

\newpage
\appendix

\section*{Appendix}

\section{Phase 3 --- Additional Operations}
\label{app-sec:Phase 3 --- Additional Operations}

This section develops methods for fully secure operations on AP list and another for fetching multiple files using a single vector of $\mathcal{O}(\delta)$, where $\delta$ is the number of files.

\medskip
\subsection{Fully Secure Operations on AP List}
\label{app_subsec:phase3 Fully Secure Operation}

Recall that based on AP list (see {\color{blue}Table~\ref{tab:ss_file}}), the client will learn the position of the keywords in AC matrix and the number of keywords appearing in a file (see {\color{blue}\S\ref{sec:Data Outsourcing}}). To overcome this, we develop a method below that requires us to change the structure of AP list.

\medskip\medskip
\noindent
\textbf{New structure of AP list.} In the new structure of AP list for each file, we store a bit value of 1 or 0, for all keywords appearing in AC matrix, where 1 refers to the keyword appearance in the file; otherwise, 0. These values are placed according to the position of the keywords in AC matrix. 
For example, refer to the first row of {\color{blue}Table~\ref{tab:cleartext_filesgiven}} that corresponds to the first file, containing only one keyword (\textit{i}.\textit{e}., Are). The keyword `Are' appears at the first position of AC matrix and there is no other keyword of AC matrix appears in the first file; thus, we add $\langle1,0,0\rangle$ in the first row of AP list. Further, the hash digest has been added as usual, like the old structure of AP list ({\color{blue}Table~\ref{tab:The current AP list}}).
DBO creates SSS of this new AP list and outsources them with files. 

\emph{Advantage of the new AP structure.} The new structure of AP list prevents an adversary from knowing the maximum number of keywords appearing in a file. Now, the adversary learns that all the keywords of AC matrix appear/disappear in each file.

\medskip\medskip
\noindent
\textbf{Query execution.} Recall that we used the old structure of AP list in Phase~3 \textsc{Step 6, 8a} (see {\color{blue}\S\ref{subsec:phase3}}). We use the new structure of AP list in the same steps, and below, we reproduce the same steps with the new AP list, denoted by $\mathit{mAP}$.

\medskip\medskip%
\noindent
\textbf{\textsc{Step 6}: \textit{Server:}} receive the vector $\mathbb{M}(v)$ from the client and has three objectives:
(\textit{i})~whether the vector contains all zero and except for a single one; 
(\textit{ii})~the vector contains one at the position of the file-id that was sent in Phase~2 by servers; and 
(\textit{iii})~the requested file does not contain a keyword to which the client has no access. Test~A and Test~B of~{\color{blue}\S\ref{Sec:Verification of the Client's Vector}} achieve the first objective. Test~2 achieves the second objective, as mentioned in~{\color{blue}\S\ref{subsec:phase3}}. Below, we produce Test~3 for achieving the third objective.

{\centerline{\color{myblue}
Test 3 on new AP list: $\mathbb{M}(\mathit{test}_3)[]\leftarrow 
(\mathbb{M}(v)\odot \mathbb{M}(mAP)) + \textnormal{\hl{$\mathbb{M}(\mathsf{RN})_z$}}
 $}}

Test~3, on the success of Test~2, performs a dot product between $\mathbb{M}(v)$ and the new AP list ($\mathbb{M}(mAP)$ (see the second column of {\color{blue}Table~\ref{tab:The new AP list}}) at $\mathcal{S}_z$. This results in a vector of size $\beta$ (where $\beta$ is the number of keywords in AC matrix) and the hash digest. The degree of polynomial for each such element will be two. 
$\mathcal{S}_{z}$ needs to send the vector to the client to reduce the degree to one. 
Adding {\hl{$\mathbb{M}(\mathsf{RN})_z$} hides from the client the number of keywords in the file during degree reduction. 
Thus, even if a minority of servers (\textit{i}.\textit{e}., $f$ servers) collude with the client, they cannot learn any information about the number of keywords appearing in the file. Also, servers keep the hash digest (denoted by $\mathbb{M}(\mathsf{H}(AP))$) at their end.

\medskip\medskip%
\noindent
\textbf{\textsc{Step 7}: \textit{Client:}} interpolates the received values and creates SSS of the same vector with a polynomial of degree one and sends the share to servers. We denote this vector by $\mathbb{M}(\mathit{vAP})$.

\medskip\medskip%
\noindent
\textbf{\textsc{Step 8a}: \textit{Server:}} checks 
(\textit{i})~$\mathbb{M}(\mathit{vAP})$ contains only zeros and ones using Test~C of {\color{blue}\S\ref{Sec:Verification of the Client's Vector}};
(\textit{ii})~$\mathbb{M}(\mathit{vAP})$ is created for positions of keywords appear in the vector sent in \textsc{Step 6} using Test~4 of~{\color{blue}\S\ref{subsubsec:algo_in_phase3}};
(\textit{iii})~finally, the file does not contain any keyword to which access is denied to the client, using Test~5 of~{\color{blue}\S\ref{subsubsec:algo_in_phase3}}. Note that in Test~4 and Test~5, servers will use the vector $\mathbb{M}(\mathit{vAP})$.

If the above tests succeed, then servers will execute \textsc{Step 8b} to send the file.

\begin{figure}[!t]
\BBB
\begin{minipage}[t]{0.99\linewidth} 
\centering
\scriptsize
 \begin{tabular}{|p{0.65cm}|p{1.05cm}|p{0.7cm}|p{0.70cm}|p{0.5cm}|p{0.5cm}|}   \hline
    File-ids & File content \\ \hline\hline
    1 & How are you\\ \hline
    2 & Are you Ana\\ \hline
    3 & Fig is a fruit \\ \hline
  \end{tabular}
\captionof{table}{Cleartext files.}
\label{tab:cleartext_filesgiven}
\BBB
\end{minipage}

\begin{minipage}[t]{0.99\linewidth}
\centering
\scriptsize
 \begin{tabular}{|@{}p{3.1cm}|p{0.29cm}|p{0.29cm}|p{0.29cm}|p{0.5cm}|p{0.5cm}|l|l|}    \hline
   Keywords $\rightarrow$ & Are &  Ana & Fig \\ \hline\hline
   Starting address in the inverted list $\rightarrow$  & 1 & 2 & 3 \\ \hline
   Hash values $\rightarrow$  & $\mathsf{H}(1)$ & $\mathsf{H}(2)$ & $\mathsf{H}(3)$ \\ \hline
   \multicolumn{4}{|l|}{Clients' information $\downarrow$   }
      \\ \hline
   Lisa & {\cellcolor[HTML]{FFFF33}{\textbf{0}}} & 
   {\cellcolor[HTML]{FFFF33}{\textbf{1}}} & {\cellcolor[HTML]{FFFF33}{\textbf{2}}} \\\hline
   
   Ava  & {\cellcolor[HTML]{FFFF33}{\textbf{3}}} & 
   {\cellcolor[HTML]{FFFF33}{\textbf{0}}} & {\cellcolor[HTML]{FFFF33}{\textbf{0}}} \\\hline
  \end{tabular}
\captionof{table}{Access control matrix.}
\label{tab:ACT_tableappendix}
\BBB
\end{minipage}

\begin{minipage}[t]{0.99\linewidth}
\centering
\scriptsize
 \begin{tabular}{|l|l|l|l|}     \hline
    {\cellcolor[HTML]{ffffff}{F-ids}} & AP list & File content  with digest\\ \hline\hline
 {\cellcolor[HTML]{ffffff}{1}}&1,0,$\mathsf{H}(1)$ & How are you, $\mathsf{H}(\textnormal{how are you},\mathsf{H}(1))$\\ \hline

    {\cellcolor[HTML]{ffffff}{2}} & 1,2,$\mathsf{H}(1){+}\mathsf{H}(2)$ &Are you Ana, $\mathsf{H}(\textnormal{are you Ana},\mathsf{H}(2))$\\ \hline
    
 {\cellcolor[HTML]{ffffff}{3}} & 3,0,$\mathsf{H}(3)$ & Fig is a fruit, $\mathsf{H}(\textnormal{Fig is a fruit},\mathsf{H}(3))$ \\ \hline  
 
    {\cellcolor[HTML]{ffffff}{0}} & 0,0,0 & Dummy, $\mathsf{H}(\textnormal{Dummy},\mathsf{H}(0))$ \\\hline 

  \end{tabular}
\captionof{table}{The current AP list.}
\label{tab:The current AP list}
\BBB
\end{minipage}

\begin{minipage}[t]{0.99\linewidth}
\centering
\scriptsize
 \begin{tabular}{|l|l|l|l|}   \hline
    {\cellcolor[HTML]{ffffff}{F-ids}} & AP list & File content  with digest\\ \hline\hline
    
 {\cellcolor[HTML]{ffffff}{1}}&1,0,0,$\mathsf{H}(1)$ & How are you, $\mathsf{H}(\textnormal{how are you},\mathsf{H}(1))$\\ \hline

    {\cellcolor[HTML]{ffffff}{2}} & 1,1,0,$\mathsf{H}(1){+}\mathsf{H}(2)$ &Are you Ana, $\mathsf{H}(\textnormal{are you Ana},\mathsf{H}(2))$\\ \hline
    
 {\cellcolor[HTML]{ffffff}{3}} & 0,0,1,$\mathsf{H}(3)$ & Fig is a fruit, $\mathsf{H}(\textnormal{Fig is a fruit},\mathsf{H}(3))$ \\ \hline  
 
    {\cellcolor[HTML]{ffffff}{0}} & 0,0,0 & Dummy, $\mathsf{H}(\textnormal{Dummy},\mathsf{H}(0))$ \\\hline 

  \end{tabular}
\captionof{table}{The new AP list.}
\label{tab:The new AP list}

\BBB
\end{minipage}
\end{figure}

\medskip
\subsection{Phase 3: Fetching Multiple Files}
\label{app_subsec:phase3 Fetching Multiple Files}

A query keyword can appear in multiple files. {\color{blue}\S\ref{subsec:phase3}} has shown how we can a single file and server can verify the vector. To fetch $\gamma$ files, we need to send $\gamma$ vectors, each of size $\delta$, where 
$\gamma$ is the maximum number of files in which a keyword appears and $\delta$ is the number of files. This results in the communication overhead of 
$\gamma\times \delta$ between a server and the client. 
To reduce such a communication overhead, this section develops a method to fetch $\gamma$ files \emph{without} sending $\gamma$ vectors, each of size equal to the number of files $\delta$.



\medskip\medskip
\noindent
\textbf{Disjoint Partitioning/Binning of Files.}
This method partitions the files into multiple blocks/bins, such that all the desired file-ids and files belong to different blocks/bins. For example, if the client wants to fetch $\gamma$ files, the client can request the servers to create $\gamma$ partitions/bins of the file set, such that each of the desired $\gamma$ files stay in a different partition. Then, the client will create $\gamma$ bit-vectors, each of size $\delta/\gamma$, such that each vector contains zeros except only one that corresponds to the desired file-id's position in the partition/bin. Note that the total elements in this vector are equal to the number of files. The client creates secret-shares of the vector. Servers will perform a dot product over each vector and do the verification before sending the files to the client, using the method developed in~{\color{blue}\S\ref{subsec:phase3}}. 

\medskip\medskip
\textbf{Example.} For the three files and one dummy file (see {\color{blue}Table~\ref{tab:The new file structure}}) and for a query to fetch files $f_1$ and $f_2$, we can create two joint bins as shown in {\color{blue}Table~\ref{tab:bin1}} and {\color{blue}Table~\ref{tab:bin2}}.

\begin{figure}
\begin{minipage}[t]{0.99\linewidth}
\centering
\scriptsize
 \begin{tabular}{|l|l|l|l|}   \hline
    {\cellcolor[HTML]{ffffff}{F-ids}} & AP list & File content  with digest\\ \hline\hline
    
 {\cellcolor[HTML]{ffffff}{1}}&1,0,0,$\mathsf{H}(1)$ & How are you, $\mathsf{H}(\textnormal{how are you},\mathsf{H}(1))$\\ \hline

    {\cellcolor[HTML]{ffffff}{2}} & 1,1,0,$\mathsf{H}(1){+}\mathsf{H}(2)$ &Are you Ana, $\mathsf{H}(\textnormal{are you Ana},\mathsf{H}(2))$\\ \hline
    
 {\cellcolor[HTML]{ffffff}{3}} & 0,0,1,$\mathsf{H}(3)$ & Fig is a fruit, $\mathsf{H}(\textnormal{Fig is a fruit},\mathsf{H}(3))$ \\ \hline  
 
    {\cellcolor[HTML]{ffffff}{0}} & 0,0,0 & Dummy, $\mathsf{H}(\textnormal{Dummy},\mathsf{H}(0))$ \\\hline\end{tabular}
\captionof{table}{The new AP list.}
\label{tab:The new file structure}
\BBB
\end{minipage}

\begin{minipage}[t]{0.99\linewidth}
\centering
\scriptsize
 \begin{tabular}{|l|l|l|l|}   \hline
    {\cellcolor[HTML]{ffffff}{F-ids}} & AP list & File content  with digest\\ \hline\hline
    
 {\cellcolor[HTML]{ffffff}{1}}&1,0,0,$\mathsf{H}(1)$ & How are you, $\mathsf{H}(\textnormal{how are you},\mathsf{H}(1))$\\ \hline

 {\cellcolor[HTML]{ffffff}{3}} & 0,0,1,$\mathsf{H}(3)$ & Fig is a fruit, $\mathsf{H}(\textnormal{Fig is a fruit},\mathsf{H}(3))$ \\ \hline

  \end{tabular}
\captionof{table}{The first bin.}
\label{tab:bin1}
\BBB
\end{minipage}

\begin{minipage}[t]{0.99\linewidth}
\centering
\scriptsize
 \begin{tabular}{|l|l|l|l|}   \hline
    {\cellcolor[HTML]{ffffff}{F-ids}} & AP list & File content  with digest\\ \hline\hline

    {\cellcolor[HTML]{ffffff}{2}} & 1,1,0,$\mathsf{H}(1){+}\mathsf{H}(2)$ &Are you Ana, $\mathsf{H}(\textnormal{are you Ana},\mathsf{H}(2))$\\ \hline
    
    {\cellcolor[HTML]{ffffff}{0}} & 0,0,0 & Dummy, $\mathsf{H}(\textnormal{Dummy},\mathsf{H}(0))$ \\\hline  \end{tabular}
\captionof{table}{The second bin.}
\label{tab:bin2}
\BBB\BB
\end{minipage}
\end{figure}

\medskip\medskip%
\noindent\textbf{Problems of disjoint partitioning.} 
Note that if a keyword is associated with $x {\ll} \gamma$ files, then all the $\gamma{-}x$ vectors will contain only zero. However, since DBO has placed only a single fake file (see~{\color{blue}Table~\ref{tab:The new file structure}}), this method will fail. Furthermore, this method has another problem: if there are many requests for the same files, say $f_1$ and $f_2$, then these two files are always placed into two different bins. This will enable the adversary to observe which files the clients are trying to fetch after many queries. To overcome this problem, we develop the following method.

\medskip\medskip
\noindent
\textbf{Overlapped Partitioning of Files.}
Instead of creating disjoint bins of the files, this method creates bins such that the intersection of two or more bins never be disjoint. Thus, a file-id/file will appear in multiple blocks. In what follows, the fake file, with content zero, will also appear in multiple or all blocks --- overcomes the first problem of the disjoint partitioning, as mentioned in the last paragraph. Furthermore, due to replicating the files in multiple bins, the same file may appear always in the same bin, and here, the same file can be fetched from different bins in different queries --- preventing the adversary from knowing anything about the requested files even after observing many queries. While this method overcomes the problem of disjoining partitioning, it comes with additional communication cost, as files are replicated in multiple bins, increasing the total number of files in the system. Important to note that such increased files are never written to the disk --- thus, the space overhead is only during query execution.



\medskip
\section{Dynamic Operations}
\label{sec:dynamic_operation}

This section explains dynamic operations --- add and delete. 

\medskip
\subsection{Add Operation}
\label{dynamic_operation:Add Operation}

The add operation allows insertion of new files with existing keywords in AC matrix or with new keywords that need to be added to AC matrix. Below, we develop algorithms for both cases.

\medskip
\subsubsection{\textbf{Case 1: Adding Files having Existing keywords.}}
\label{app_subsubsec:Case 1: Adding Files having Existing keywords}

We first consider the case of adding a new file with existing keywords, in an inverted list ({\color{blue}Table~\ref{tab:inverted_list}}) or in \textsf{OptInv} ({\color{blue}Table~\ref{table:cleartext_OptInv}}). Suppose there is a free slot available; see the second last positions with keywords Ana and Fig, in {\color{blue}Table~\ref{tab:inverted_list}} and the 7th slot of {\color{blue}Table~\ref{table:cleartext_OptInv}}.

\medskip\medskip
\noindent
\textbf{{Addition operation on the inverted list.}}
To add the new file having an existing keyword in AC matrix, DBO needs to obliviously learn the row of the inverted list corresponding to the keyword, and then, the hash digest and the empty positions in the row, which can hold the new file-ids. 

For example, DBO wants to add a file-id to the third row with the keyword `Fig,' then DBO needs to know the existing hash digest $\mathsf{H}(3,\mathsf{H}(\textnormal{Fig}))$ and the position (\textit{i}.\textit{e}., the second position in the third row) of the free slot, where a new file-id can be inserted.

\begin{table}
 \begin{tabular}{|l|l|l|l|l|}    \hline
   {\cellcolor[HTML]{B8B8B8}{Positions}}  & File-ids \\\hline\hline
  
 {\cellcolor[HTML]{B8B8B8}{1 (are)}} & 1, 2, {\color{orange}3},
 $\mathsf{H}(2,\mathsf{H}(1,\mathsf{H}(\textnormal{are})))$ \\\hline
 
  
   {\cellcolor[HTML]{B8B8B8}{2 (Ana)}} & 2, 0, {\color{orange}2}, $\mathsf{H}(0,\mathsf{H}(2,\mathsf{H}(\textnormal{ana})))$ \\\hline
   
   {\cellcolor[HTML]{B8B8B8}{3 (Fig)}} & 3, 0, {\color{orange}2}, $\mathsf{H}(3,\mathsf{H}(\textnormal{Fig}))$ \\\hline

   {\cellcolor[HTML]{B8B8B8}{0 (Fake)}} & 0, 0, 0, $\mathsf{H}(0,\mathsf{H}(\textnormal{Fake}))$ \\\hline
   
  \end{tabular}
\captionof{table}{Inverted list (modified).}
\label{tab:inverted_list_app}
\end{table}

\medskip\medskip
\noindent \textbf{\emph{When free slots are available in an inverted list.}} Now, let us consider that the inverted list contains free slots; see {\color{blue}Table~\ref{tab:inverted_list_app}} where 0 refers to free slot. Further, note that we modified the inverted list having the position number where a new file can be added (see orange-colored text). 
If this newly added orange-colored value equals to the position where it is written, then this shows that there is no free slot --- for this case, we develop an algorithm soon. For example, in  the row for `are,'
the position is three for the orange-colored value and it is written at the third position of the list, showing there is no free slot for new file-id having `are.'

After knowing the hash digest and the empty positions in the row of the inverted list,  DBO sends three vectors in SSS form, each of length $\beta$ (the number of keywords). Vectors are constructed as follows: all positions are set to zero except for the desired position, which contains the following values:
\begin{enumerate}[noitemsep,nolistsep,leftmargin=0.01in]
\item \emph{Vector for file-id} --- the first vector contains the new file-id, 

\item \emph{Vector for counting free slots} --- the second vector contains  one, and 

\item \emph{Vector for hash digest} --- the third contains the difference between the new hash digest and the old hash digest. 
\end{enumerate}
Upon receiving these vectors, servers add each vector to the desired position in the inverted list.

 Suppose, we want to add a file to the entry corresponding to Fig, \textit{i}.\textit{e}., $\langle3, 0, 2, \mathsf{H}(3,\mathsf{H}(\textnormal{Fig}))\rangle$. Here, the first vector will look like $\langle$0,0,new\_file-id,0$\rangle$ and will be added to the second position of the entire inverted list. This will replace 0 at the second position in the row for `Fig' with a new file-id. 
 The second vector will look like $\langle$0,0,1,0$\rangle$  and  will be added to the third position of the entire inverted list, making it 2+1=3 for the row of `Fig,' showing that now this row cannot hold any new file-id. 
 The third vector will look like
 $\langle$0,0,new\_hash-digest,0$\rangle$ 
 and will be added to the last position of the entire inverted list.
 
Note that by knowing the position of zero and fetching the existing hash digest, we avoid the communication cost of fetching one of the entire rows of the inverted list. 
Also, \emph{\textbf{chaining of hash digest}} avoids fetching all entries and recomputing the hash over file-ids.

\medskip
\noindent
\textbf{\emph{No free slots in an inverted list.}} Now, let us consider that there is no free slot in the inverted list to add a new file-id. For example, the first row $\langle 1,2,{\color{orange}3}, \mathsf{H}(2,\mathsf{H}(1,\mathsf{H}(\textnormal{are})))\rangle $ for the keyword `Are' in {\color{blue}Table~\ref{tab:inverted_list_app}}. 

In this case, we need to adjust the inverted list. Particularly, for each row of the inverted list, we need to adjust (\textit{i})~the last value (\textit{i}.\textit{e}., the hash digest) by shifting towards one position to the right, and
(\textit{ii})~the count of the free slot by shifting towards one position to the right. This will increase the size of the entire inverted list and make free slots. 

For example, $\langle 1,2,{\color{orange}3},\mathsf{H}(2,\mathsf{H}(1,\mathsf{H}(\textnormal{are})))\rangle $ will appear as $\langle 1,2,\ast,{\color{orange}3}, \mathsf{H}(2,\mathsf{H}(1,\mathsf{H}(\textnormal{are})))\rangle$ in secret-shared form.

Here, the first vector will look like $\langle$new\_file-id,0,0,0$\rangle$ and will be added to the third position of the entire inverted list. This will replace $\ast$ at the third position in the row for `are' with a new file-id. 
 The second vector will look like $\langle$1,0,0,0$\rangle$  and  will be added to the third position of the entire inverted list, making it 3+1=4 for the row of `are,' showing that now this row cannot hold any new file-id. 
 The third vector will look like
 $\langle$new\_hash-digest,0,0,0$\rangle$ 
 and will be added to the last position of the entire inverted list.

The addition operation does not reveal to servers the updated row of in the inverted list, but reveals the column number of the inverted list. To avoid this, we can mimic the operation by adding vectors having all zeros in several columns.

\medskip\medskip
\noindent
\textbf{{Addition operation over \textsf{OptInv}.}}
To perform the add operation on \textsf{OptInv}, DBO needs to update \textsf{AddrList} and \textsf{OptInv}.
DBO prepares a vector of size equal to the number of keywords with the desired keyword index set to one to fetch keyword's \textsf{AddrList} values of SiP, CuT, {\color{purple}HD} and {\color{orange}HdV} from the server. For example, for the keyword `ana', DBO will fetch the value as: SiP as 4, CuT as 3, and {\color{purple}$H_2$} and ${\color{orange}h_2}$. The update of \textsf{AddrList} is done similarly to the add operation on the inverted list. 

Now, to add new file-ids in \textsf{OptInv}, DBO prepares $\mathit{row\_vector}$ and $\mathit{pos\_vec}$ based on the \textsf{AddrList}'s SiP and CuT values. With these vectors, DBO fetches the file-ids currently associated with the keyword and their hash digest value. For keyword `ana', to add a new file-id $\mathit{f}_{\mathit{new}}$, DBO needs to update the value of hash digest and fill the empty slot with the new file-id, \textit{i}.\textit{e}., the slot 6 and slot 7 in {\color{blue}Table~\ref{table:cleartext_OptInv}} need to be updated.

To do so, DBO sends a vector in SSS form of length \textsf{OptInv} with all zeros, except for the desired positions containing the new file-id and hash digest. For example, Slot 6 of the vector will contain the new file-id, and slot 7 will store the new hash digest. 
Servers add this vector to the existing \textsf{OptInv}.DBO can also use this method to insert multiple new file-ids in a single round if there are several desired number of empty slots. Note that this method is completely oblivious, but incurs communication cost. Further, when there is no empty slot to insert the file-ids for a keyword, this  method will not work.

To address the above problems, we propose the following method. We organize \textsf{OptInv} in the form of a matrix, as in~{\color{blue}\S\ref{sec:Inverted Index Optimization}}. Let $i$ be the row of the matrix, where we want to insert new file-ids, but there is no free slot. To insert the item in the $i^{\mathit{th}}$ row, DBO fetches the entire 
$i^{\mathit{th}}$ row and inserts the new file-ids, as they want. Note that this may increase the number of elements in the row. To have the row with the same number of elements as other rows of the matrix, all the exceeding elements are placed into a new row. This new row is padded with fake items to have the same size. Finally, DBO sends two rows in SSS form to servers that replace the old $i^{\mathit{th}}$ row with the new row and place the new row at the $(i+1)^{\mathit{th}}$ place in the matrix. 
Note that while this method enables inserting new file-ids, it reveals the row-ids of the matrix where the elements are inserted. However, servers do not know the exact place in the row where new file-ids are inserted.

Furthermore, \textsf{AddrList} is  updated for each keyword. In case of the presence of a free slot in OptInv, updating SiP, CuT, {\color{purple}HD}, {\color{orange}HdV} for each keyword is easy. DBO sends four vectors, each of length $\beta$. 

All these vectors contain zero, except at the desired position. Particularly, the first vector contains zero at the desired position also. The second vector contains $x$ at the desired position, where $x$ is number of files got inserted. 
The third vector contains $\mathsf{H}(\textnormal{SiP},\textnormal{CuT}+x,\textnormal{keyword})-\mathsf{H}(\textnormal{SiP},\textnormal{CuT},\textnormal{keyword})$ at the desired position.
The last vector will contain 
${\color{orange}
\mathsf{H}(\textnormal{CuT+1})+\ldots+\mathsf{H}(\textnormal{CuT}+x)}$ at the desired position. Servers will add such vector to \textsf{AddrList}.

In case of no free slots in \textsf{OptInv}, however, updating HD and HdV is not trivial, since the addition of a new row in \textsf{OptInv} shifts the proceeding values of \textsf{OptInv}.
In this case, when a new row has been added, DBO reconstructs the entire \textsf{AddrList}.

\medskip
\subsubsection{
\textbf{Case 2: Adding Files with New Keywords.}} 
\label{app_subsubsec:Case 2: Adding Files with New Keywords}
{\color{black} To add a new keyword $\mathcal{K}$, which is not present in the AC matrix, we need to update AC matrix, inverted list or \textsf{AddrList} and \textsf{OptInv}. DBO performs a simple append operation to update each of these lists. 

\medskip\medskip%
\noindent\textbf{Updating AC matrix}. To update AC matrix, we prepare a column vector with the first value as the keyword $\mathcal{K}$, followed by the keyword position value in \textsf{AddrList} or inverted list, following by the hash digest of the position, and finally, followed by the access value for each client. 

For instance, for a new keyword say \texttt{is}, at the 5th position, for two clients (Lisa and Ava in~{\color{blue}Table~\ref{tab:ACT_table}}) each of them having access to the keyword, DBO prepares a vector as $\langle \textnormal{is}, \mathsf{H}(5), 0, 0\rangle$. Then, DBO creates shares of the vector and sends to the servers that append to the vector at the end of AC matrix

\medskip\medskip%
\noindent\textbf{Updating inverted list.} In a similar manner, DBO updates {inverted list} with a new row of file-ids containing the new keyword and their hash digest, and servers append at the end. 

\medskip\medskip%
\noindent\textbf{Updating \textsf{AddrList}.} \textsf{AddrList}
 is updated similarly to the inverted list with a vector having values for SiP, CuT, {\color{purple}HD} and {\color{orange}HdV} for the keyword {Bob}. 
 
\medskip\medskip%
 \noindent\textbf{Updating \textsf{OptInv}.} 
 To update \textsf{OptInv}, DBO creates a vector like the inverted list 
 and servers append it to the end of \textsf{OptInv}. It may be possible that the newly added keyword could occur in files added previously. However, in this case, \emph{we only consider updating the new keyword for newly added files.} To update the new keyword with previous file-ids, DBO needs to know the files containing the keyword and then update AP list also, along with AC matrix, \textsf{AddrList}, \textsc{OptInv}. However, adding a new keyword to  existing files' AP list is a very costly operation, which may require the owner to fetch all the files, update the AP list, and then recreate the shares of all the files. 

\medskip
\subsection{Delete Operation}
\label{subsec:Delete Operation}


DBO can delete a keyword or delete a file-id associated a keyword.

\medskip\medskip
\noindent\textbf{Case 1: Delete a keyword.}  DBO needs to update AC matrix to delete a keyword. To do so, DBO has two ways: one, which is completely secure but requires huge communication overhead, and second, which has low communication overhead but suffers from access pattern leakage. 

(\textit{i}) To update AC matrix, DBO prepares a matrix of dimensions that is the same as the AC matrix, filled with zeros for all places except for the capability of the keyword for all clients. The values corresponding to the deleted keyword will store the random number, 
representing no access for the keyword for any client. DBO sends the matrix in SSS form to the servers, which adds the newly received AC matrix to the existing AC matrix. Since each client now has no search access to the keyword, the keyword is considered to be deleted from AC matrix.


(\textit{ii}) To update AC matrix, DBO prepares a vector of size equal to the number of clients. The vector will contain random number, 
representing no access to the keyword. DBO outsources the vector in secret shares to the servers, which adds this vector to the AC matrix at the position of the keyword to be deleted. 


Although the approach is faster than the previous approach, the method suffers from access-pattern leakage as the position of the deleted keyword is revealed to the servers.

\medskip\medskip
\noindent
\textbf{Case 2: Delete a file-id associated with a keyword.}  DBO needs to update the inverted list or \textsf{OptInv} to delete the file-id associated with a keyword. DBO can perform the deletion of file-ids in two ways --- one provides complete security but suffers from huge communication overhead, and the second leaks access pattern but has low communication overhead. 

(\textit{i}) To update the inverted list, DBO first needs to obliviously know the $i^{\mathit{th}}$ row of the inverted list associated with the keyword. Then, DBO prepares a new inverted list of size the same as the inverted list filled with zeros except for the $i^{\mathit{th}}$ row. The $i^{\mathit{th}}$  row is filled with zeros except for the $j^{\mathit{th}}$ position containing the negative value of the file-id to be deleted and the last position to have the difference of the new hash digest and the previous hash digest. Then, DBO will create the shares of this new list and send to servers, which will add the new list to the existing inverted list.  


Likewise, DBO updates \textsf{OptInv} after organizing \textsf{OptInv} in a matrix form, obliviously fetching the $i^{\mathit{th}}$ row of the matrix, and finally, obliviously updating the entire  \textsf{OptInv} along with new hash digest. This method, although completely secure, requires huge communication cost, since the data of size equals to the size of an inverted list or \textsf{(OptInv)} is transmitted to the servers. 

({\textit{ii}}) To update the inverted list, after knowing $i^{\mathit{th}}$ row of the inverted list associated with the keyword, DBO prepares a vector with deleted file-id and new hash digest. Then, DBO creates shares of the vector and sends to servers, which replaces the vector with the existing $i^{\mathit{th}}$ row of the inverted list. Likewise, DBO can fetch $i^{\mathit{th}}$ row by organizing \textsf{OptInv} in matrix form and asks the servers to replace the $i^{\mathit{th}}$  row. While this method is efficient, this method reveals access-patterns to the servers.

Furthermore, there could be another possibility by fetching the $j^{\mathit{th}}$ file to be deleted at DBO and replacing the  $j^{\mathit{th}}$ file at the server with dummy data.

\medskip
\section{Granting or Revoking Access Rights }
\label{app_sec:Access Rights Revocation}

Granting or revoking access rights for a client in AC matrix ({\color{blue}Table~\ref{tab:ACT_table}}) is done by DBO. A trivial way to do this is to download all the keywords and the client's access rights at DBO, then update the access rights, and finally, re-create shares of the keywords and access rights. This method requires $4\beta$ communication cost. Instead, DBO can grant/revoke access rights with communication cost of $2\beta$ using the following method.

First, DBO needs to know the position of the desired keyword and the current access rights for the keyword, and then, needs to change the right for the keyword. 



Recall that to show a no access for a keyword for a client, AC matrix contains a large random number; as mentioned in~{\color{blue}\S\ref{sec:Data Outsourcing}}. To make grant and revoke operations easy, DBO can select such random numbers using 
$\mathcal{PRG}()$ which can take a secret seed, the keyword, and the client-id as input. Now, to grant or revoke access, DBO can work as follows:

{\color{black} To know the index (\textit{i}.\textit{e}., the position of the keyword in AC matrix) and the access rights of the keyword for a client, DBO asks servers to perform Phase~1 without performing multiplication with {\hl{$\mathbb{M}(\mathsf{RN}_z)$}}. DBO learns the output is either zero or a random number. Since DBO has allocated random numbers of each keyword for the client using a function, such as $\mathcal{PRG}($seed, keyword, client$)$, DBO can learn the index of the keyword in AC matrix.

Now, DBO prepares a vector of size equal to the number of keywords filled with all zeros except for the index of the keyword, which is replaced with value $\mathcal{PRG}($seed, keyword, client$)$ (or negative value of $\mathcal{PRG}($seed, keyword, client$)$) for changing from access (or no access) to no access (or access). DBO creates shares of this vector and sends to servers, which perform an add operation with the vector corresponding to the client in AC matrix that updates the access rights for the client.  }

\medskip
\section{Verification Operations at Clients}
\label{sec:verification operation at client}

Three phases of \textsc{Doc}$^\star$ come with three different verification objectives for the client. In a nutshell, \emph{{clients wish to verify the {correctness of the access rights in Phase~1} and to verify the {correctness and completeness (of file-id and files) in Phase~2 and Phase~3}}}. 

\medskip\medskip
\noindent
\textbf{Verification in Phase~1.} Recall that Phase~1 executes at three servers. If one of the servers does not perform computation (\textit{i}.\textit{e}., checking access rights) correctly on AC matrix, the client will always receive random numbers, showing 
no access to search for the keyword. Thus, in Phase~1, the client's objective is to verify that they really have no access to search the keyword,  when receiving random numbers. 
For this, the client executes the computation of Phase~1 at four servers, among them, only one can be malicious. Then, the client interpolates the received vectors in triplets, (as  
$\langle\mathcal{S}_1,\mathcal{S}_2,\mathcal{S}_3\rangle$,
$\langle\mathcal{S}_1,\mathcal{S}_2,\mathcal{S}_4\rangle$,  $\langle\mathcal{S}_1,\mathcal{S}_3,\mathcal{S}_4\rangle$, and  
$\langle\mathcal{S}_2,\mathcal{S}_3,\mathcal{S}_4\rangle$)
and compares the answer for each triplet. If one of the servers has not performed the work correctly, all four values will not be identical. 

\medskip\medskip
\noindent
\textbf{Verification in Phase 2.} The client's objective is to verify that they receive all the correct file-ids, associated with the keyword. To do so, in \textsc{Step 5a}, the client computes the hash function over all the received file-ids identically to DBO (see {\color{blue}Table~\ref{tab:inverted_list}}). If \emph{the servers have returned all the file-ids associated with the keyword, the computed hash digest will match against the received hash digest}.

\medskip\medskip
\noindent
\textbf{Verification in Phase 3.} The client's objective is to verify that they receive the entire file, containing the keyword. To do so, they perform similar processing as for the verification in Phase~2.



\medskip\medskip
\textbf{Security discussion.}
Servers do not learn anything, since verification does not involve any additional steps at the server. 
Clients also learn nothing extra from the hash digest they compute.

\medskip
\section{Additional Experiments}
\label{app_sec:additional exps}

This section evaluates \textsc{Doc}$^{\star}$ on additional criteria, as follows:
\begin{enumerate}[noitemsep,nolistsep,leftmargin=0.01in]
    \item Different implementation of AP list for Phase~3 --- Exp A1.
    \item Processing time for granting or revoking access rights --- Exp A2.
    \item Processing time for adding new file-id(s) to an existing keyword --- Exp A3.
    \item Processing time for deleting an existing keyword --- Exp A4.
    \item Processing time for deleting an existing file --- Exp A5.
    \item Processing time on increasing the number of servers --- Exp A6.
   
\end{enumerate}

\medskip\medskip
\noindent\textbf{Exp A1: Different implementation of AP list.}  This experiment shows the impact of using  different implementations of AP list in Phase~3. The one we developed in~{\color{blue}\S\ref{subsubsec:algo_in_phase3}}, denoted by \emph{reduced-size AP list} in {\color{blue}Figure~\ref{fig:impact of AP list size}}, reveals the number of keywords in a file and their positions in AC matrix to the client, while another method developed in~{\color{blue}Appendix~\ref{app_subsec:phase3 Fully Secure Operation}}, denoted by \emph{large-size AP list} in~{\color{blue}Figure~\ref{fig:impact of AP list size}}, does not reveal anything to the client. Recall that the method for the large-size AP list of~{\color{blue}Appendix~\ref{app_subsec:phase3 Fully Secure Operation}} places 0 or 1, equals to the number of keywords $\beta$ in AC matrix, in each entry of AP list in share form, and this results in space overhead compared to reduced-size AP list. In particular, the large-size AP list at each server required 7.31GB for the database having 5K keywords and 500K files, while the reduced-size AP list at each server took 381MB for the same database.

Due to the less data in the reduced-size AP list, it also performs better compared to the large-size AP list; see {\color{blue}Figure~\ref{fig:impact of AP list size}}. 


\begin{figure}[!h]
\BBB
\begin{center}
\includegraphics[scale=0.6]{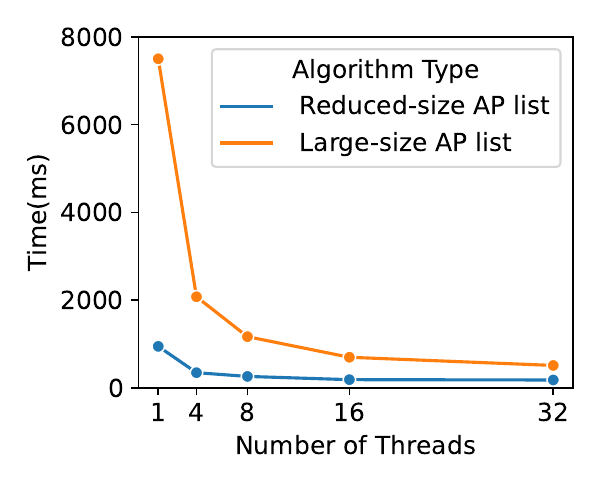}
\end{center}
\BBB\BBB
\caption{Exp A1: Different implementations of the AP list in Phase~3.}
\label{fig:impact of AP list size}
\BB
\end{figure}

\medskip\medskip
\noindent
\textbf{Exp A2: Processing time for granting or revoking access rights.} To grant/revoke access to a keyword to a client, DBO needs to update the row corresponding to the client in AC matrix. To do so, DBO creates a vector of size equals to the number of keywords in AC matrix. Recall that as mentioned in~{\color{blue}Appendix~\ref{app_sec:Access Rights Revocation}}, the vector will contain all zeros except for the index of the keywords, which is replaced by a random number (\textit{i}.\textit{e}., $\mathcal{PRG}($seed, keyword, client$)$) in case of revoking access rights; otherwise 
the negative random number (\textit{i}.\textit{e}., $-\mathcal{PRG}($seed, keyword, client$)$) will appear at the index corresponding to the keyword in case of granting access.

For 5,000 keywords, the entire process took $\approx$25.9ms using a single thread. Particularly, DBO took 18.3ms to prepare the vector with updated access rights, servers took 5ms to update a row of AC matrix, and the communication time between the server and DBO was $\approx$2.6ms.

\medskip\medskip
\noindent
\textbf{Exp A3: Processing time for adding new file-id(s) to an existing keyword.} We performed this experiment only over \textsf{AddrList} and \textsf{OptInv}, since such data structures perform well compared to the inverted list, as shown in Exp~4 in~{\color{blue}\S\ref{sec:experiments}}. To add new file-id(s) to an existing keyword requires updating both \textsf{OptInv} and \textsf{AddrList}. There are two cases depending on the availability of free slots in \textsf{OptInv}; as mentioned in {\color{blue}Appendix~\ref{dynamic_operation:Add Operation}}.

First, when \textsf{OptInv} contains a free slot to accommodate the new file-id(s), and second is one when \textsf{OptInv} has no free slot to accomodate any new file-id. 

In the first case, on a dataset of 5K keywords and 500K files, the total time taken to update \textsf{AddrList} and \textsf{OptInv} with the new file-id(s) for a keyword was 2821.3ms using a single thread.  Specifically, DBO took 1637ms to create vectors, while the server took 173ms to update the existing \textsf{AddrList} and \textsf{OptInv}, and the time taken to transfer the updated vector from DBO to the server was 1011.3ms. 

In the second case, the total time to add new file-id(s) was 18091.6ms using a single thread. Particularly, DBO took 16732ms to create updated information for \textsf{AddrList} and \textsf{OptInv}, whereas the server took 304ms to update these data structures. The network transfer time was 1055.6ms. 
Observe that DBO is taking more time, due to first reconstructing \textsf{AddrList} at their end, then updating the values of SiP, CuT, and hash digests, and finally, recreating the shares of \textsf{AddrList}.

\medskip\medskip
\noindent
\textbf{Exp A4: Processing time for deleting an existing keyword.} The deletion of an existing keyword using the \emph{secure 
 approach} of {\color{blue}Appendix~\ref{subsec:Delete Operation}} requires an update of the capability for all clients in AC matrix. 
For 4,096 clients and 5K keywords, the total time taken to perform the keyword deletion operation was 6620.1ms using a single thread. Specifically, DBO took 3209ms to prepare the vector, a server took 3158ms, and the network transmission time was 253.1ms.  

\medskip\medskip
\noindent
\textbf{Exp A5: Processing time for deleting an existing file.} The deletion of an existing file for a keyword using the \emph{secure method} requires an update of \textsf{OptInv} only to delete the file-id, see~{\color{blue}Appendix~\ref{subsec:Delete Operation}}.
For 5K keywords and 500K files, the overall time taken to delete a file for a keyword was 3117.ms using a single thread. Particularly, DBO took 1761 ms to prepare and share the updated vector, while the server took 276ms to perform addition operation on existing \textsf{OptInv}. The total network time incurred in the deletion operation was 1080ms. 

\medskip\medskip
\noindent
\textbf{Exp A6: Processing time on increasing the number of servers.} Our experiments in {\color{blue}\S\ref{sec:experiments}}
used four servers for achieving the functionality of \textsc{Doc}$^\star$.  This experiment evaluates how  \textsc{Doc}$^\star$ performs when increasing the number of servers from four to eight. To do so, we created two groups/clusters, each having four servers. 

Each cluster stores a subset of \textsc{Doc}$^\star$'s data structures. We divide \textsf{AddrList} and Files equally between the two clusters. 
Based on the last row of \textsf{AddrList} and its SiP and CuT values at the first cluster, \textsf{OptInv} is partitioned appropriately such that the file-id(s) associated with the last keyword is included within the corresponding partition of \textsf{OptInv} in the first cluster. AC matrix remains undivided as required for the operation \textsc{Doc}$^\star$ protocol, to execute Test~5 in Phase~3.

{\color{blue}Table~\ref{tab:Sharding}} shows the result of this experiment. As we increase the number of servers, the processing time decreases, since each cluster of four servers processes only a subset of \textsf{OptInv}, \textsf{AddrList}, and Files.  For simplicity of comparison, we assume both client and server are trusted and hence eliminated the execution of the tests. For a database 500K files and 5K keywords using a single thread, the total processing time decreases from 1045.1ms to 653ms. The breakdown of time taken by each phase is mentioned in {\color{blue}Table~\ref{tab:Sharding}}.

The size of \textsf{AddrList} and Files is reduced by half when using two clusters.  
However, the size of \textsf{OptInv} reduces from 138.6MB in the case of a single cluster to 110.6MB in the case of two clusters. Observe that the size of \textsf{OptInv} does not reduce by half, due to the distribution of keywords appearing in the file.  

\bgroup
\def\arraystretch{1}
\begin{table}[!h]
\begin{tabular}
{|l||l|l|l|l|}\hline
     \# servers& Phase 1 & Phase 2  & Phase 3 & Total   \\\hline\hline
{\cellcolor[HTML]{C1FD8E}{4}} & {\cellcolor[HTML]{C1FD8E}{\textbf{16.6ms}}} & 
{\cellcolor[HTML]{C1FD8E}{\textbf{78.7ms}}} & 
{\cellcolor[HTML]{C1FD8E}{\textbf{949.8ms}}} & 
{\cellcolor[HTML]{C1FD8E}{\textbf{1045.1ms}}}  \\\hline

{\cellcolor[HTML]{FFE4C4}{8}} & {\cellcolor[HTML]{FFE4C4}{\textbf{16.4ms}}} & 
{\cellcolor[HTML]{FFE4C4}{\textbf{63.9ms}}} & 
{\cellcolor[HTML]{FFE4C4}{\textbf{574.5ms}}} & 
{\cellcolor[HTML]{FFE4C4}{\textbf{654.8ms}}}  \\\hline
\end{tabular}
\caption{Exp A6: The impact of increasing the number of servers.}
\BBB\BB
\label{tab:Sharding}

\end{table}
\egroup

\section{Security Proofs}
\label{app_sec:security_proofs}

This section provides the security of  {\hl{${\mathsf{RN}}$}}, security of randomization, and formal security proof.

\medskip
\subsection{Verification of Random number Generation}
\label{app_sec:Verification of Random number Generation}

According to the description of distributed secret-shared random number generation in~{\color{blue}\S\ref{subsec:Random Number Generation and Verification}}, we can see that $\mathsf{RN}$ is produced by adding three random numbers generated by $\mathcal{S}_{z\in\{1,2,3\}}$,
respectively. However, a malicious server can do the following behaviors:
\begin{itemize}
\item Do not send anything to other servers, this malicious can be detected immediately;
\item Sent arbitrary numbers instead of shares to other servers. This is what we should verify.
\end{itemize}

Note that due to incorporating the client-side verification as mentioned in~{\color{blue}\S\ref{subsec:Random Number Generation and Verification}}, the secret-shared random numbers will be generated at four servers, and they will verify the random number. Below, we provide a case where three servers verify random numbers. The following case also generalizes to the four servers. 

According to the Lagrange interpolation ($LI$), to generate $\mathbb{M}(\mathsf{RN})[i])$, no matter which random number is sent by the malicious server, the aggregated shares will produce a polynomial of degree less than two. We classify these results into two cases: (\textit{i}) polynomial of degree one, (we recognize it is legal); and (\textit{ii}) polynomial of degree two.
In this case, the computation of $\sum_{1\leq i \leq \beta}(PRG(seed)\times{\mathbb{M}(\mathsf{RN})[i])}\bmod{p}$ will also produce a polynomial of degree two. Please note that multiplying with $PRG(seed)$ can prevent malicious servers from passing this test, as no server can predict which number will multiply with $\mathbb{M}(\mathsf{RN})[i])$. Before proving Theorem~1, 
let us introduce a lemma.

\medskip\medskip%
\noindent
\textbf{Lemma 1: 
Let $F(x)$ be a polynomial of degree at most two over $\mathbb{F}_p$, $f_1(x)=(x-c_1)(x-c_2)$ and $f_2(x)=(x-c_1)(x-c_3)$, where $c_1,c_2,c_3$ are pairwise different. Then $F(x)\bmod{f_1(x)}\neq F(x)\bmod{f_1(x)}$.}

\begin{proof}
Provide that $F(x)=h_1(x)f_1(x)+r_1(x)$ and $F(x)=h_2(x)f_2(x)+r_2(x)$. Obviously, if $r_1(x)=r_2(x)$, then we can deduce that $h_1(x)f_1(x)=h_2(x)f_2(x)$. Recall that the degree of $F(x)$ is at most two. Thus both $h_1(x)$ and $h_2(x)$ are constants. Without loss of generality, let $h_1(x)=\alpha, h_2(x)=\beta$, we have
$\alpha\cdot(x-c_1)(x-c_2)=\beta\cdot(x-c_1)(x-c_3)$. Note that $c_1,c_2,c_3$ are all different, no matter what $\alpha, \beta$ are, the equation $\alpha\cdot(x-c_2)$ is not equal to $\beta\cdot(x-c_3)$. Thus, the above equation will never hold; this is a contradiction.
\end{proof}

\medskip\medskip
\noindent
\textbf{\emph{Theorem 1: If malicious servers have not created the shares of random numbers correctly, $LI(a_{z},a_{z+1})$,
$LI(a_{z+1},a_{z+2})$, $LI(a_{z+2},a_{z})$, at $\mathcal{S}_z$ will produce different results, where $LI(*)$ means Lagrange interpolation.}} 


\begin{proof}
Let $F(x)$ denote the secret polynomial of $\sum_{1\leq i \leq \beta}(PRG(seed){\times}{\mathbb{M}(\mathsf{RN})[i])}\bmod{p}$. If malicious servers create shares incorrectly, the degree of $F(x)$ will be two. When performing interpolation for every two shares, 
$LI(a_z,a_{z+1}){=}F(x)\bmod{(x-c_1)(x-c_2)}$, $LI(a_{z+1},a_{z+2}){=}F(x)\bmod{(x-c_2)(x-c_3)}$ and $LI(a_{z+2},a_{z}){=}F(x)\bmod{(x-c_3)(x-c_1)}$, 
{\color{black} where $c_1\ldots,c_3$ are the values used in polynomials (\textit{e}.\textit{g}., $f(x{=}c_1){=}(k{\times} c_1+\mathit{secret})\bmod p$) to create Shamir's share}, are pairwise different. According to Lemma~1, 
$LI(a_z,a_{z+1})\neq LI(a_{z+2},a_{z})\neq LI(a_{z+1},a_{z+2})$; thus, we conclude Theorem~1. 
\end{proof}

\medskip
\subsection{Security of Random Numbers}
\label{app_subsec:Security of Random Numbers}
We further prove the security of Phase~1, as it involves an addition step of multiplication of {\hl{$\mathsf{RN}$}}. A similar step is also used in \textsc{Step~B} of server processing over \textsf{OptInv} in~{\color{blue}\S\ref{subsec:Inverted Index Optimization query exection}}. 

\medskip\medskip
\noindent
\textbf{Theorem 2.}
\emph{\textbf{If a malicious client colludes with a minority of malicious servers, such malicious entities cannot deduce any
extra information for the shares they obtained, except for $\mathit{ans}\mathcal{S}_{z}$, in Phase~1.}}

\begin{proof}
\noindent
We first recall the following keyword search formulation in Phase~1:

{\color{myblue}
\begin{equation*}
\begin{split}
\mathbb{M}(\mathit{ans}\mathcal{S}_{z})[i] \leftarrow [&(\mathbb{M}(\mathit{sw})[i]{-}\mathbb{M}(\mathit{uw}) + \mathbb{M}(AC)[i])  \\[-0.7ex] & \times 
{\textnormal{\hl{$\mathbb{M}(\mathsf{RN})[i]]+\mathbb{M}(0)$}}} 
{\bmod} p, \forall i \in \{1,\ldots,\beta\}
\end{split}
\end{equation*}
}
For the $i$-th answer, the malicious client has the share of $\mathit{ans}\mathcal{S}_{z}$, also the malicious server can provide its shares of each operand. 

Without loss of generality, we can recognize the formula $\mathit{sw}[i]{-}\mathit{uw}+AC[i]$ as a integral, denoted by $\theta_i$. Now, $\theta_i$ is secret-shared using a linear polynomial $f(x)=a_1x+\theta_i$. Similarly, {\hl{$\mathsf{RN}[i]$}} is shared by another linear polynomial, denoted by $g(x)=b_1x+$ {\hl{$\mathsf{RN}[i]$}}.
Now, we assume that $\mathcal{S}_1$ is malicious, while $\mathcal{S}_2$ and $\mathcal{S}_3$ are honest. According to SSS, each of the servers keeps the shares of $\theta_i$, {\hl{$\mathsf{RN}[i]$}}, as follows:\\
$\mathcal{S}_1$: $a_1+\theta_i, b_1+$ {\hl{$\mathsf{RN}[i]$}}; for $x=1$,\\
$\mathcal{S}_2$: $2a_1+\theta_i, 2b_1+$ {\hl{$\mathsf{RN}[i]$}}; for $x=2$,\\
$\mathcal{S}_3$: $3a_1+\theta_i, 3b_1+$ {\hl{$\mathsf{RN}[i]$}}, for $x=3$.

The multiplications between shares of $\theta_i$ and {\hl{$\mathsf{RN}[i]$}} corresponds the polynomial multiplication of $f(x)\times g(x)$, \textit{i}.\textit{e}., 
$$f(x)\times g(x)=(a_1x+\theta_i)\times(b_1x+{\textnormal{\hl{$\mathsf{RN}[i]$}}})$$ 
$$= a_1b_1x^2+(\theta_i b_1+{\textnormal{\hl{$\mathsf{RN}[i]$}}} a_1)x+\theta_i{\textnormal{\hl{$\mathsf{RN}[i]$}}}$$
Also note that this product should be add with an extra polynomial $h(x)=h_1x^2+h_2x$, which corresponds to the secret polynomial of degree two for $\mathbb{M}(0)$.

The malicious client can obtain all the shares stored at the malicious server $\mathcal{S}_1$ and coefficients of $f(x)\times g(x)+h(x)$. Here, the malicious client wants to recover unknown $\theta_i$, which may reveal the keyword. Then, malicious client has to solve a non-linear equation set:
\begin{equation}\label{eq:nonlin}
\begin{split}
&a_1+\theta_i  = v_1\\
&b_1+{\textnormal{\hl{$\mathsf{RN}[i]$}}} = v_2\\
&a_1b_1+h_1=v_3\\
&\theta_i\times b_1+{\textnormal{\hl{$\mathsf{RN}[i]$}}}\times a_1+h_2=v_4\\
&\theta_i\times{\textnormal{\hl{$\mathsf{RN}[i]$}}}=v_5\\
&h_1+h_2=v_6
\end{split}
\end{equation}
The parameters $\{a_1,b_1, \theta_i,{\textnormal{\hl{$\mathsf{RN}[i]$}}}, h_1, h_2\}$ are all unknown. When we plug the first two equations into the remaining one, we can obtain
\begin{equation*}
\begin{split}
&a_1b_1+h_1=v_3\\
&(v_1-a_1)\times b_1+(v_2-b_1)\times a_1+h_2=v_4\\
&(v_1-a_1)\times (v_2-b_1)=v_5\\
& h_1+h_2=v_6
\end{split}
\end{equation*}
Further, simplifying the above three equations results in an equation without any unknowns $v_1v_2+v_6=v_3+v_4+v_5$.
For the next query, as servers choose different polynomials for $\mathbb{M}(0)$ to randomize the secret polynomial, even malicious client searches the same $uw$, the client obtains similar equations with the different parameter set $\{h'_1, h'_2, b'_1, \mathsf{RN}'[i]\}$ ($a_1, \theta_i$ is unchanged), which are independent of the previous $\{h_1, h_2, b_1, \mathsf{RN}[i]\}$. Therefore, new equations cannot help the client knowing more about $\theta_i$.
That is to say, the malicious client cannot deduce $\theta_i$ according to the  equation set (\ref{eq:nonlin}), which concludes the theorem. $\blacksquare$
\end{proof}

\medskip
\subsection{Security of Randomization}
\label{app_subsec:Security of Randomization}
We now consider another kind of malicious behavior of the client. 
Assume that server $\mathcal{S}_1$ is malicious. 
Note that $\mathbb{M}(\mathit{sw}[i])$ and $\mathbb{M}(AC)[i]$ are fixed. If a \emph{malicious client generates $\mathbb{M}(\mathit{uw})$ using the same secret polynomial for multiple queries}, the secret polynomial related to $\mathbb{M}(\mathit{sw})[i]{-}\mathbb{M}(\mathit{uw}) + \mathbb{M}(AC)[i]$ is fixed, and denote such a polynomial by $f(x)=a_1x+\theta_i$. Meanwhile, the polynomials related to $\mathbb{M}(\mathsf{RN})[i]$ for two queries are chosen as $b_1x+\mathsf{RN}[i]$ and $b_2x+\mathsf{RN}'[i]$. 
Therefore, when the client collects and interpolates shares from all the servers, they will obtain:
$$
\begin{array}{rl}
F_1(x)\!\!\!\!\!&\!=a_1 b_1 {x}^{2}+(a_1\mathsf{RN}[i]+ b_1\theta_i)x+ \mathsf{RN}[i]\theta_i\\
\!&\!= a_1b_1(x+\theta_i a_1^{-1})(x+ \mathsf{RN}[i] \times b_1^{-1}),\\
F_2(x)\!\!\!\!\!&\!=a_1 b_2 {x}^{2}+(a_1\mathsf{RN}'[i]+ b_2\theta_i)x+ \mathsf{RN}'[i]\theta_i \\
\!&\!= a_1b_2(x+\theta_i a_1^{-1})(x+ \mathsf{RN}'[i]\times b_2^{-1}).\\
\end{array}
$$
Then, the client can find the common divisor between $F_1(x)$ and $F_2(x)$ 
and guess the explicit formula of $a_1x+\theta_i$. 
It is easy to know ${(x+\theta a_1^{-1})}$ is one of its divisors. But its constant divisor is hard to guess, as in the finite field $\mathbb{Z}_p$, $a_1b_1$ and $a_1b_2$ can have arbitrary nonzero elements to be their common factors.

In this context, $\mathcal{S}_1$ send its shares, \textit{i}.\textit{e}., $f(1)=a_1+\theta_i$ to the client. The client can now compute $(1+\theta a_1^{-1})$ and compare it with $f(1)$ to determine the explicit formula of $f(x)$.
Therefore, the client can know $\mathit{sw}[i]+\mathit{AC}[i]$. Combined with the result of Test~1 of {\color{blue}\S\ref{subsec:phase2}} (in the case of an illegal vector that produces the value for the non-access right), a malicious client can know $\mathit{sw}[i]$.

If \emph{\textbf{the servers randomize the share product by adding $\mathbb{M}(0)$, such an attack is avoided.}}

\medskip\medskip
\noindent
\textbf{Theorem~3.}
\emph{\textbf{Randomization can prevent malicious clients from deducing $\mathit{sw}[i]+AC[i]$ over multiple queries, even colluding with a minority of malicious servers.}}
\begin{proof}
Since the servers add $\mathbb{M}(0)$ to $(\mathbb{M}(\mathit{sw})-\mathbb{M}(\mathit{uw})+\mathbb{AC}[i])\times \mathbb{M}(\mathsf{RN})[i]$ for each queries, the secret polynomials related previous shares are not identical. 
Let $f(x)=a_1x+\theta_i$ denote its former meaning.
When such polynomials multiply with $b_1x+\mathsf{RN}[i]$, $b_2x+\mathsf{RN}'[i]$, and then add with $h_1x^2+h_2x$, $h_3x^2+h_4x$, respectively, 
the client can obtain two polynomials as:
$$
\begin{array}{rl}
F_1(x)\!\!\!\!\!&\!=(a_1 b_1+h_1) {x}^{2}+(a_1\mathsf{RN}[i]+ b_1\theta_i+h_2)x+ \mathsf{RN}[i]\theta_i\\
F_2(x)\!\!\!\!\!&\!=(a_1 b_2+h_2) {x}^{2}+(a_1\mathsf{RN}'[i]+ b_2\theta_i+h_3)x+ \mathsf{RN}'[i]\theta_i  \\
\end{array}
$$
At this time, there is no common divisor between $F_1(x)$ and $F_2(x)$ and the shares of $\mathcal{S}_1$ can be the output of any these divisors. Thus, malicious client cannot utilize polynomial factorization to obtain extra information.
\end{proof}

\medskip
\subsection{Security Proof of \textsc{Doc}$^\star$}

We prove the
security of \textsc{Doc}$^\star$ by following real-ideal paradigm. For the details of the real-ideal paradigm, interested readers may refer to Chapter~7 of~\cite{DBLP:books/cu/Goldreich2004}, Chapters~1 and~4~of~\cite{DBLP:journals/iacr/Escudero22}, or~\cite{DBLP:journals/iacr/AsharovL11}. 
\begin{itemize}[leftmargin=0.01in,nolistsep,noitemsep]
\item 
In the real world, the adversary corrupts a subset of the servers of cardinality $f$, interacts with the honest servers, and learns the result as per the protocol at the end of the execution of \textsc{Doc}$^\star$. In other words, the real world executes  \textsc{Doc}$^\star$ that reveals the result to the client and reveals the desired results of the tests to malicious servers.

\item 
In the ideal world, an ideal functionality $F_{\mathit{Doc}}$ realizes
\textsc{Doc}$^\star$ with the desired security guarantees. The ideal world consists of the client and ``non-malicious'' servers sending their inputs to a third trusted party that computes the functionality and returns only the desired results to the client, without leaking any additional information. The functionality has access to the access matrix, inverted list, and files. 
\end{itemize}

\medskip\medskip
\noindent
\textbf{Ideal Functionality $\boldsymbol{F_{\mathit{Doc}}}$.} Below, we define the ideal functionality. 

\noindent
\textit{Initialization.} $F_{\mathit{Doc}}$ stores three data structures: an AC matrix, an inverted list, and  file data structure (as explained in~{\color{blue}\S\ref{sec:Data Outsourcing}}).


\noindent
\textit{Query request for asking files having keyword $\mathit{uw}$.} The client sends a request for a keyword $\mathit{uw}$ with their identity. The functionality checks the access right for the keyword. If the client has access to search for the keyword, the functionality returns all the files having the queried keyword to the client. The client does not receive any file containing at least a single keyword to which they do not have permission to search.

\noindent
\textit{Grant and revoke permission.}
If the DBO
wishes to change the access rights to a client, the DBO sends
the client-id along with 0 (if granting accesses) or random
number (for revocation) to the functionality that takes action appropriately. 





\medskip\medskip%
Now, we are ready to define the formal security property.

\begin{theorem}
Let $F_{\mathit{Doc}}$ be the ideal functionality. Let $n$ be the number of servers that are connected over a secure point-to-point network among themselves and with the client. 
Let $\pi$ be an $n$-server protocol for computing $F_{\mathit{Doc}}$. 
Each $i^{\mathit{th}}$ server $\mathcal{S}_i$ holds input data 
and executes the protocol on the inputs. Let $\mathsf{F}\subset \{f_1, ..., f_f\} \subseteq [n]$, be the set of static malicious adversaries. 


The execution of $\pi$ under $\mathsf{F}$ and a probabilistic adversary $Adv$ in the real world 
is denoted by $real_{\pi, \mathsf{F}, Adv}$. 

The execution of $F_{\mathit{Doc}}$ under $\mathsf{F}$ and a probabilistic algorithm of comparable complexity $\mathit{Sim}$ (representing an ideal world adversarial strategy --- comparable with $\mathit{Adv}$) in the ideal world 
is denoted by $ideal_{F_{\mathit{Doc}}, \mathsf{F}, \mathit{Sim}}$. 

We say that $\pi$ $f$-securely computes $F_{\mathit{Doc}}$ under $\mathsf{F}$ of cardinality $f$ if for every probabilistic algorithm $\mathit{Adv}$ (representing a real would adversary strategy), there exists a probabilistic algorithm of comparable complexity $\mathit{Sim}$, the following hold
$$ideal_{F_{\mathit{Doc}}, \mathsf{F}, \mathit{Sim}}\textnormal{ and } real_{\pi, \mathsf{F}, Adv} \textnormal{ are indistinguishable. }$$
\end{theorem}

The above theorem states that the adversary cannot distinguish between the real world and the ideal world.

\medskip
In the ideal world, we define a coordinator or simulator $\mathit{Sim}$ that interacts with the real adversary, \textit{i}.\textit{e}., $\mathsf{F}$ servers, and the functionality. 
The simulator also provides an interface for the adversary to interact with ``virtual'' honest servers that do not have real input data. If the above theorem holds to be true, then the adversary cannot distinguish between its interaction/working in the real world's servers and the ideal world's simulator.

As per the discussion given in~\cite{DBLP:journals/iacr/AsharovL11} on the security of the addition and multiplication operations, which are also involved in \textsc{Doc}$^\star$, over SSS, intuitively, \textsc{Doc}$^\star$ is $f$-private because the values that the $\mathsf{F}$ servers will see during the
computation of each phase are random, \textit{i}.\textit{e}., unrelated to any dataset, and eventually, $\mathit{Sim}$ can generate the output of the tests at $\mathsf{F}$ servers whatever $\mathit{Sim}$ wants, \textit{i}.\textit{e}., unrelated to the random data and computation at $\mathsf{F}$ servers. Now, we need to describe the simulator $\mathit{Sim}$.

\medskip\medskip
\noindent
\textbf{The Simulator, $\boldsymbol{\mathit{Sim}}$.} 
Below, to make proof simple, we first discuss the case, where the client and $n\backslash \mathsf{F}$ servers are honest. 
Here, the job of $\mathit{Sim}$  is to ``simulate'' messages to the adversary that are similar to what the client and
honest servers send to the malicious server in the real world. However, $\mathit{Sim}$ can only interact with the malicious server and $F_{\mathit{Doc}}$. Also, $\mathit{Sim}$ does not know the actual real inputs at the honest server. 



\begin{itemize}[noitemsep,nolistsep,leftmargin=0.01in]
\item 
\textbf{Inputs.}  $\mathit{Sim}$ learns from the functionality $F_{\mathit{Doc}}$ certain information about the query keyword $\mathit{sw}$ and knows the schema of access control matrix, inverted list, and files ---
the number of clients in AC matrix, the number of searchable keywords in AC matrix, the size of each entry in the inverted list, the number of entries in the inverted list, the number of files, the size of files, and the maximum number of files associated with a keyword. 

\item \textbf{Simulation.} 

\begin{itemize}[noitemsep,nolistsep,leftmargin=0.15in]

\item \textbf{Initialization.} 
Without loss of generality, we assume that the server $\mathcal{S}_4$ is a malicious server, \textit{i}.\textit{e}., $\mathsf{F}=\{\mathcal{S}_4\}$, and thus, $f{=}1$. Based on the inputs received from $F_{\mathit{Doc}}$, $\mathit{Sim}$ creates fake shares of the access matrix, inverted list, and files, and such shares are given to $\mathcal{S}_4$. Note that at this time, only the malicious server has data. 


\item
\textbf{Access a file.} On receiving a query for a keyword $\mathit{sw}$ from $F_{\mathit{Doc}}$, $\mathit{Sim}$ executes the three phases. Below, let us see how $\mathit{Sim}$ will execute Phase~1 and Test~A in Phase~2, and the other phases and tests will be executed in the same manner.

In Phase~1, $\mathit{Sim}$ 
receives the query keyword $sw$ and the client identity from $F_{\mathit{Doc}}$. Then, $\mathit{Sim}$ generates $f{=}1$ fake/dummy shares of the query keyword $sw$ and sends them to malicious server $\mathcal{S}_4$. Note that at this time, the remaining three servers do not have any share of the query keyword as well as any share of the data. $\mathcal{S}_4$ executes Phase~1 on fake data of AC matrix and returns fake results to $\mathit{Sim}$. At this time, $\mathcal{S}_4$ is not aware of the  computation at other servers. 
Since $\mathit{Sim}$ knows the query keyword, $\mathit{Sim}$ can execute the query using $F_{\mathit{Doc}}$ and provide the correct answer of Phase~1 to the client, nothing to $\mathcal{S}_4$. 

In Phase~2, $\mathit{Sim}$ can again generate a fake vector for the dot product operation at $\mathcal{S}_4$. This vector can have either all zeros, all ones, or any random numbers. Here, Phase~2 will execute Tests~1,~A,~B,~C. Let us see Test~A, \textit{i}.\textit{e}., $\mathbb{M}(\mathit{test}_A)\leftarrow \sum \mathbb{M}(v)$. Here, $\mathcal{S}_4$ will do addition on the vector and return the result to $\mathit{Sim}$. At this time, for the interpolation of the result of Test~A, $\mathit{Sim}$ needs to provide additional shares, corresponding to ``virtual'' honest servers, to $\mathcal{S}_4$. Here, $\mathit{Sim}$ will create a polynomial of degree $f$ for each of the ``virtual'' honest servers, such that the constant of the polynomial is either 0 or a random number, depending on what $\mathit{Sim}$ wants.

Other steps of Phase~2 and Phase~3 will be simulated in an identical way to Phase~1, and all the tests will also be simulated identically to Test~A.

\end{itemize}
\end{itemize}
In the simulated world, we need to show two things: first, the data at the $\mathsf{F}$ malicious server does not enable the adversary to distinguish between real and fake data/world, and second, the computation output (particularly the tests) does not reveal anything to the adversary.  

In the initialization, the adversary receives $f{=}1$ random values in AC matrix, inverted list, and files. Due to the properties of SSS, the shares of a secret are uniformly distributed at random in the finite field, and any subset of at most $f$ shares does not reveal any information about the secret (see the proof of Claim~1 and Claim~2 in~\cite{DBLP:journals/iacr/AsharovL11}). Thus, the adversary cannot learn any information from the stored data at $\mathsf{F}$ servers.

During the computation, the malicious server $\mathcal{S}_4$ provides either a list of $\beta$ numbers in Phase~1, a row of the inverted list in Phase~2, a file in Phase~3, or the output of all tests, which will include one number or a vector. First note that whatever $\mathcal{S}_4$ will provide as the final answer to each phase does not reveal anything to $\mathcal{S}_4$, as storing only one share, which corresponds to nothing, due to the initialization phase. 
Now, let us discuss the outputs of all the tests, which are revealed to each server. Since $\mathit{Sim}$ knows the answer to the query keyword using $F_{\mathit{Doc}}$, based on that $\mathit{Sim}$ can simulate the entire output of the tests at malicious servers as $\mathit{Sim}$ wants by the following steps: Recall that the malicious servers hold only $f$ shares, which are also known to $\mathit{Sim}$ due to the communication involved in the tests for result interpolation. 
At this time, $\mathit{Sim}$ knows the $f$ shares from $f$ malicious servers and know the output of the tests as per the desire of $\mathit{Sim}$. Thus, $\mathit{Sim}$ can generate shares corresponding to ``virtual'' honest servers and send them to the $f$ malicious servers. On interpolation, the malicious servers will obtain the answer that $\mathit{Sim}$ wants. This is just the outputs that correspond to the outputs of the malicious servers, and nothing else. Thus, based on the computation output, $f$ malicious servers do not learn additional information.

Combining the above two arguments creates a view that is indistinguishable from the real world's protocol execution. $\blacksquare$

\medskip
\section{Hiding Test~1 Output for Incorrect Vector and Degree Reduction}
\label{app_sec:Hiding Test1 Output for Incorrect Vector}

Recall that in Test~1, a malicious client can learn the value of non-access by creating a wrong vector in \textsc{Step~3b}.
 
 {\color{myblue}
{\centerline{\noindent
 Test~1: $\mathbb{M}(\mathit{test}_1) \leftarrow \mathbb{M}(AC)\odot \mathbb{M}(v)
   $
}}}

 If we want to hide non-access value, \textit{i}.\textit{e}., the output of  Test~1, we can modify Test~1 as follows: 

{\centerline{\noindent
 Modified Test~1: 
 $
 [\mathit{DegR}(\mathbb{M}(AC)\odot \mathbb{M}(v))] \times 
\textnormal{ \hl{$\mathbb{M}(\mathsf{RN})$}}
   $.}}
   
Here, $\mathcal{S}_{z}$ first reduces the degree, denoted by $\mathit{DegR}(\ast)$, of the output of the dot product, then multiplies a random number, and finally, sends the output of the test to other servers. Then, 
each server interpolates the values. 
If the vector $\mathbb{M}(v)$ is correct, then the output of modified test~1 will be 0; otherwise, a random number, which will be different from the non-access value.

\medskip\medskip
{\color{black}\noindent
{\large\textbf{Degree reduction.}}
For degree reduction of a value, say $\mathbb{M}(\mathit{obj})$, of degree two  to degree one, we do the following

1. $\mathbb{M}(\mathit{op})\leftarrow \mathbb{M}(\mathit{obj})+\mathbb{M}(\mathsf{RN})$, where $z=1,2,3$ and $\mathbb{M}(\mathsf{RN})$ is of degree one.

2. $\mathcal{S}_2$ and $\mathcal{S}_3$ sends $\mathbb{M}(\mathit{op}_2)$ and  $\mathbb{M}(\mathit{op}_3)$to $\mathcal{S}_1$.

3. $\mathcal{S}_1$ interpolates $op\leftarrow \mathsf{Interpolate}(\mathbb{M}(\mathit{op}_1),\mathbb{M}(\mathit{op}_2),\mathbb{M}(\mathit{op}_3))$. 

4. $\mathcal{S}_1$ creates shares of $op$, denoted by $op'$ using a polynomial of degree one. 

5. $\mathcal{S}_1$ sends shares to $\mathcal{S}_2$ and $\mathcal{S}_3$.

6. $\mathbb{M}(\mathit{obj}_z^{\prime}) \leftarrow \mathbb{M}(\mathit{op}_z^{\prime})-\mathbb{M}(\mathsf{RN})$, where $z=1,2,3$,  $\mathbb{M}(\mathsf{RN})$ is of degree one used in step 1, and $\mathbb{M}(\mathit{obj}^{\prime})$ is of degree one.

}


\begin{thebibliography}{10}
\bibitem{multicloud4}
Multi-cloud mature organizations are 6.3 times more likely to go to market and succeed before their competition. Available at: {tinyurl.com/2ym8xcx7}.
\end{thebibliography}
\end{document}